\documentclass[11pt,a4paper]{article}
\pdfoutput=1
\usepackage{jheppub}
\usepackage[T1]{fontenc}
\usepackage{fix-cm}
\usepackage{lmodern}
\usepackage{amsmath,amssymb,mathtools,mathrsfs,empheq}

\usepackage{tikz}

\usepackage{xcolor}

\newcommand{\sigg}{\Sigma_{\mathfrak{g}}}

\newcommand{\mfrk}{\mathfrak}

\def\slash#1{\rlap{\hbox{$\mskip 1 mu /$}}#1}      % good slash for lower case
\def\Slash#1{\rlap{\hbox{$\mskip 3 mu /$}}#1}      % " upper
\usepackage{tensor}
\newcommand{\ind}[1]{\indices{#1}}

\title{Higher-Derivative Supergravity, AdS$_4$ Holography, and Black Holes}
\author[\rm H]{Nikolay Bobev,}
\author[\rm H]{Anthony M. Charles,}
\author[\rm D]{Kiril Hristov,}
\author[\rm H]{and Valentin Reys}

\affiliation[\rm H]{Instituut voor Theoretische Fysica, KU Leuven, \\
Celestijnenlaan 200D, B-3001 Leuven, Belgium}

\affiliation[\rm D]{Faculty of Physics, Sofia University,\\
J. Bourchier Blvd. 5, 1164 Sofia, Bulgaria}

\affiliation[\rm D]{INRNE, Bulgarian Academy of Sciences,\\
Tsarigradsko Chaussee 72, 1784 Sofia, Bulgaria}

\emailAdd{nikolay.bobev@kuleuven.be}
\emailAdd{anthony.charles@kuleuven.be}
\emailAdd{khristov@phys.uni-sofia.bg}
\emailAdd{valentin.reys@kuleuven.be}

\abstract{We use conformal supergravity techniques to study four-derivative corrections in four-dimensional gauged supergravity. We show that the four-derivative Lagrangian for the propagating degrees of freedom of the $\mathcal{N}=2$ gravity multiplet is determined by two real dimensionless constants. We demonstrate that all solutions of the two-derivative equations of motion in the supergravity theory also solve the four-derivative equations of motion. These results are then applied to explicitly calculate the regularized on-shell action for any asymptotically locally AdS$_4$ solution of the two-derivative equations of motion. The four-derivative terms in the supergravity Lagrangian modify the entropy and other thermodynamic observables for the black hole solutions of the theory. We calculate these corrections explicitly and demonstrate that the quantum statistical relation holds for general stationary black holes in the presence of the four-derivative corrections. Employing an embedding of this supergravity model in M-theory we show how to use supersymmetric localization results in the holographically dual three-dimensional SCFT to determine the unknown coefficients in the four-derivative supergravity action. This in turn leads to new detailed results for the first subleading $N^{\frac{1}{2}}$ correction to the large $N$ partition function of a class of three-dimensional SCFTs on compact Euclidean manifolds. In addition, we calculate explicitly the first subleading correction to the Bekenstein-Hawking entropy of asymptotically AdS$_4$ black holes in M-theory. We also discuss how to add matter multiplets to the supergravity theory in the presence of four-derivative terms and to generalize some of these results to six- and higher-derivative supergravity.}

\begin{document}

\maketitle

%%%%%%%%%%%%%%%%%%%%%%%%
\section{Introduction}
\label{sec:Introduction}
%%%%%%%%%%%%%%%%%%%%%%%%

Supergravity theories that arise in the low-energy limit of string or M-theory provide an accessible arena on which to explore the dynamics of a UV complete theory of quantum gravity. The well-known two-derivative actions of 10d supergravity are corrected by stringy and quantum effects to yield an infinite series of higher-derivative (HD) corrections. Computing these HD corrections can in principle be done by using the microscopic formulation of string theory but in practice it is cumbersome to do this efficiently. In the absence of a first-principle microscopic formulation of M-theory the HD corrections to 11d supergravity are usually determined by exploiting string dualities which makes them even harder to access. In many situations one is interested in string or M-theory reduced on a compact manifold to a lower-dimensional supergravity theory. In favorable circumstances one can systematically describe the HD corrections to the lower-dimensional supergravity theory resulting in explicit Lagrangians that can be applied to the physical system of interest. A well-studied example of this approach is the 4d $\mathcal{N}=2$ compactifications of string or M-theory to asymptotically flat supergravity. Many tools have been developed in this context and have been successfully applied to uncover important aspects of the microscopic dynamics of black holes in string and M-theory, see \cite{Mohaupt:2000mj} for a review.

The AdS/CFT correspondence provides another strong incentive to systematically study HD corrections to supergravity. The classical two-derivative supergravity action allows for the calculations of physical observables in the dual CFT to leading order in the large $N$ and large 't Hooft coupling approximation. There are at least three important reasons to go beyond this leading order approximation. First, it is desirable to test and extend the AdS/CFT duality beyond the two-derivative supergravity approximation where it has been mainly explored. Second, the HD corrections to supergravity offer explicit calculational access to many CFT observables that may be hard to compute by other means. Third, the physics of black holes is modified by the HD corrections which in turn offers a window into the quantum gravity corrections to asymptotically AdS black holes. Given this status quo it is imperative to develop new tools to access the HD corrections to string and M-theory and their reductions to asymptotically AdS backgrounds in lower dimensions.  Our goal in this paper is to make progress in this direction in the context of 4d $\mathcal{N}=2$ gauged supergravity.

In the formalism of conformal supergravity, which we employ in this work, specifying the HD action of a 4d $\mathcal{N}=2$ supergravity theory involves making several choices. One has to specify the matter content of the theory, the prepotential which captures all F-term type contributions to the Lagrangian, as well as the choice of gauging. In general one may also have the freedom to specify additional D-term type supersymmetry invariants. For many 4d $\mathcal{N}=2$ asymptotically flat compactifications of string and M-theory it is known how to determine this data from geometric and topological properties of the internal manifold. Much less is known for asymptotically AdS 4d $\mathcal{N}=2$ compactifications. There are two main reasons for this. First, all explicitly known $\mathcal{N}=2$ AdS$_4$ compactifications of string and M-theory do not have scale separation between the AdS and Kaluza-Klein (KK) scales. This is in contrast to asymptotically flat compactifications of string theory for which typically the internal compact manifold can be made parametrically small, justifying the use of a 4d effective action. This in turn implies that the 4d actions used in AdS$_4$ compactifications of string and M-theory arise from a consistent truncation to a finite subset of the infinite KK modes. Such 4d $\mathcal{N}=2$ consistent truncations are relatively rare and it is not generally understood how and when they arise.  The second reason is that the internal manifolds for $\mathcal{N}=2$ AdS$_4$ compactifications are not Ricci flat which makes it harder to identify the structure of the 4d $\mathcal{N}=2$ supergravity theory in terms of their topological and geometric data. To bypass these difficulties we focus on the so-called minimal 4d $\mathcal{N}=2$ supergravity which describes the gravity multiplet and has been shown to arise as a universal consistent truncation for many $\mathcal{N}=2$ AdS$_4$ compactifications of string and M-theory, see \cite{Gauntlett:2007ma} and references thereof. This theory is also interesting from a holographic perspective where, as emphasized in \cite{Azzurli:2017kxo,Bobev:2017uzs,Bobev:2019zmz}, it captures the universal dynamics of the stress-energy tensor in the planar limit of the dual 3d $\mathcal{N}=2$ SCFT and many of its deformations.

Focusing on minimal 4d $\mathcal{N}=2$ supergravity and using the tools of conformal supergravity, see \cite{Lauria:2020rhc} for a review, we explicitly construct the general form of the four-derivative gauged supergravity Lagrangian. In particular, we show that it is determined by two F-term type superspace invariants given by the square of the Weyl multiplet and the  $\mathbb{T}$-log multiplet discussed in~\cite{Butter:2013lta}. After gauge fixing the conformal and gauge symmetries of the theory one obtains a four-derivative Lagrangian for the propagating degrees of freedom of the supergravity theory. We explicitly construct this Lagrangian and show that it is fully determined up to three dimensionless constants. The two-derivative Lagrangian is controlled by the dimensionless ratio of the AdS$_4$ length scale $L$ and the Newton constant $G_{N}$. In addition, there are two independent dimensionless coefficients, $c_1$ and $c_2$, that determine the four-derivative terms in the Lagrangian. These three dimensionless parameters should be determined by a consistent embedding of the 4d minimal supergravity model in string or M-theory. The general consistency of the HD expansion of the supergravity theory dictates that given such an embedding the coefficients $c_{1,2}$ are parametrically smaller than the dimensionless ratio $L^2/G_N$. As a by-product of our general analysis we also notice the possibility of adding two additional imaginary terms, one at two-derivatives and another one at four-derivatives, to the Euclidean minimal supergravity action that have a topological nature and may have applications to holographic setups that involve breaking of parity.

Equipped with this four-derivative action we proceed to analyze its detailed properties. Interestingly, one can show that all solutions of the equations of motion of the two-derivative minimal supergravity theory also solve the equations of motion derived from the more involved four-derivative Lagrangian. In addition, we show that the number of supercharges preserved by a given two-derivative solution remains the same after including the four-derivative corrections. Motivated by black hole physics and holographic applications we also study in detail the on-shell action of the four-derivative supergravity theory. In order to find finite and well-behaved results we work with smooth Euclidean supergravity solutions and employ holographic renormalization adapted to the four-derivative context. We derive a simple compact formula for the regularized four-derivative on-shell action of any solution of the two-derivative equations of motion that reads
\begin{equation}
	I_\text{on-shell}^{(\text{HD})} = \left[1 + \frac{64 \pi G_N}{L^2}(c_2 - c_1)\right] \frac{\pi L^2}{2 G_N} \mathcal{F}(\mathbb{S}) + 32 \pi^2 c_1 \chi(\mathbb{S})~. 
\label{eq:ionshellintro} 
\end{equation}
For a given Euclidean solution $\mathbb{S}$, $\mathcal{F}(\mathbb{S})$ is determined by the regularized on-shell action of the two-derivative supergravity theory calculated by the standard rules of holographic renormalization and $\chi(\mathbb{S})$ is the regularized Euler number of the four-dimensional Euclidean metric. We show how to apply this formula in detail for a number of well-known solutions of the minimal supergravity theory and determine $\mathcal{F}(\mathbb{S})$ and $\chi(\mathbb{S})$ for each of them. By studying the linearized spectrum of excitations around the AdS$_4$ vacuum of the theory it is also possible to calculate the coefficient of the two-point function of the stress-energy tensor in the dual CFT which takes the simple form
\begin{equation}
	C_T = \frac{ 32 L^2}{\pi G_N} + 2048 (c_2 - c_1)~.
\label{eq:cttintro}
\end{equation}

It is well-known that black hole thermodynamics is modified in the presence of HD corrections. We explore this in the context of our four-derivative supergravity model and calculate explicitly the HD corrections to black hole charges and thermodynamic potentials. In particular, we use the Wald formalism to show that the four-derivative black hole entropy in 4d $\mathcal{N}=2$ minimal supergravity takes the form
\begin{equation}
	S = \left[1 + \frac{64 \pi G_N (c_2 - c_1)}{L^2}\right] \frac{A_H}{4 G_N} - 32 \pi^2 c_1 \chi(H)~,
\label{eq:entropyintro}
\end{equation}
where $A_H$ is the area of the horizon and $\chi(H)$ is its Euler number. We also discuss a number of pertinent questions that relate our results to the recent literature on the weak gravity conjecture and related developments.

The dimensionless coefficients in the four-derivative supergravity Lagrangian are determined by the details of the string theory compactification that leads to the minimal 4d $\mathcal{N}=2$ supergravity as a consistent truncation. Unfortunately, it is technically challenging to perform such a truncation explicitly in the presence of HD corrections to type II or 11d supergravity. For instance, the leading HD correction to 11d supergravity comes at the eight-derivative order and it is not known how to reduce these eight-derivative terms to four-dimensions for $\mathcal{N}=2$ AdS$_4$ compactifications. To break this impasse we employ the holographic dictionary and recent results on supersymmetric localization in 3d SCFTs. For concreteness we focus on 3d SCFTs arising on the world-volume of M2-branes in M-theory. In this context the dimensionless constants that determine the 4d supergravity Lagrangian have the following form
\begin{equation}
\label{eq:constintro}
	\frac{L^2}{2G_N} =  A\,N^{\frac{3}{2}} + a\,N^{\frac{1}{2}}\,, \qquad c_i =  v_i\,\frac{N^{\frac{1}{2}}}{32\pi} \, .
\end{equation}
Here $N$ is the number of M2-branes, $(A,a,v_1,v_2)$ are real numbers of order one, and we have included the leading two terms in the large $N$ approximation. Since the regularized on-shell action of a supergravity solution is mapped to the logarithm of the partition function in the dual SCFT we can combine \eqref{eq:ionshellintro} and \eqref{eq:constintro} above to derive the following holographic prediction for the leading terms of the large $N$ SCFT partition function
\begin{equation}
\label{eq:IAv1v2intro}
	\log Z = \pi\,\mathcal{F}\,\left[ A\,N^{\frac{3}{2}} + (a+v_2)\,N^{\frac{1}{2}}\right] - \pi\,(\mathcal{F} - \chi)\,v_1\,N^{\frac{1}{2}} \, .
\end{equation}
From the perspective of the CFT the quantities $\mathcal{F}$ and $\chi$ are determined by the Euclidean background metric and $U(1)$ R-symmetry gauge field and the constants $(A,a,v_1,v_2)$ contain information about the details of the theory at hand. For certain choices of 3d SCFTs one can derive these quantities by using results from supersymmetric localization. In particular, we show in detail how this can be done for the $\mathcal{N}=6$ ABJM theory holographically dual to M-theory on AdS$_4\times S^7/\mathbb{Z}_k$ as well as for the 3d $\mathcal{N}=4$ SYM theory with one adjoint and $N_f$ fundamental hypermultiplets which is dual to M-theory on AdS$_4\times S^7/\mathbb{Z}_{N_f}$. These two classes of SCFTs are distinguished by the difference between the $\mathbb{Z}_k$ and $\mathbb{Z}_{N_f}$ orbifold actions. To determine the unknown coefficients in \eqref{eq:IAv1v2intro} for these two families of SCFTs we use only supersymmetric localization results for the round $S^3$ partition function and the coefficient $C_T$ of the two-point stress-energy correlator. The holographic result in \eqref{eq:IAv1v2intro} then leads to new predictions for the large $N$ partition function of the SCFT on many compact three-dimensional manifolds. Notably, this includes a number of supersymmetric partition functions like the superconformal index, the topologically twisted index, and the squashed $S^3$ partition function. Moreover, we are able to use \eqref{eq:entropyintro} to calculate explicitly the first subleading correction to the black hole entropy of any asymptotically AdS$_4$ black hole dual to states in these SCFTs, including examples with no supersymmetry like the AdS-Schwarzschild solution.

While the main focus of our work is on four-derivative 4d $\mathcal{N}=2$ minimal supergravity our approach can be extended also to 4d supergravity coupled to matter multiplets as well as to supergravity Lagrangians with six or higher-order derivatives. In Section~\ref{sec:ExtGen} below we discuss these generalizations in some detail and after making several well-justified assumptions we arrive at two intriguing conjectures. First, we derive a formula for the regularized AdS$_4$ on-shell action of 4d $\mathcal{N}=2$ minimal supergravity in the presence of any number of HD corrections. Second, we study the four-derivative $STU$ model of 4d $\mathcal{N}=2$ supergravity and use supersymmetric localization results for the ABJM theory with real mass deformations to derive a conjecture for the four-derivative correction to the prepotential of the theory.

We organize our presentation by starting in the next section with a detailed analysis of the four-derivative corrections to the 4d $\mathcal{N}=2$ minimal supergravity action using the conformal supergravity formalism. We proceed in Section~\ref{sec:onshell} with a discussion of the solutions in this four-derivative theory and how to evaluate their on-shell action using holographic renormalization. We illustrate this with several explicit examples. In Section~\ref{sec:univ} we show how to calculate the spectrum of linearized fluctuations around the AdS$_4$ vacuum solution of the supergravity theory and how to compute the coefficient in the two-point function of the stress-energy tensor in the dual 3d $\mathcal{N}=2$ SCFT. Section~\ref{sec:BHTD} is devoted to the analysis of black hole thermodynamics in our four-derivative theory. In particular, we calculate the leading correction to the Bekenstein-Hawking entropy of any stationary black hole solution in the supergravity theory. In Section~\ref{sec:susyloc} we show how to use our holographic results in conjunction with supersymmetric localization to find the first subleading correction to the supersymmetric partition function on compact closed three-manifolds of two classes of 3d SCFTs arising from M2-branes. In Section~\ref{sec:ExtGen} we study two possible generalizations of our results by adding vector multiplets to the 4d supergravity theory and by showing how to construct supergravity Lagrangians involving six- and higher-order derivative terms. We conclude in Section~\ref{sec:discussions} with a discussion on several possible directions for further study. In the five appendices we present some of the details of the technical calculations that form the backbone of our work. This paper is a longer and extended version of \cite{Bobev:2020egg} and \cite{Bobev:2020zov} where we summarized part of the results discussed below. In addition to deriving in detail all the results announced in \cite{Bobev:2020egg} we also present many new results that can be found in Sections~\ref{sec:theta},~\ref{subsec:linspec}, and \ref{sec:ExtGen}.

%%%%%%%%%%%%%%%%%
\section{Conformal supergravity and higher-derivative actions}
\label{sec:HD}
%%%%%%%%%%%%%%%%%

In this section we review the conformal supergravity formalism and explain how to build the various invariants that enter the higher-derivative action we consider. We work in Euclidean signature for most of the presentation, and briefly comment on Lorentzian signature towards the end of the section. For a review of~$\mathcal{N}=2$ conformal supergravity in four dimensions and Lorentzian signature, we refer the reader to~\cite{Lauria:2020rhc}. The Euclidean version was obtained in~\cite{deWit:2017cle} by means of an off-shell time-like dimensional reduction from five dimensions, and we will follow the conventions of the latter. Most of the technical details are relegated to Appendix~\ref{app:sugra}. Here we simply recall that the Euclidean conformal supergravity theory is obtained by gauging the full superconformal algebra~$\mathrm{SU}^*(4|2)$, which contains the bosonic subgroup~$\mathrm{SO}(5,1) \times \mathrm{SU}(2)_R \times \mathrm{SO}(1,1)_R$. The first factor is the conformal group of four-dimensional Euclidean space, and the last two make up the R-symmetry group (which contains a non-compact factor in Euclidean signature). The gauge fields of the various transformations are gathered in the so-called Weyl multiplet, together with a number of auxiliary fields required for off-shell closure of the superconformal algebra. Upon gauge-fixing the extra superconformal symmetries and eliminating the auxiliary fields, the superconformal theory reduces to the usual Poincar\'{e} supergravity theory. This procedure can only be carried out consistently provided the Weyl multiplet is supplemented by two compensating multiplets in order to fix the gauge degrees of freedom. One of these must be a vector multiplet, and it contains the vector field that will become the graviphoton in the Poincar\'{e} theory. There are three known choices for the other compensator which lead to different formulations of Poincar\'{e} supergravity~\cite{deWit:1982na}. For most of the presentation we will choose a hypermultiplet, although we will also comment about the formulation using a compensating tensor multiplet below and in Appendix~\ref{app:R2}.

Since in this paper we mostly focus on minimal gauged supergravity, we do not consider additional matter multiplets.\footnote{We discuss the addition of vector multiplets to the theory in Section~\ref{sec:ExtGen}.} Thus, the field content of the theory is that of one Weyl multiplet, one vector multiplet, and one hypermultiplet. As mentioned above, this theory is gauge-equivalent to the Poincar\'{e} supergravity theory describing the dynamics of the metric, two gravitini and a graviphoton field, and the gauging will allow for a Euclidean AdS vacuum with a negative cosmological constant. The first step to show this is the construction of the action for the theory, which is greatly simplified in the superconformal formalism. In fact, the formalism also allows for a straightforward construction of higher-derivative invariants, as we now review.

%%%%%%%%%%%%%%%%%%%%%%%%%%%%%%%%%%%%%%%%
\subsection{Superconformal action}
\label{sec:superconf}
%%%%%%%%%%%%%%%%%%%%%%%%%%%%%%%%%%%%%%%%

The starting point to build superconformally invariant actions in~$\mathcal{N}=2$,~$d=4$ supergravity is the following chiral density formula in Euclidean signature,
\begin{equation}
\label{eq:chiral-superspace}
\int d^4x\,d^4\theta\,\mathcal{E}_\pm\,\mathscr{L}_\pm = \int d^4x\,\mathcal{L}_\pm \, ,
\end{equation}
with
\begin{equation}
\label{eq:chiral-density}
e^{-1}\mathcal{L}_\pm = C_\pm\big\vert_{\mathscr{L}_\pm} + \frac1{16}\,(T_{ab}^\pm)^2\,A_\pm\big\vert_{\mathscr{L}_\pm} + \text{fermions} \, .
\end{equation}
Here,~$\mathscr{L}_\pm$ denotes a chiral~$(+)$ or anti-chiral~$(-)$ multiplet with Weyl weight~$w=2$,~$\mathcal{E}_\pm$ is the (anti-)chiral superspace measure, and~$C_\pm\big\vert_{\mathscr{L}_\pm} $,~$A_\pm\big\vert_{\mathscr{L}_\pm} $ denote the highest and lowest components of~$\mathscr{L}_\pm$, respectively. The tensor~$T_{ab}$ is one of the auxiliary fields belonging to the Weyl multiplet, see Appendix~\ref{app:sugra}. We have refrained from displaying the fermionic terms.  The action on the right-hand side of~\eqref{eq:chiral-superspace} is superconformally invariant by construction~\cite{Butter:2011sr}. We now briefly review the explicit construction of various actions based on the (anti-)chiral multiplets available in minimal conformal supergravity.

A first possibility is to consider the compensating vector multiplet. In Euclidean signature, this multiplet is related to the reducible combination of a chiral and an anti-chiral scalar multiplet -- denoted by~$\mathcal{X}_\pm$ -- on which one imposes a supersymmetric constraint. This can be done provided~$\mathcal{X}_\pm$ carry Weyl weight~$w=1$. Using the superconformal multiplet calculus~\cite{deWit:2017cle}, one can square these multiplets to obtain the desired (anti-)chiral multiplets with~$w=2$. The lowest and highest components read
\begin{equation}
A_\pm\big\vert_{\mathcal{X}_\pm^2} = (X_\pm)^2 \, , \quad C_\pm\big\vert_{\mathcal{X}_\pm^2} = 2\,X_\pm\Bigl(2\,\square_c X_\mp + \frac14\,\widehat{F}_{ab}^\pm\,T^{ab\,\pm}\Bigr)+ \frac12\,Y^{ij}Y_{ij} - (\widehat{F}_{ab}^\mp)^2 \, ,
\end{equation}
where~$X_\pm$ are real scalar fields belonging to~$\mathcal{X}_\pm$, the combination~$\widehat{F}_{ab}^\pm \equiv F_{ab}^\pm - \tfrac14 X_\pm T_{ab}^\pm$ contains the gauge field strength and the~$T$-tensor,~$Y_{ij}$ is an~$\mathrm{SU}(2)_R$ triplet of auxiliary fields, and~$\square_c = D^a D_a$ is the superconformal d'Alembertian. We now consider the following superspace integral, 
\begin{equation}
\label{eq:V-super-pm}
\frac12\,\int d^4x\,d^4\theta\,\mathcal{E}_\pm\,(\mathcal{X}_\pm)^2 \equiv  \int d^4x\,\mathcal{L}_{\mathrm{V}\pm} \, .
\end{equation}
The bosonic terms of the Lagrangian densities on the right-hand side are\footnote{To write the result in this form, we make use of the rules for superconformally covariant derivatives given in Appendix~\ref{app:sugra}. This makes explicit the presence of the Ricci scalar and the auxiliary scalar field~$D$.}
\begin{equation}
\begin{split}
\label{eq:V-Lag-pm}
e^{-1}\mathcal{L}_{\mathrm{V}\pm} =&\; 2\,X_\pm X_\mp\,\Bigl(\frac16\,R - D\Bigr) - 2\,\mathcal{D}_\mu X_\pm\mathcal{D}^\mu X_\mp - \frac12\,(\widehat{F}_{ab}^\mp)^2 \\
&+ \frac14\,X_\pm\,\widehat{F}_{ab}^\pm\,T^{ab\,\pm} + \frac14\,Y^{ij}Y_{ij} + \frac{1}{32}\,(X_\pm)^2\,(T_{ab}^\pm)^2 \, .
\end{split}
\end{equation}

Both the chiral and anti-chiral Lagrangian densities~\eqref{eq:V-Lag-pm} are manifestly real and invariant under superconformal transformations. To construct a suitable real Lagrangian, we can therefore consider an arbitrary linear combination,
\begin{equation}
\label{eq:V-Lag-lin-comb}
\mathcal{L}_\mathrm{V} = \alpha\,\mathcal{L}_{\mathrm{V}+} + \beta\,\mathcal{L}_{\mathrm{V}-} \, , \quad \text{with} \quad \alpha,\,\beta \in \mathbb{R} \, .
\end{equation}
This amounts to rescaling the multiplets~$\mathcal{X}_+$ and~$\mathcal{X}_-$ by different factors in~\eqref{eq:V-super-pm}. In the Lorentzian theory, these multiplets are related by complex conjugation~\cite{Lauria:2020rhc} and we must therefore have~$\beta = \bar{\alpha} = \alpha$. We also make this choice in Euclidean signature so that our theory is related to the standard Lorentzian one by a simple Wick rotation, as discussed below in Section~\ref{sec:Lorentzian}. We further fix~$\alpha = 1$ since it can be reabsorbed by a simple redefinition of the fields in~$\mathcal{X}_\pm$, leading to the Lagrangian density\footnote{In Section~\ref{sec:theta}, we discuss a generalization to complex actions where we allow~$\alpha,\,\beta \in \mathbb{C}$ in~\eqref{eq:V-Lag-lin-comb}.\label{foot:theta-tease}}
\begin{equation}
\label{eq:V-Lag}
\mathcal{L}_\mathrm{V} = \mathcal{L}_{\mathrm{V}+} + \mathcal{L}_{\mathrm{V}-} \, .
\end{equation}
At this stage, the first term in~\eqref{eq:V-Lag-pm} appears problematic since the auxiliary field~$D$ acts as a Lagrange multiplier imposing the constraint~$X_+X_- = 0$, which in turn suppresses the Einstein-Hilbert term in the action. However, the fix is well known~\cite{Lauria:2020rhc}: one adds to~$\mathcal{L}_\mathrm{V}$ the Lagrangian density for a hypermultiplet compensator coupled to conformal supergravity~\cite{deWit:2017cle}, 
\begin{equation}
\begin{split}
\label{eq:H-Lag}
e^{-1}\mathcal{L}_\mathrm{H} =&\; \chi_\mathrm{H}\,\Bigl(\frac16\,R + \frac12\,D\Bigr) - \frac12\,\varepsilon^{ij}\Omega_{\alpha\beta}\,\mathcal{D}_\mu A_i{}^\alpha \mathcal{D}^\mu A_j{}^\beta \\
& + g\,\Omega_{\alpha\beta}\,\Bigl(2\,g\,X_+ X_-\,\varepsilon^{ij} A_i{}^\alpha\,t^\beta{}_\gamma\,t^\gamma{}_\delta\,A_j{}^\delta - \frac12\,Y^{ij} A_i{}^\alpha\,t^\beta{}_\gamma\,A_j{}^\gamma\Bigr) \, .
\end{split}
\end{equation}
Here, Greek indices~$\alpha,\beta,\ldots$ are~$\mathrm{Sp}(1) = \mathrm{SU}(2)$ indices,~$\Omega$ is the invariant anti-symmetric tensor for~$\mathrm{Sp}(1)$, and we have included a coupling to the compensating vector multiplet which generates local gauge transformations with coupling constant~$g$ and generators~$t^\alpha{}_\beta$. This will effect the gauging in the Poincar\'{e} theory. We also introduced the usual hyper-K\"{a}hler potential in terms of the scalars in the hypermultiplet,
\begin{equation}
\chi_\mathrm{H} \equiv \tfrac12\,\varepsilon^{ij}\Omega_{\alpha\beta}\,A_i{}^\alpha A_j{}^\beta \, .
\end{equation} 
Adding~\eqref{eq:H-Lag} to~\eqref{eq:V-Lag} shows that the~$D$ field now acts as a Lagrange multiplier relating the combination of vector multiplet scalars~$X_+ X_-$ to~$\chi_\mathrm{H}$. We thus obtain the following consistent Lagrangian density,
\begin{equation}
\label{eq:VH-Lag}
\mathcal{L}_{2\partial} = \mathcal{L}_{\text{V}} + \mathcal{L}_{\text{H}} \, .
\end{equation}
The subscript on the left hand side indicates that it contains terms that are at most second order in derivatives. \\

To build additional superconformal invariants, we can use other (anti-)chiral multiplets of weight~$w=2$. The first is built out of the Weyl multiplet, which in Euclidean signature is related to a reducible combination of a chiral anti-self-dual tensor multiplet~$\mathcal{W}^-_{ab+}$ and an anti-chiral self-dual tensor multiplet~$\mathcal{W}^+_{ab-}$~\cite{deWit:2017cle}. As with the vector multiplet above, one can square each multiplet to construct scalar (anti-)chiral multiplets of weight~$w=2$ as $\mathcal{W}_\pm^2 \equiv (\mathcal{W}_{ab\pm}^\mp)^2$, whose lowest and highest components are given by,
\begin{equation}
\label{eq:W2-comp}
\begin{split}
A_\pm\big\vert_{\mathcal{W}_\pm^2} =&\; (T_{ab}^\mp)^2 \, , \\
C_\pm\big\vert_{\mathcal{W}_\pm^2} =&\; -64\,(\mathcal{R}(M)^\mp_{ab}{}^{cd})^2 - 32\,(R(\mathcal{V})^\mp_{ab}{}^i{}_j)^2 + 16\,T^{ab\,\mp}D_a D^c T_{cb}{\!}^\pm \, , 
\end{split}
\end{equation}
modulo fermions. The curvatures appearing in these expressions are defined in Appendix~\ref{app:sugra}. We now consider the superspace integral,
\begin{equation}
-\frac1{64}\,\int d^4x\,d^4\theta\,\mathcal{E}_+\,\mathcal{W}_+^2 -\frac1{64}\,\int d^4x\,d^4\theta\,\mathcal{E}_-\,\mathcal{W}_-^2 \equiv \int d^4x\,\mathcal{L}_{\mathrm{W}^2} \, .
\end{equation}
Unpacking the bosonic terms in the curvature~$\mathcal{R}(M)_{ab}{}^{cd}$, we have
\begin{equation}
\bigl(\mathcal{R}(M)_{ab}{}^{cd}\bigr)^2 = \bigl(C(\omega,e)_{ab}{}^{cd}\bigr)^2 + 2\,\bigl(R(A)_{ab}\bigr)^2 + 6\,D^2 \, ,
\end{equation}
with
\begin{equation}
C(\omega,e)_{ab}{}^{cd} \equiv R(\omega)_{ab}{}^{cd} - 2\,\delta_{[a}{}^{[c}\,R(\omega,e)_{b]}{}^{d]} + \frac13\,\delta_{[a}{}^{[c}\,\delta_{b]}{}^{d]}\,R(\omega,e) \, .
\end{equation}
Note that this is not (yet) the Weyl tensor, since the curvature~$R(\omega)$ still contains the dilatation gauge field~$b_\mu$ through the spin-connection~$\omega_\mu{}^{ab}$. This will be remedied when gauge-fixing to Poincar\'{e} supergravity. Staying in the superconformal frame for the moment, the Weyl-squared density contains the following bosonic terms,
\begin{equation}
\begin{split}
\label{eq:W2-Lag}
e^{-1}\mathcal{L}_{\mathrm{W}^2} =&\; \bigl(C(\omega)_{ab}{}^{cd}\bigr)^2 + 2\,\bigl(R(A)_{ab}\bigr)^2 + 6\,D^2 + \frac12\,\bigl(R(\mathcal{V})_{ab}{}^i{}_j\bigr)^2 \\
&\; - \frac14\,T^{ab\,-}D_a D^c T_{cb}{\!}^+ - \frac14\,T^{ab\,+}D_a D^c T_{cb}{\!}^- - \frac1{512}\,(T_{ab}^-)^2 (T_{cd}^+)^2 \, .
\end{split}
\end{equation}
This Lagrangian density contains terms with up to four derivatives. Another invariant of the same order in derivatives can be built using the so-called kinetic multiplet~\cite{deWit:1980lyi}. Starting from an arbitrary (anti-)chiral multiplet~$\Phi_\mp$ of weight~$w \neq 0$, one can construct the multiplet~$\ln(\Phi_\mp)$. One then applies four superspace derivatives to this multiplet and constructs an (anti-)chiral multiplet of weight 2, denoted~$\mathbb{T}_\pm$ and with highest component~\cite{Butter:2013lta},
\begin{equation}
\label{eq:T-comp}
\begin{split}
C_\pm\big\vert_{\mathbb{T}_\pm} = w\Bigl[&\,2\,\bigl(R(\omega,e)_{ab}\bigr)^2 - \frac23\,\bigl(R(\omega,e)\bigr)^2 + 2\,\bigl(R(A)_{ab}\bigr)^2 + 6\,D^2 + \bigl(R(\mathcal{V})^\pm_{ab}{}^i{}_j\bigr)^2 \\
& - \frac12\,T^{ab\,\mp}D_a D^c T_{cb}{\!}^\pm - \frac{1}{512}\,(T_{ab}^\mp)^2(T_{cd}^\pm)^2\,\Bigr] - \frac1{16}\,(T_{ab}^\pm)^2 A_\pm\big\vert_{\mathbb{T}_\pm} + \mathcal{D}_a V_\pm^a \, ,
\end{split}
\end{equation}
again suppressing fermions. Here~$V_\pm^a$ is a vector built out of the components of the~$\ln(\Phi_\mp)$ multiplet, whose explicit form will not be needed in what follows. We now construct the superspace integral
\begin{equation}
-\frac1{2w}\,\int d^4x\,d^4\theta\,\mathcal{E}_+\,\mathbb{T}_+{\!} - \frac1{2w}\,\int d^4x\,d^4\theta\,\mathcal{E}_-\,\mathbb{T}_- \equiv \int d^4x\,\mathcal{L}_\mathbb{T} \, ,
\end{equation}
where the bosonic terms of the Lagrangian density on the right-hand side are
\begin{align}
\label{eq:Tlog-Lag}
e^{-1}\mathcal{L}_\mathbb{T} =&\; \frac23\,\bigl(R(\omega,e)\bigr)^2 - 2\,\bigl(R(\omega,e)_{ab}\bigr)^2 - 2\,\bigl(R(A)_{ab}\bigr)^2 - 6\,D^2 - \frac12\,\bigl(R(\mathcal{V})_{ab}{}^i{}_j\bigr)^2 \\
&+ \frac14\,T^{ab\,-}D_a D^c T_{cb}{\!}^+ + \frac14\,T^{ab\,+}D_a D^c T_{cb}{\!}^- + \frac{1}{512}\,(T_{ab}^-)^2(T_{cd}^+)^2 - \frac1{2w}\,\mathcal{D}_a\bigl(V_+^a + V_-^a\bigr) \, . \nonumber
\end{align}
This form of the Lagrangian density makes it clear that the precise choice of~$\Phi_\mp$ does not matter in this construction, since the explicit components associated to the multiplet enter via a total derivative. It is also clear that~$\mathcal{L}_\mathbb{T}$ contains terms with up to four derivatives.\\

An important question is whether the above multiplets exhaust all possible weight 2 (anti-)chiral multiplets we can construct in minimal conformal supergravity. The answer is of course negative in general. For instance, using the superconformal multiplet calculus, we can build additional chiral multiplets with weight 2 of the form~$\Phi'_+\mathcal{X}_+^2$,~$\Phi'_+\mathcal{W}_+^2$ or~$\Phi'_+\mathbb{T}_+$ where~$\Phi'_+$ has weight zero. In minimal supergravity~$\Phi'_+$ can be obtained as a composite of the Weyl, the compensating vector or the~$\mathbb{T}_+$ multiplets. From the weight zero requirement, we see that it can take the following form,
\begin{equation}
\Phi'_+ = \sum_{n_1,n_2\geq 0} a(n_1,n_2)\,\left(\frac{\mathcal{W}_+^2}{\mathcal{X}_+^2}\right)^{n_1}\left(\frac{\mathbb{T}_+}{\mathcal{X}_+^2}\right)^{n_2} \, ,
\end{equation}
for some (possibly zero) real coefficients~$a(n_1,n_2)$. Here we have assumed that~$\Phi'_+$ has a polynomial form in the composite multiplets~$\mathcal{W}_+^2\mathcal{X}_+^{-2}$ and~$\mathbb{T}_+\mathcal{X}_+^{-2}$. Negative powers would result in Lagrangians with fields of the Weyl or~$\mathbb{T}_+$ multiplets in the denominator, which should not be allowed for dynamical fields. In contrast, the fields of the vector multiplet will eventually be fixed by a combination of gauge-fixing and equations of motion, so using~$\mathcal{X}_+$ to compensate for the Weyl weight is a priori allowed. Consider now the~$\Phi'_+\mathcal{X}_+^2$ multiplet. Its highest component~$C_+\big\vert_{\Phi'_+\mathcal{X}_+^2}$ contains terms of the form
\begin{equation}
\label{eq:C-phi-prime}
\begin{split}
\sum_{n_1,n_2\geq 0} a(n_1,n_2)\,\Bigl[&\,(1 - n_1 - n_2)\,\bigl(T_{ab}^-\bigr)^{2n_1}\,\bigl(A_+\vert_{\mathbb{T}_+}\bigr)^{n_2}\,(X_+)^{-2(n_1 + n_2)}\, C_+\big\vert_{\mathcal{X}_+^2} \\
& + n_1\,\bigl(T_{ab}^-\bigr)^{2(n_1 - 1)}\,\bigl(A_+\vert_{\mathbb{T}_+}\bigr)^{n_2}\,(X_+)^{-2(n_1 + n_2 - 1)}\,C_+\big\vert_{\mathcal{W}_+^2} \\[1mm]
& + n_2\,\bigl(T_{ab}^-\bigr)^{2n_1}\,\bigl(A_+\vert_{\mathbb{T}_+}\bigr)^{n_2 - 1}\,(X_+)^{-2(n_1 + n_2 - 1)}\,C_+\big\vert_{\mathbb{T}_+}\Bigr] \, .
\end{split}
\end{equation}
Note that when~$(n_1, n_2) = (0,0)$ we simply recover~$C_+\big\vert_{\mathcal{X}_+^2}$ up to an overall constant~$a(0,0)$, while for~$(n_1, n_2) = (1,0)$ and~$(n_1,n_2) = (0,1)$ we recover~$C_+\big\vert_{\mathcal{W}_+^2}$ and~$C_+\big\vert_{\mathbb{T}_+}$, respectively (also up to overall constants~$a(1,0)$ and~$a(0,1)$). For all other values of~$(n_1,n_2)$, we get non-zero powers of~$(T_{ab}^-)^2$ or~$A_+\big\vert_{\mathbb{T}_+}$ multiplying the highest components of the~$\mathcal{X}_+^2$,~$\mathcal{W}_+^2$ and~$\mathbb{T}_+$ multiplets. The latter contain two or four derivatives, as discussed previously. Anticipating slightly the formulation in the Poincar\'{e} frame detailed in Section~\ref{sec:Poincare}, the~$T$-tensor will eventually be proportional to the graviphoton field strength and thus also increases the order of derivatives. Finally, the lowest component of the~$\mathbb{T}_+$ multiplet is given in terms of the highest component of the~$\ln(\Phi_-)$ multiplet, and therefore contains a term of the form~$C_-\big\vert_{\Phi_-}(A_-\big\vert_{\Phi_-})^{-1}$. For any choice of~$\Phi_-$, it will involve two or more derivatives. By a simple counting, we thus see that the terms in~\eqref{eq:C-phi-prime} are \emph{at least} of order~$2n_1 + 2n_2 + 2$ in derivatives. For~$(n_1,n_2) \geq (1,1)$, this involves at least six derivatives. The same arguments apply to the other bosonic piece~$(T_{ab}^-)^2\,A_+\big\vert_{\Phi'_+\mathcal{X}_+^2}$ of the chiral density formula~\eqref{eq:chiral-density}. The situation becomes even more dire when considering the combinations~$\Phi'_+\,\mathcal{W}_+^2$ or~$\Phi'_+\,\mathbb{T}_+$, where we obtain six-derivative terms already for~$(n_1,n_2) \neq (0,0)$. A way out would be to build~$\Phi'_+$ out of ratios of vector multiplets, which would lead to ratios in~\eqref{eq:C-phi-prime} without derivatives. However, this is not possible in minimal supergravity since we have a single compensating vector multiplet which can only lead to trivial (constant) ratios. 

Even at the four-derivative order, there are additional superconformal invariants besides the Weyl-squared and~$\mathbb{T}$ ones. They are built from tensor multiplets~\cite{deWit:2006gn,Kuzenko:2015jxa,Hegde:2019ioy} and contain terms quadratic in the Ricci scalar~$R$. Although we have not considered tensor multiplets so far, we already mentioned that there exists an alternative formulation of Poincar\'{e} gauged supergravity that makes use of a tensor multiplet compensator instead of the hypermultiplet introduced above~\cite{deWit:1982na}. We discuss this formulation and the associated four-derivative invariant in more details in Appendix~\ref{app:R2}. There, we show that introducing higher-derivative terms for the \emph{compensating} tensor multiplet in the superconformal frame leads to pathologies in the Poincar\'{e} frame.\footnote{There might exist a mechanism to fix these pathologies, but finding it or ruling it out is outside the scope of this paper.} We will therefore not consider such~$R^2$ invariants in what follows. \\

In conclusion, the most general \emph{four-derivative} Lagrangian density that we can write using the minimal conformal supergravity field content under consideration is given by a linear combination of~\eqref{eq:VH-Lag},~\eqref{eq:W2-Lag} and~\eqref{eq:Tlog-Lag},
\begin{equation}
\label{eq:SCHD-Tlog}
\mathcal{L}_{\text{HD}} = \mathcal{L}_{2\partial} + c_1\,\mathcal{L}_{\mathrm{W}^2} + c_2\,\mathcal{L}_{\mathbb{T}} \, , 
\end{equation}
where~$c_1$ and~$c_2$ are arbitrary real constants. We will refer to~\eqref{eq:SCHD-Tlog} as the superconformal higher-derivative (SCHD) Lagrangian, to highlight the fact that it involves all the bosonic fields of conformal supergravity. We will soon discuss how the theory based on the SCHD Lagrangian is gauge-equivalent to a higher-derivative Poincar\'{e} theory, but before doing so we point out that there is a useful rewriting of~\eqref{eq:SCHD-Tlog} based on the following identity, see~\cite{Butter:2013lta}:
\begin{equation}
\label{eq:HD-basis-change}
\mathcal{L}_{\mathrm{W}^2} + \mathcal{L}_{\mathbb{T}} = \mathcal{L}_\mathrm{GB} \, ,
\end{equation}
where we define
\begin{equation}
e^{-1}\,\mathcal{L}_\mathrm{GB} \equiv C(\omega,e)_{ab}{}^{cd}\,C(\omega,e)^{ab}{}_{cd} - 2\,R(\omega,e)^{ab} R(\omega,e)_{ab} + \frac23\,R(\omega,e)^2 \, .
\end{equation}
In the Poincar\'{e} frame where~$b_\mu = 0$, this reduces to the familiar Gauss-Bonnet density. Using~\eqref{eq:HD-basis-change}, we can eliminate the density~\eqref{eq:Tlog-Lag} and write the SCHD Lagrangian as
\begin{equation}
\label{eq:SCHD-chi}
\mathcal{L}_{\text{HD}} = \mathcal{L}_{2\partial} + (c_1 - c_2)\,\mathcal{L}_{\mathrm{W}^2} + c_2\,\mathcal{L}_\mathrm{GB} \, .
\end{equation}
This form of the SCHD Lagrangian will be particularly useful in the Poincar\'{e} frame, since it is known that the Gauss-Bonnet density is topological in four dimensions and therefore will not affect the equations of motion (EoMs) in the gauge-fixed theory. 

%%%%%%%%%%%%%%%%%%%%%%%%%%%%%%%%%%%%%%%%
\subsection{Poincar\'{e} action}
\label{sec:Poincare}
%%%%%%%%%%%%%%%%%%%%%%%%%%%%%%%%%%%%%%%%

We now recall how the SCHD Lagrangian~\eqref{eq:SCHD-chi} reduces to the Poincar\'{e} supergravity Lagrangian upon gauge-fixing the extra superconformal symmetries and eliminating the superconformal fields by putting them on-shell. We begin by fixing the local conformal boost and~$\mathrm{SU}(2)$ R-symmetry transformations by means of the  
\begin{equation}
\label{eq:K-gauge}
\text{K-gauge} \;\; : \quad b_\mu = 0 \, ,
\end{equation}
and the
\begin{equation}
\label{eq:V-gauge}
\text{V-gauge} \;\; : \quad A_i{}^\alpha = \chi_\mathrm{H}^{1/2}\,\delta_i{}^\alpha \, .
\end{equation}
We further gauge-fix the local dilatation symmetry by imposing the
\begin{equation}
\label{eq:D-gauge}
\text{D-gauge} \;\; : \quad \chi_\mathrm{H} = 2\,\kappa^{-2} \, , \quad \text{where} \quad \kappa^2 \equiv 8\pi\,G_N \, ,
\end{equation}
and the local~$\mathrm{SO}(1,1)$ R-symmetry using the
\begin{equation}
\label{eq:A-gauge}
\text{A-gauge} \;\; : \quad X_+ = X_- \equiv X \, .
\end{equation}
In this gauge, the equation of motion (EoM) for the auxiliary triplet~$Y_{ij}$ derived from the higher-derivative Lagrangian~$\mathcal{L}_{\text{HD}}$ fixes
\begin{equation}
\label{eq:Y-EoM}
Y_{ij} = \frac{g}{\kappa^2}\,\varepsilon_{ik}\,t^k{}_j \, ,
\end{equation}
and the EoM for the real scalar field~$X$ is given by,
\begin{equation}
\label{eq:X-EoM}
\square X + X\,\Bigl(\frac16\,R - \frac{2g^2}{\kappa^2} - D + A^\mu A_\mu\Bigr) + \frac1{16}\,\Bigl(F_{ab} - \frac14\,X\,T_{ab}\Bigr)T^{ab} = 0 \, .
\end{equation}
To eliminate the remaining superconformal fields, we now turn to the EoMs for the R-symmetry connections~$\mathcal{V}_\mu{}^i{}_j$ and~$A_\mu$, and for the (anti-)self-dual projections of the~$T$-tensor. From~\eqref{eq:SCHD-chi} they are, respectively,
\begin{align}
\label{eq:dyn-EoM}
0 =&\; \kappa^{-2}\,\bigl(\mathcal{V}_\nu{}^i{}_j + 2\,g\,W_\nu\,t^i{}_j\bigr) + 4\,(c_2 - c_1)\,\mathcal{D}^\mu R(\mathcal{V})_{\mu\nu}{}^i{}_j \, , \nonumber \\
0 =&\; X^2 A_\nu + (c_2 - c_1)\,\Bigl(\nabla^\mu R(A)_{\mu\nu} + \frac1{16}\,T^+_\nu{}^\rho\,\mathcal{D}^\mu T_{\mu\rho}^- - \frac1{16}\,T^-_\nu{}^\rho\,\mathcal{D}^\mu T_{\mu\rho}^+\Bigr) \, , \\
0 =&\; X\Bigl(F_{ab}^\pm - \frac14\,X\,T_{ab}^\pm\Bigr) + (c_2 - c_1)\Bigl(\Pi^{ef}_{\pm\,ab}\Bigl[\mathcal{D}_e\mathcal{D}^c T_{cf}^\mp + \frac12\,T^{\mp c}{}_e R_{cf}\Bigr] - \frac1{128}\,T_{ab}^\pm\,(T_{cd}^\mp)^2\Bigr) \, , \nonumber
\end{align}
where~$\Pi_\pm^{ab}{}_{cd} \equiv \tfrac12\bigl(\delta_{[c}^a\delta_{d]}^b \pm \tfrac12\varepsilon^{ab}{}_{cd}\bigr)$. In contrast to~\eqref{eq:Y-EoM} and~\eqref{eq:X-EoM}, these equations explicitly depend on the higher-derivative couplings. Moreover, the above fields are no longer proper auxiliary fields from the perspective of the SCHD Lagrangian due to the curvature terms in~$\mathcal{L}_{\mathrm{W}^2}$. Note that~$\mathcal{L}_\text{GB}$ does not affect the EoMs, in accordance with the topological nature of the Gauss-Bonnet density in four dimensions. The last field to eliminate is the Weyl multiplet scalar~$D$, whose EoM is 
\begin{equation}
\label{eq:D-EoM}
4\,X^2 - \kappa^{-2} + 12\,(c_2 - c_1)\,D = 0 \, .
\end{equation}
This shows that~$D$ is a Lagrange multiplier in the two-derivative theory, and a proper auxiliary field in the higher-derivative theory. \\

Solving the EoMs~\eqref{eq:X-EoM},~\eqref{eq:dyn-EoM}, and~\eqref{eq:D-EoM} in full generality is a complicated task. We can, however, make the following crucial observation: in the two-derivative theory, i.e. when~$c_1 = c_2 = 0$, the solutions are simply given by
\begin{equation}
\label{eq:2der-aux-sol}
X = \frac{1}{2\kappa} \, , \quad D = \frac16\,R - \frac{2g^2}{\kappa^2} \, , \quad \mathcal{V}_\mu{}^i{}_j = -2\,g\,W_\mu\,t^i{}_j \, , \quad A_\mu = 0 \, , \quad T_{ab} = 8\,\kappa\,F_{ab} \, .
\end{equation}
Using this in~\eqref{eq:VH-Lag} gives the two-derivative piece of the Poincar\'{e} Lagrangian as expected,
\begin{equation}
\label{eq:2der-P}
e^{-1}\mathcal{L}_{2\partial} = \frac{1}{16\pi G_N}\,\Bigl(R - 6\,g^2\kappa^{-2}\Bigr) + \frac12\,F_{\mu\nu}F^{\mu\nu} \, .
\end{equation}
The two-derivative Einstein equations derived from this Lagrangian density are
\begin{equation}
\label{eq:2der-EE}
R_{\mu\nu} - \frac12\,g_{\mu\nu}\,R + 3\,g^2\kappa^{-2}\,g_{\mu\nu} = 4\,\kappa^2\,F_{\mu\rho}^-\,F^{+\rho}{}_\nu \, ,
\end{equation}
and the Maxwell equation and Bianchi identity for the graviphoton amount to
\begin{equation}
\label{eq:2der-MB}
\nabla^\mu F_{\mu\nu}^+ = \nabla^\mu F_{\mu\nu}^- = 0 \, .
\end{equation}
It is now straightforward to check that~\eqref{eq:2der-aux-sol} also solves the higher-derivative EoMs~\eqref{eq:dyn-EoM} and~\eqref{eq:D-EoM} for arbitrary non-zero~$c_1$ and~$c_2$, by making use of~\eqref{eq:2der-EE} and~\eqref{eq:2der-MB}. In other words, it is consistent to eliminate all extra superconformal fields using their two-derivative on-shell values, even in the presence of the higher-derivative couplings!

Following this procedure for the SCHD density given in~\eqref{eq:SCHD-chi} yields the bosonic Lagrangian density for the physical metric and graviphoton fields,\footnote{Here and below, we switch to a more usual convention where the AdS vacuum of the theory has negative constant curvature. This is ensured by the redefinition~$R_{abcd} \rightarrow -R_{abcd}$. We also canonically normalize the graviphoton term by sending~$F_{ab} \rightarrow (2\kappa)^{-1}F_{ab}$. \label{foot:redef}} 
\begin{equation}
\begin{split}
\label{eq:PHD}
e^{-1}\mathcal{L}_\text{PHD} =&\; -\frac{1}{16\pi G_N}\,\Bigl(R + 6\,L^{-2} - \frac14\,F_{\mu\nu}F^{\mu\nu}\Bigr) \\
&\; + (c_1 - c_2)\,\Bigl[\bigl(C_{abcd}\bigr)^2 - L^{-2}\,F_{ab}\,F^{ab} + \frac12\,\bigl(F_{ab}^+\bigr)^2\,\bigl(F_{ab}^-\bigr)^2 \\
&\qquad\qquad\qquad - 4\,F_{ab}^-\,R^{ac}\,F^+_c{}^b + 8\,\bigl(\nabla^a F_{ab}^-\bigr)\,\bigl(\nabla^c F^+_c{}^b\bigr)\,\Bigr] \\
&\; + c_2\,\Bigl[\bigl(R_{abcd}\bigr)^2 - 4\,R^{ab} R_{ab} + R^2\,\Bigr] \, ,
\end{split}
\end{equation} 
where~$L \equiv g^{-1}\kappa$ and~$C_{abcd}$ is the Weyl tensor. We will refer to~\eqref{eq:PHD} as the Poincar\'{e} higher-derivative (PHD) Lagrangian, to emphasize that it is obtained from the SCHD Lagrangian by fixing the superconformal gauges~\eqref{eq:K-gauge},~\eqref{eq:V-gauge},~\eqref{eq:D-gauge} and~\eqref{eq:A-gauge}, and by consistently eliminating the superconformal fields using their two-derivative on-shell values. In fact, we can go even further: the four-derivative corrections to the Einstein equations derived from~\eqref{eq:PHD} read
\begin{equation}
\label{eq:4der-EE}
E_{\mu\nu} = \mathcal{T}_{\mu\nu} \, ,
\end{equation}
with
\begin{equation}
\begin{split}
E_{\mu\nu} =&\; -4\,\Bigl(R_{\mu\rho}R_\nu{}^\rho - \frac14\,R^{\rho\sigma}R_{\rho\sigma}\,g_{\mu\nu}\Bigr) + \frac43\,R\,\Bigl(R_{\mu\nu} - \frac14\,R\,g_{\mu\nu}\Bigr) \\
&\; - 2\,\Bigl(\square R_{\mu\nu} + \frac12\,g_{\mu\nu}\,\square R - 4\,\nabla_\rho\nabla_{(\mu}R_{\nu)}{}^\rho\Bigr) + \frac43\,\Bigl(g_{\mu\nu}\,\square R - \nabla_\mu\nabla_\nu R\Bigr) \, . \\
\mathcal{T}_{\mu\nu} =&\; \frac14\,g_{\mu\nu}\,(F^+)^2(F^-)^2 + F_{\mu\rho}^+ F^+_\nu{}^\rho (F^-)^2 + F_{\mu\rho}^- F^-_\nu{}^\rho (F^+)^2 - 4\,L^{-2}\,F^-_{\mu\rho}F^+_\nu{}^\rho \\
&\; + 4\,F^-_\mu{}^\rho R_{\rho\sigma} F^+_\nu{}^\sigma - 2\,g_{\mu\nu}\,F^-_{\rho\sigma} R^{\rho\lambda} F^+_{\lambda}{}^\sigma - 8\,R_{\rho(\mu} F^-_{\nu)\sigma} F^{+ \rho\sigma} \\
&\; + 2\,\square\bigl(F^-_\mu{}^\rho F^+_{\nu\rho}\bigr) - 8\,\nabla^\rho\nabla_{(\mu}\bigl(F^-_{\nu)}{}^\sigma F^+_{\sigma\rho}\bigr) \, .
\end{split}
\end{equation}
It is a (slightly) tedious but straightforward exercise to check that the two-derivative solutions for the metric and graviphoton also solve~\eqref{eq:4der-EE}. We can proceed analogously with the higher-derivative generalization of the Maxwell equation. The four-derivative equation of motion for the Maxwell field reads
\begin{equation}\label{eq:4derMaxwell}
\nabla^\mu[2L^{-2} F_{\mu\nu} - F^+_{\mu\nu} (F^-_{\rho\sigma})^2 - F^-_{\mu\nu} (F^+_{\rho\sigma})^2 + 3 R_{\mu\rho}F^\rho{}_\nu - R_{\nu\rho} F^\rho{}_\mu - R F_{\mu\nu}] = 0 \, .
\end{equation}
To simplify this equation we can take the two-derivative Einstein equation, multiply it with $F$ and find the relation
\begin{equation}
R_{\mu\rho} F^\rho{}_\nu = - 3 L^{-2} F_{\mu\nu} + \frac14 F^+_{\mu\nu} (F^-_{\rho\sigma})^2 + \frac14 F^-_{\mu\nu} (F^+_{\rho\sigma})^2 \, .
\end{equation}
Using this relation in \eqref{eq:4derMaxwell} we find that the four-derivative Maxwell equation reduces to the two-derivative Maxwell equation, i.e. $\nabla^\mu F_{\mu\nu} = 0$. We have therefore demonstrated that there exists a consistent subset of solutions to the higher-derivative EoMs derived from~\eqref{eq:SCHD-chi}, and that this subset simply consists of the two-derivative on-shell values for all physical and auxiliary fields in the theory. We note that such a phenomenon has also been exhibited in 4d $\mathcal{N}=2$ minimal ungauged supergravity \cite{Charles:2016wjs,Charles:2017dbr} as well as in the non-supersymmetric context, see for instance \cite{Smolic:2013gz}.\\

\begin{figure}
\centering
\begin{tikzpicture}

\begin{scope}[blend group=soft light]
   \fill[red!30!white]   ( 75:1.2) circle (2);
    \fill[green!30!white] (75:0.5) circle (1.2);
	\fill[blue!30!white] (75:0.2) circle (0.5);
	\fill[blue!30!white] (40:2.2) circle (0.5);
 \end{scope}

 \node at ( 80:2.2)    {4-der. solutions};
  \node at (75:1.4)    {2-der.};
\node at (75:1.1)    {solutions};
\node at (75:0.2)  {BPS};
\node at (42:2.3)  {BPS};
\node at (35:2.1) {(?)};

\end{tikzpicture}

\caption{Schematic diagram of the spectrum of solutions of minimal supergravity with arbitrary four-derivative terms. The subset of two-derivative solutions and their supersymmetric properties remain unchanged within the set of all possible four-derivative solutions.}
\label{fig:sol-subset}
\end{figure}
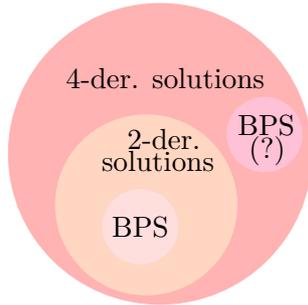

We now come to an analysis of the preserved supersymmetries in the Poincar\'{e} frame. For this, we need one more (fermionic) gauge-fixing obtained as the supersymmetry variation of the D-gauge~\eqref{eq:D-gauge}. This fixes the
\begin{equation}
\label{eq:S-gauge}
\text{S-gauge} \;\; : \quad \zeta^\alpha = 0 \, .
\end{equation}
The supersymmetry variation of the S-gauge then fixes the parameter of the conformal S-supersymmetry~$\eta^i$ in terms of the parameter of the Q-supersymmetry~$\epsilon^i$ as
\begin{equation}
\eta^i = 2\,g\,X\,t^i{}_j\,\epsilon^j \, .
\end{equation}
We can now write down the variation of the gravitino in the Poincar\'{e} frame:
\begin{equation}
\label{eq:E-gravitino-var}
\delta \psi_\mu{\!}^i = 2\,\nabla_\mu\epsilon^i - L^{-1}\,W_\mu\,t^i{}_j\,\epsilon^j + \frac14\mathrm{i}\,F_{ab}\,\gamma^{ab}\gamma_\mu\epsilon^i - \mathrm{i}\,L^{-1}\,t^i{}_j\,\gamma_\mu\epsilon^j \, ,
\end{equation}
where~$\nabla_\mu$ contains the spin-connection, and we have used the values~\eqref{eq:2der-aux-sol} for the superconformal fields. This is precisely the two-derivative gravitino variation in minimal gauged supergravity~\cite{Lauria:2020rhc}. Thus, if the two-derivative solution~$(g_{\mu\nu},\,F_{\mu\nu})$ is supersymmetric, so is the four-derivative solution, and it preserves the same amount of supersymmetry. For convenience we present a pictorial summary of the discussion in this section in Figure~\ref{fig:sol-subset}. As a final remark on this construction, observe that all the steps leading to~\eqref{eq:PHD} can be repeated after setting the gauge coupling $g$ to zero, which amounts to sending the cosmological constant $L^{-1}$ to zero. In this limit, we obtain a four-derivative action for \emph{ungauged} Poincar\'{e} supergravity, suitable to study e.g. asymptotically flat supergravity solutions.

%%%%%%%%%%%%%%%%%%%%
%%%%%%%%%%%%%%%%%%%%

%%%%%%%%%%%%%%%%%%%%%%%%%%%%%%%%%%%%%%%%
\subsubsection{Comments on Lorentzian signature}
\label{sec:Lorentzian}
%%%%%%%%%%%%%%%%%%%%%%%%%%%%%%%%%%%%%%%%

The construction of the previous sections made use of Euclidean conformal supergravity. Of course, an analogous construction goes through in Lorentzian signature. We work in mostly plus Lorentzian signature and discuss the minor differences present in the construction when changing the signature from Euclidean to Lorentzian below. The starting point is the Lorentzian version of the chiral density formula~\eqref{eq:chiral-density} for the Lorentzian vector multiplet~$\mathcal{X}$,
\begin{equation}
\frac12\,\int d^4 x\,d^4\theta\,\mathcal{E}\,\mathcal{X}^2 \; + \; \text{h.c.} = \int d^4 x\,\mathcal{L}^\mathrm{(L)}_\mathrm{V} \, .
\end{equation}
Note that in Lorentzian the Hermitian conjugate of a chiral multiplet is an anti-chiral multiplet. We add to this the Lagrangian density for the hypermultiplet compensator in Lorentzian signature to obtain the two-derivative Lorentizan Lagrangian density
\begin{equation}
\mathcal{L}^\mathrm{(L)}_{2\partial} = \mathcal{L}^\mathrm{(L)}_\mathrm{V} + \mathcal{L}^\mathrm{(L)}_\mathrm{H} \, .
\end{equation}
This can be gauge-fixed just as before to obtain the Poincar\'{e} action
\begin{equation}
e^{-1}\mathcal{L}^{\mathrm{(L)}}_{2\partial} = \frac{1}{16\pi G_N}\,\Bigl(R + 6\,L^{-2} - \frac14\,F_{\mu\nu}F^{\mu\nu}\Bigr) \, , 
\end{equation}
where now the D-gauge is fixed by imposing~$\chi_\mathrm{H} = -2\,\kappa^{-2}$, and we have implemented the redefinitions mentioned in Footnote~\ref{foot:redef}. Notice that the change of signature is reflected in an overall sign for the two-derivative action, see the two-derivative terms in \eqref{eq:PHD}. The same is true for the four-derivative invariants, and the end result for the PHD Lagrangian in Lorentzian signature is simply
\begin{equation}
\label{eq:Wick-rotation}
\mathcal{L}^{\mathrm{(L)}}_{\text{PHD}} = - \mathcal{L}_{\text{PHD}} \, .
\end{equation}
This simple relation allows us to change signatures in the action at will. For the fermions, things are a bit more subtle due to the change of reality conditions in Euclidean and Lorentzian signatures, and we refer to~\cite{deWit:2017cle} for a more detailed discussion on this point. Here we simply record the Lorentzian gravitino variation in the Poincar\'{e} frame,
\begin{equation}
\label{eq:L-gravitino-var}
\delta\psi_\mu{\!}^i = 2\,\nabla_\mu\epsilon^i - L^{-1}\,W_\mu\,t^i{}_j\,\epsilon^j - \frac14\,F_{ab}\,\gamma^{ab}\gamma_\mu\varepsilon^{ij}\epsilon_j + L^{-1}\,t^i{}_j\,\varepsilon^{jk}\,\gamma_\mu\epsilon_k \, ,
\end{equation}
where the position of the~$\mathrm{SU}(2)_R$ index on the spinors encodes their chirality~\cite{Lauria:2020rhc}.

%%%%%%%%%%%%%%%%%%%%%%%%%%%%%%%%%%%%%%%%
\subsection{Reality conditions and topological terms}
\label{sec:theta}
%%%%%%%%%%%%%%%%%%%%%%%%%%%%%%%%%%%%%%%%

As alluded to in Footnote~\ref{foot:theta-tease}, it is possible to construct an action more general than the PHD action~\eqref{eq:PHD} with the technology presented so far, at the expense of relaxing reality condition requirements. In view of holographic applications, we are interested in evaluating the higher-derivative minimal gauged supergravity action on-shell for a variety of two-derivative solutions. According to the AdS/CFT correspondence these Euclidean on-shell actions are dual to the (logarithm of) partition functions for various three-dimensional theories on curved backgrounds. The latter can in principle be complex quantities, as emphasized in e.g.~\cite{Closset:2012vp,Closset:2012vg}, and so it is also natural to consider \emph{complex} actions in Euclidean signature in order to compare to the SCFT partition functions. \\

To this end we consider the Lagrangian in~\eqref{eq:V-Lag-lin-comb} and now take~$\alpha,\beta \in \mathbb{C}$. Having relaxed the reality condition of the action, it seems that there is a priori no relation between the coefficients~$\alpha$ and~$\beta$. Note however that the (anti-)chiral densities~\eqref{eq:V-Lag-pm} involve the (anti-) self-dual parts of the field strength and the~$T$-tensor in an asymmetric manner. In order to build an action consistent with the~$\mathrm{Sp}(2,\mathbb{R})$ electric-magnetic duality of the theory~\cite{deWit:2017cle}, we thus restrict to~$\beta = \bar{\alpha}$ in what follows.\footnote{We do not dwell here on the subtleties related to electric-magnetic duality on spaces with a boundary, like AdS$_4$.} This leads to the following complex Lagrangian for Euclidean vector multiplets coupled to the Weyl multiplet,
\begin{equation}
\label{eq:V-Lag-complex}
\mathcal{L}_\mathrm{V}{}^\mathbb{C} = \mathrm{Re}(\alpha)\,\bigl(\mathcal{L}_{\mathrm{V}+} + \mathcal{L}_{\mathrm{V}-}\bigr) + \mathrm{i}\,\mathrm{Im}(\alpha)\,\bigl(\mathcal{L}_{\mathrm{V}+} - \mathcal{L}_{\mathrm{V}-}\bigr) \, .
\end{equation}
The real part of~$\alpha$ can be absorbed by simple field redefinitions of the vector multiplets~$\mathcal{X}_\pm$ and the first bracket on the r.h.s. of~\eqref{eq:V-Lag-complex} is just the Lagrangian~$\mathcal{L}_\mathrm{V}$ discussed in Section~\ref{sec:superconf}. Using~\eqref{eq:V-Lag-pm} one finds that the second bracket is given by
\begin{equation}
e^{-1}\bigl(\mathcal{L}_{\mathrm{V}+} - \mathcal{L}_{\mathrm{V}-}\bigr) = \frac12\,F_{ab}\,\widetilde{F}^{ab} \, , \quad \text{with} \quad \widetilde{F}_{ab} = \frac12\,\varepsilon_{abcd}\,F^{cd} \, .
\end{equation}
Thus, the complex Euclidean vector multiplet Lagrangian involves a topological term,
\begin{equation}
\mathcal{L}_\mathrm{V}{}^\mathbb{C} = \mathcal{L}_\mathrm{V} + \mathrm{i}\,\theta\,e\,F_{ab}\,\widetilde{F}^{ab} \, ,
\end{equation}
where we defined the real number~$\theta \equiv \mathrm{Im}(\alpha)/(2\mathrm{Re}(\alpha))$.

An analogous construction involves the Weyl-squared and~$\mathbb{T}$ multiplets. In particular, using~\eqref{eq:W2-comp} and~\eqref{eq:T-comp}, one finds
\begin{equation}
\begin{split}
e^{-1}\bigl(\mathcal{L}_{\mathcal{W}^2_+} - \mathcal{L}_{\mathcal{W}^2_-}\bigr) =&\; C_{abcd}\,\widetilde{C}^{abcd} + 2\,R(A)_{ab}\,\widetilde{R}(A)^{ab} \\
&\; + \frac12\,R(\mathcal{V})_{ab}{}^i{}_j\,\widetilde{R}(\mathcal{V})^{ab}{}_i{}^j + \frac14\,T_{ab}^-\,\widetilde{R}(A)^{ac}\,T^+_c{}^b \, , \\
e^{-1}\bigl(\mathcal{L}_{\mathbb{T}_+} - \mathcal{L}_{\mathbb{T}_-}\bigr) =&\;  \frac12\,R(\mathcal{V})_{ab}{}^i{}_j\,\widetilde{R}(\mathcal{V})^{ab}{}_i{}^j - \frac14\,T_{ab}^-\,\widetilde{R}(A)^{ac}\,T^+_c{}^b + \frac{1}{2w}\mathcal{D}_a(V^a_+ - V^a_-) \, . 
\end{split}
\end{equation}
These densities involve a gravitational topological term, as well as topological terms for the auxiliary~$\mathrm{SU}(2)_R$ and~$\mathrm{SO}(1,1)_R$ gauge fields. \\

Due to their topological nature, none of the above terms affect the EoMs for the superconformal fields. We can therefore proceed to the Poincar\'{e} frame just as in Section~\ref{sec:Poincare}. The end result is the following complex Euclidean four-derivative Lagrangian density,
\begin{equation}
\label{eq:PHD-complex}
\begin{split}
e^{-1}\mathcal{L}_{\text{PHD}}{}^\mathbb{C} =&\; -\frac{1}{16\pi G_N}\Bigl(R + 6\,L^{-2} - \frac14\,F_{ab}\,F^{ab} - \mathrm{i}\,\frac{\theta}{2}\,F_{ab}\,\widetilde{F}^{ab}\Bigr) \\
&\; + (c_1 + c_2)\,e^{-1}\mathcal{L}_{\mathrm{W}^2} + c_2\,e^{-1}\mathcal{L}_{\text{GB}} \\[1mm]
&\; + \mathrm{i}\,c_3\,C_{abcd}\,\widetilde{C}^{abcd} + 2\mathrm{i}\,(c_3 + c_4)\,L^{-2}\,F_{ab}\,\widetilde{F}^{ab} \, ,
\end{split}
\end{equation}
where the real PHD Lagrangian~\eqref{eq:PHD} has been supplemented by two new topological terms, with imaginary coefficients parametrized by the real constants~$\theta$,~$c_3$ and~$c_4$. Note that the new gravitational term is the Pontryagin density~$C_{abcd}\,\widetilde{C}^{abcd} = R_{abcd}\,\widetilde{R}^{abcd}$ since the Weyl tensor is traceless. We will not study this complexified four-derivative Lagrangian further in this work. It will be interesting to understand how it can be used in a holographic setting to calculate the complexified partition function of the dual 3d SCFT on compact Euclidean manifolds, see \cite{Choi:2020baw} for a recent discussion on this in the two-derivative context.

%%%%%%%%%%%%%%%%%
\section{Solutions and the on-shell action}
\label{sec:onshell}
%%%%%%%%%%%%%%%%%

The upshot of the discussion so far is that the full bosonic four-derivative action of the minimal supergravity theory can be obtained by integrating the Lagrangian density $\mathcal{L}_\text{PHD}$ in \eqref{eq:PHD}. The action then reads
\begin{equation}
	S_{\text{HD}} = S_{2\partial} + (c_1 - c_2)\,S_{\text{W}^2} + c_2\,S_\text{GB}~,
\label{eq:ihd}
\end{equation}
for some real constants $c_1$ and $c_2$, where the two-derivative action $S_{2\partial}$ is the familiar Einstein-Maxwell action
\begin{equation}
	S_{2\partial} = -\frac{1}{16\pi G_N}\int d^4x\,\sqrt{g}\,\left(R - \frac{1}{4} F_{\mu\nu} F^{\mu\nu} + \frac{6}{L^2}\right)~,
\label{eq:i2deriv}
\end{equation}
the four-derivative action $S_{\text{W}^2}$ is given by
\begin{equation}\begin{aligned}
	S_{\text{W}^2} &= \int d^4x\,\sqrt{g}\,\left( C_{\mu\nu\rho\sigma}C^{\mu\nu\rho\sigma} - \frac{1}{L^2} F_{\mu\nu} F^{\mu\nu} - \frac{1}{8} \left(F_{\mu\nu} F^{\mu\nu}\right)^2 + \frac{1}{2} F_{\mu\nu} F^{\nu\rho}F_{\rho\sigma}F^{\sigma\mu}\right. \\
	&\quad\quad\quad\quad\quad\quad \left. - 2 R\ind{_\mu^\nu}F^{\mu\rho}F_{\nu\rho} + \frac{1}{2} R F_{\mu\nu}F^{\mu\nu} + 2 (\nabla^\mu F_{\mu\rho})(\nabla_\nu F^{\nu\rho}) \right)~,
\label{eq:iw2}
\end{aligned}\end{equation}
and the action $S_\text{GB}$ is simply the integrated Gauss-Bonnet density
\begin{equation}
	S_\text{GB} = \int d^4x \, \sqrt{g}\,\left(R_{\mu\nu\rho\sigma}R^{\mu\nu\rho\sigma} - 4 R_{\mu\nu}R^{\mu\nu} + R^2 \right)~.
\label{eq:igb}
\end{equation}
As already established in Section~\ref{sec:Poincare}, any solution to the two-derivative equations of motion derived from $S_{2\partial}$ is also a solution to the equations of motion for the full higher-derivative action $S_\text{HD}$. Our goal here is to evaluate the action $S_\text{HD}$ on-shell for arbitrary two-derivative solutions. In the following, we will use $\mathbb{S} = (g_{\mu\nu}, W_\mu)$ to denote any two-derivative solution of interest with metric $g_{\mu\nu}$ and graviphoton $W_{\mu}$.  We will consider general asymptotically locally AdS$_4$ solutions and will not assume anything about the amount of supersymmetry they preserve, i.e. our on-shell action discussion applies also for non-supersymmetric solutions.

%%%%%%%%%%%%%%%%%%%%%
\subsection{Holographic renormalization}
\label{sec:holorenorm}
%%%%%%%%%%%%%%%%%%%%%

We first consider the two-derivative action $S_{2\partial}$ in \eqref{eq:i2deriv} and note that it is generically divergent when evaluated on an asymptotically AdS$_4$ solution $\mathbb{S}$.  It is well known how to regulate this divergence, see \cite{Skenderis:2002wp} for a review. We have to put a boundary cut-off surface on the spacetime at some finite radial coordinate, add the appropriate boundary counterterms to the action, and then send the radial coordinate to infinity.  In particular, the counterterms needed to regularize the two-derivative action are the usual Gibbons-Hawking, curvature, and cosmological constant terms given by the boundary action, see for instance \cite{Chamblin:1998pz},
\begin{equation}\label{eq:2derSct}
	S_{2\partial}^{\text{CT}} = \frac{1}{8\pi G_N} \int d^3x\,\sqrt{h}\,\left(-K + \frac{L}{2} \mathcal{R} + \frac{2}{L}\right)~,
\end{equation}
where $h_{ab}$ is the induced metric on the conformal boundary, $K_{ab}$ is the extrinsic curvature, and $\mathcal{R}_{abcd}$ is the Riemann tensor of this induced metric.  The Gibbons-Hawking term is required not only to remove divergences in the Einstein-Hilbert action but also to have a well-posed variational principle for the action.  The second term in \eqref{eq:2derSct} is associated with the curvature of the boundary, and the third term is simply the standard cosmological constant term.  All three pieces in the counterterm action \eqref{eq:2derSct} are in general divergent, but they can also yield finite contributions to the action in the limit where the radial cut-off goes to infinity.\footnote{An alternate way of regularizing the two-derivative action is to use the Gauss-Bonnet density as a counterterm, as discussed in~\cite{Miskovic:2009bm}.}

The regularized two-derivative bosonic on-shell action is then simply the sum of the two-derivative action and the counterterm action, i.e. $I_{\text{on-shell}}^{(2\partial)} = S_{2\partial} + S_{2\partial}^{\text{CT}}$.  We will denote this two-derivative on-shell action by
\begin{equation}\label{eq:I2deronshell}
	I_{\text{on-shell}}^{(2\partial)} \equiv \frac{\pi L^2}{2 G_N}\,\mathcal{F}(\mathbb{S})~,
\end{equation}
where $\mathcal{F}$ is a quantity that depends on the two-derivative solution $\mathbb{S}$.

The Gauss-Bonnet term \eqref{eq:igb} is similarly divergent, and it can be regularized via the counterterm
\begin{equation}\label{eq:4derSct}
	S^\text{CT}_{\text{GB}} = 4 \int d^3x\,\sqrt{h}\,\left( \mathcal{J} - 2 \mathcal{G}_{ab} K^{ab}\right)~,
\end{equation}
where $\mathcal{J}$ is the trace of the tensor $\mathcal{J}_{ab}$, defined by
\begin{equation}
	\mathcal{J}_{ab} = \frac{1}{3} \left(2 K K_{ac}K\ind{^c_b} + K_{ab} K_{cd}K^{cd} - 2 K_{ac}K^{cd}K_{db} - K^2 K_{ab}\right)~,
\label{eq:jtens}
\end{equation}
and $\mathcal{G}_{ab} = \mathcal{R}_{ab} - \frac{1}{2} h_{ab} \mathcal{R}$ is the boundary Einstein tensor.  This counterterm is also sometimes referred to as the Gibbons-Hawking-Myers counterterm, as it is precisely the boundary term needed to make the variational problem associated with the Gauss-Bonnet action well posed \cite{Myers:1987yn}.  In contrast to the two-derivative counterterm, where extra counterterms beyond the Gibbons-Hawking term are required by holographic renormalization, the Gibbons-Hawking-Myers term in \eqref{eq:4derSct} is the full boundary counterterm.

The regularized Gauss-Bonnet on-shell action is then given by $I_{\text{on-shell}}^{(\text{GB})} = S_{\text{GB}} + S^\text{CT}_{\text{GB}}$.  The Gauss-Bonnet theorem then immediately tells us that this on-shell action is given entirely in terms of topological data, i.e.
\begin{equation}\label{eq:IGBonshell} 
	I_{\text{on-shell}}^{(\text{GB})} = 32 \pi^2 \chi(\mathbb{S})~,
\end{equation}
where $\chi(\mathbb{S})$ is the Euler characteristic of the full spacetime.

We now move on to consider evaluating the $S_{\text{W}^2}$ action on-shell.  As a first step, we choose judiciously to rewrite the action \eqref{eq:iw2} by eliminating all instances of the Weyl tensor with the Gauss-Bonnet density instead, which leaves us with
\begin{equation}\begin{aligned}
	S_{\text{W}^2} &= S_\text{GB} + \int d^4x\,\sqrt{g}\,\left( 2 R_{\mu\nu}R^{\mu\nu} - \frac{2}{3} R^2 - \frac{1}{L^2} F_{\mu\nu} F^{\mu\nu} - \frac{1}{8} \left(F_{\mu\nu} F^{\mu\nu}\right)^2 \right. \\
	&\quad\left. + \frac{1}{2} F_{\mu\nu} F^{\nu\rho}F_{\rho\sigma}F^{\sigma\mu} - 2 R\ind{_\mu^\nu}F^{\mu\rho}F_{\nu\rho} + \frac{1}{2} R F_{\mu\nu}F^{\mu\nu} + 2 (\nabla^\mu F_{\mu\rho})(\nabla_\nu F^{\nu\rho}) \right)~.
\label{eq:iw2_2}
\end{aligned}\end{equation}
We are interested in evaluating $S_{\text{W}^2}$ on solutions to the two-derivative equations of motion, which allows us to eliminate all instances of the Ricci tensor $R_{\mu\nu}$, the Ricci scalar $R$, as well as any term with derivatives acting on the field strength tensor, e.g. $\nabla^\mu F_{\mu\nu}$.  If we keep the Gauss-Bonnet contribution separated out explicitly, then we find by explicit computation that all terms that are quartic in the field strength cancel, and thus the on-shell value of $S_{\text{W}^2}$, which we denote as $I_{\text{W}^2}$, can be written as
\begin{equation}
	I_{\text{W}^2} = I_\text{GB} + \int d^4x\,\sqrt{g}\,\left(-\frac{24}{L^4} - \frac{1}{L^2} F_{\mu\nu} F^{\mu\nu} \right)~,
\label{eq:iw2_3}
\end{equation}
where $I_\text{GB}$ is the on-shell value of $S_\text{GB}$ in \eqref{eq:igb}.

Finally, we note that the two-derivative equations of motion set $R = - \frac{12}{L^2}$, which allows us to rewrite the result \eqref{eq:iw2_3} as
\begin{equation}
		I_{\text{W}^2} = I_\text{GB} + \frac{4}{L^2}\int d^4x\,\sqrt{g}\left(R - \frac{1}{4} F_{\mu\nu} F^{\mu\nu} + \frac{6}{L^2}\right) = I_\text{GB} - \frac{64 \pi G_N}{L^2} I_{2 \partial}~.
\label{eq:w2toothers}
\end{equation}
That is, by a careful reorganization of terms, we find that the $\text{W}^2$ action in \eqref{eq:iw2} coincides on-shell for any two-derivative solution $\mathbb{S}$ with a particular combination of the two-derivative and Gauss-Bonnet actions.  This is a highly non-trivial and somewhat unexpected consequence of supersymmetry, as it relies heavily on the precise combination of four-derivative terms that appear in $I_{\text{W}^2}$ in order to arrive at such a simple final result.  Moreover, we stress that simply adding the Weyl invariant $C_{\mu\nu\rho\sigma}C^{\mu\nu\rho\sigma}$ to a general gravitational action introduces the Bach tensor into the equations of motion, which then generically become difficult to manage.  The supersymmetric completion of the Weyl invariant given in \eqref{eq:iw2}, however, appears to make these difficulties manageable as well as yielding a very simple final on-shell result.

However, we are not yet done; the two-derivative action and the Gauss-Bonnet action both can have divergences when evaluated on-shell, and thus those divergences can in principle be present in our on-shell $\text{W}^2$ action~\eqref{eq:w2toothers}.  However, since we know how to apply holographic renormalization to remove these divergences in these actions, this means that any divergences that arise in the $\text{W}^2$ action can be immediately cancelled by using a combination of the boundary counterterms for the two-derivative and Gauss-Bonnet actions, \eqref{eq:2derSct} and \eqref{eq:4derSct}, respectively. That is, if we define
\begin{equation}
	S^\text{CT}_{\text{W}^2} \equiv S^\text{CT}_{\text{GB}}- \frac{64 \pi G_N}{L^2}\,S_{2\partial}^{\text{CT}}~,
\label{eq:w2ct}
\end{equation}
as a boundary counterterm action, then the full $\text{W}^2$ on-shell action $I_\text{on-shell}^{(\text{W}^2)} = S_{\text{W}^2} + S_{\text{W}^2}^{\text{CT}}$ is finite in the limit where the boundary cut-off surface is pushed off to infinity.  Using the relations \eqref{eq:w2toothers} and \eqref{eq:w2ct}, we then find that this on-shell action takes the simple form:
\begin{equation}
	I_\text{on-shell}^{(\text{W}^2)} = I_\text{on-shell}^{(\text{GB})} - \frac{64 \pi G_N}{L^2} I_\text{on-shell}^{(2\partial)} = 32 \pi^2 \left(\chi(\mathbb{S}) - \mathcal{F}(\mathbb{S})\right)~.
\end{equation}
The full on-shell action for the higher-derivative theory at hand is therefore given by
\begin{equation}
\boxed{
	I_\text{on-shell}^{(\text{HD})} = \left(1 + \frac{64 \pi G_N}{L^2}(c_2 - c_1)\right) \frac{\pi L^2}{2 G_N}\mathcal{F}(\mathbb{S}) + 32 \pi^2 c_1 \chi(\mathbb{S})~. }
\label{eq:ionshell} 
\end{equation}
This final result for the on-shell action is remarkably simple; it can be obtained purely by knowing $\mathcal{F}(\mathbb{S})$, which is proportional to the two-derivative on-shell action, as well as the Euler characteristic $\chi(\mathbb{S})$.  A priori $I_\text{on-shell}^{(\text{HD})} $ could have had much more intricate dependence on the solution $\mathbb{S}$, due to the presence of the four-derivative terms in the action, but instead the only new piece of data that needs to be computed in order to evaluate $I_\text{on-shell}^{(\text{HD})} $ is the Euler characteristic of the metric. 

A crucial point in our analysis is that we have utilized 4d $\mathcal{N}=2$ supersymmetry only when constructing the action in \eqref{eq:ihd} from conformal supergravity.  We have not assumed anything about the solution $\mathbb{S}$ to the two-derivative theory that we are looking at, and in particular we have made no assumptions that the solution preserves any of the supersymmetries of the theory.  Thus, our result \eqref{eq:ionshell} applies for both BPS and non-BPS solutions.  It is only supersymmetry at the level of the action that is responsible for the simplifications we find.  The constants $c_1$ and $c_2$ in the action \eqref{eq:ihd} should be determined by embedding the 4d $\mathcal{N}=2$ supergravity theory in string or M-theory where it should arise as a consistent truncation from 10d or 11d supergravity. An alternative way to determine the precise form of the on-shell action in \eqref{eq:ionshell} is to employ holography and supersymmetric localization results in the dual 3d $\mathcal{N}=2$ SCFT. It was shown in \cite{Bobev:2020egg} and will be discussed further in Section~\ref{sec:susyloc} below how this can be done for SCFTs arising from M2-branes. In \cite{Bobev:2020zov} the same approach was generalized to 3d theories of class $\mathcal{R}$ obtained form wrapped M5-branes. Once the dimensionless constants $(L^2/G_N, c_1,c_2)$ have been determined by one of the methods discussed above then we can use the result for the on-shell action in \eqref{eq:ionshell} to find the leading and subleading on-shell action for all non-supersymmetric solutions of the two-derivative theory, including for instance the 4d AdS-Schwarzschild black hole

It is worth noting that the on-shell action analysis above required the use of the counterterm \eqref{eq:w2ct}, which can be written explicitly as
\begin{equation}
	S_{\text{W}^2}^\text{CT} = \int d^3x\,\sqrt{h}\,\left(4 \mathcal{J} - 8 \mathcal{G}_{ab}K^{ab} + \frac{8}{L^2}K - \frac{4}{L}\mathcal{R} - \frac{16}{L^3} \right)~.
\label{eq:w2ct2}
\end{equation}
To the best of our knowledge this counterterm has not appeared in the literature before and deserves some further scrutiny. Interestingly on all of the explicit solutions we have looked at, the counterterm action $S_{\text{W}^2}^\text{CT}$ yields no contribution to the on-shell action in the limit where the radial cut-off is sent to infinity. Despite this fact, it is important to always include the $S_{\text{W}^2}^\text{CT}$ when performing general holographic renormalization calculations using our four-derivative supergravity model. For instance, in calculations of general $n$-point functions or when studying black hole thermodynamics this counterterm will prove crucial in obtaining the correct results, as we will see explicitly in Section~\ref{sec:qsr}.

%%%%%%%%%%%%%%%%%%%%
\subsection{Examples}
\label{subsec:examples}
%%%%%%%%%%%%%%%%%%%%

To illustrate how the result for the on-shell action in \eqref{eq:ionshell} works for explicit solutions we now present several well-known examples of regular Euclidean solutions of 4d $\mathcal{N}=2$ minimal supergravity and evaluate the quantities $\mathcal{F}$ and $\chi$ which enter in \eqref{eq:ionshell}. Motivated by the fact that holography relates these results for the on-shell action to the free energy of the dual SCFT on various curved manifolds we focus on supersymmetric solutions in the discussion below. This will allow us to connect our supergravity calculations with supersymmetric localization results as discussed in Section~\ref{sec:susyloc}.

%%%%%%%%%%%%%%%%%%%%
\subsubsection{Euclidean AdS$_4$}
%%%%%%%%%%%%%%%%%%%%

The simplest solution of the equations of motion of the two-derivative theory is given by the Euclidean AdS$_4$ background with an $S^3$ boundary. The metric for this solution can be written as
\begin{equation}
	ds^2 = \frac{L^2}{L^2 + r^2} dr^2 + r^2 d\Omega_3^2~,
\end{equation}
where $d\Omega_3^2$ is the metric on the round unit radius $S^3$. The gauge field for this solution vanishes, i.e. $W_\mu = 0$. One can now plug this metric and gauge field into the regularized on-shell actions in \eqref{eq:I2deronshell} and \eqref{eq:IGBonshell} to find
\begin{equation}
	\mathcal{F} = 1~, \qquad\qquad \chi = 1~.
\end{equation}
Note that the boundary topology of the background is important.  If we had chosen Euclidean AdS$_4$ with $S^1 \times S^2$ boundary, i.e. thermal AdS$_4$, the Euler characteristic $\chi$ would be different, namely $\chi=0$, which in turn will modify the four-derivative corrections to the on-shell action in \eqref{eq:ionshell}. 

%%%%%%%%%%%%%%%%%%%%
\subsubsection{Euclidean Romans solution}
\label{subsubsec:EucRomans}
%%%%%%%%%%%%%%%%%%%%

A simple solution of the two-derivative equations of motion is given by a supersymmetric Euclidean version of the dyonic Reissner-Nordstr\"om black hole. We refer to this background as the Euclidean Romans solution, see for instance \cite{Romans:1991nq,Azzurli:2017kxo,Bobev:2020pjk}. The solution takes the explicit form
\begin{equation}\begin{aligned}
	ds^2 &= U(r) d\tau^2 + \frac{dr^2}{U(r)} + r^2 ds_{\sigg}^2~, \\
	U(r) &= \left(\frac{r}{L} + \frac{\kappa L}{2r} \right)^2 - \frac{Q^2}{4r^2}~, \\
	F &= \pm\kappa L\,V_{\sigg} + \frac{Q}{r^2} d\tau \wedge dr~.
\end{aligned}\end{equation}
With $ds^2_{\Sigma_\mathfrak{g}}$ we denote the metric on a constant curvature Riemann surface of genus $\mathfrak{g}$ with normalization chosen such that the curvature $\kappa$ is given by $\kappa = 1$, $\kappa =0$, and $\kappa = -1$ for genus $\mathfrak{g} = 0$, $\mathfrak{g}=1$, and $\mathfrak{g} > 1$, respectively.\footnote{Generalization of this solution to arbitrary, not constant curvature, metrics on the Riemann surface were constructed in \cite{Bobev:2020jlb}.}  Note that supersymmetry requires the magnetic flux $P$ across the Riemann surface to have magnitude $|P| = |\kappa| L$. The electric charge $Q$ on the other hand is a free parameter and is not restricted by supersymmetry. We denote the volume form on the Riemann surface by $V_{\sigg}$, and define $\omega_{\sigg}$ to be the one-form potential for this volume form such that $d \omega_{\sigg} = V_{\sigg}$.  Integrating the volume form yields:
\begin{equation}
	\text{Vol}(\sigg) = \int_{\sigg} V_{\sigg} = 2\pi \eta~, \quad \eta \equiv \begin{dcases} 2| \mfrk{g} - 1| & \mfrk{g} \neq 1 \\ 1 & \mfrk{g} = 1 \end{dcases}~.
\end{equation}
The metric function $U(r)$ has two zeroes $r_\pm$, given by
\begin{equation}
	r_\pm = L \sqrt{-\frac{\kappa}{2} \pm \frac{|Q|}{2L}}~.
\end{equation}
We impose that the outer radius $r_+$ is real in order for the spacetime to cap off at a real value of the coordinate $r$.  Additionally, we need to ensure that $r_+ > 0$ to avoid a naked singularity where the Riemann surface shrinks down to zero size.  We therefore have to demand that
\begin{equation}
	|Q| > \kappa L~.
\end{equation}
This imposes the constraint $|Q| > L$ for $\mathfrak{g}=0$, $|Q| > 0$ for $\mathfrak{g}=1$, while imposing no constraint for a higher-genus Riemann surface. Thus, for $\kappa = 0,1$, we cannot take the $Q \to 0$ limit if we insist on having a non-singular and real metric.

Assuming that the above conditions are satisfied, then as $r \to r_+$, the metric becomes locally $\mathbb{R}^2 \times \sigg$. The $\mathbb{R}^2$ is written in polar coordinates $(r,\tau)$ where the $\tau$ coordinate has periodicity
\begin{equation}
	\beta = \frac{2 \pi L r_+}{|Q|} = \frac{2 \pi L^2}{|Q|}\sqrt{-\frac{\kappa}{2} + \frac{|Q|}{2L}}~.
\label{eq:qtob}
\end{equation}
To compute the on-shell action of this solution we integrate the radial coordinate from $r = r_+$ to a cut-off at $r = r_b$, and the time coordinate from $\tau = 0$ to $\tau = \beta$. Using the counterterms described above and taking the cutoff to infinity we find
\begin{equation}\label{eq:AdSRNFchidef}
	\mathcal{F} = 1-\mathfrak{g}~, \qquad \chi = 2(1-\mathfrak{g})~.
\end{equation}
All dependence on the electric charge $Q$ drops out of the final on-shell action, and so once we fix the genus $\mathfrak{g}$ of the Riemann surface we have a one-parameter family of solutions (labelled by the charge $Q$) all with the same on-shell action. This independence of the on-shell action on $Q$ was discussed in detail in \cite{BenettiGenolini:2019jdz,Bobev:2020pjk} and here we have shown how the result extends in the presence of higher derivative corrections.  We note that using the relation \eqref{eq:qtob} the independence of the on-shell action on $Q$ implies that it is independent of the periodicity of the Euclidean time coordinate $\tau$.  

The independence of the four-derivative on-shell action on $\beta$ has a simple interpretation in the dual 3d CFT. The on-shell action of the Euclidean Romans solutions is dual to the topologically twisted index of a 3d $\mathcal{N}=2$ SCFT on $S^1 \times \Sigma_{\mathfrak{g}}$, see for instance \cite{Benini:2015eyy}. This index takes the schematic form
\begin{equation}\label{eq:ZTIdef}
	Z = \text{Tr}\,(-1)^F e^{-\beta H}~,
\end{equation}
where $F$ is the fermion number, $H$ is the topologically twisted Hamiltonian and $\beta$ is the size of the $S^1$.  Supersymmetry guarantees that only states with zero energy contribute to this index, and so the topologically twisted index is independent of $\beta$.  Therefore the $\beta$-independence of the bulk on-shell action at the four-derivative level is consistent with the dual topologically twisted index.

There is a family of explicit supersymmetric solutions that generalizes the Euclidean Romans solution discussed above. They are referred to as AdS-Taub-Bolt solutions and were discussed in \cite{Toldo:2017qsh}, see also \cite{AlonsoAlberca:2000cs}. Instead of the $S^1\times \Sigma_\mathfrak{g}$ boundary topology of the Euclidean Romans solution these more general backgrounds have a boundary three-manifold $\mathcal{M}_{\mathfrak{g},p}$ which is smooth and has topology $\mathcal{O}(-p) \to \Sigma_{\mathfrak{g}}$. The explicit form of this solution as well as some details on the calculation of its on-shell action are discussed in \cite{Bobev:2020zov}. The end result is that the quantities $(\mathcal{F},\chi)$ take the simple form
\begin{equation}\label{eq:BoltpmFchi}
\mathcal{F} = \frac{4(1-\mathfrak{g})\mp p}{4}\,, \qquad \chi = 2(1-\mathfrak{g})~.
\end{equation}
One can simply plug this result in \eqref{eq:ionshell} to obtain the full four-derivative on-shell action of this solution. The holographic interpretation of this on-shell action is in terms of the free energy of the 3d $\mathcal{N}=2$ SCFT placed on the supersymmetric background studied in \cite{Closset:2018ghr}.

%%%%%%%%%%%%%%%%%%%%
\subsubsection{$U(1) \times U(1)$ squashed sphere}
\label{subsubsec:U1U1sq}
%%%%%%%%%%%%%%%%%%%%

We now move on to a class of solutions for which the boundary has spherical topology with a squashed metric. We start by discussing an Euclidean $\frac{1}{4}$-BPS solution which can be obtained from the Plebanski-Demianski solutions of the Einstein-Maxwell theory, \cite{Plebanski:1976gy}. This solution is holographically dual to a 3d SCFT placed on the squashed $S^3$ background with $U(1) \times U(1)$ invariance studied in \cite{Hama:2011ea}. 

The Euclidean supersymmetric gravity solution of interest is studied in some detail in~\cite{Martelli:2011fu}, see also \cite{AlonsoAlberca:2000cs}, and can be written as
\begin{equation}\begin{aligned}
	ds^2 &= f_1(x,y)^2 dx^2 + f_2(x,y)^2 dy^2 + \frac{(d\Psi + y^2 d\Phi)^2}{f_1(x,y)^2} + \frac{(d\Psi + x^2 d\Phi)^2}{f_2(x,y)^2}~, \\
	W &= \frac{s^2-1}{L(x+y)} (d\Psi - xy d\Phi)~,
\end{aligned}\end{equation}
where we have defined the metric functions $f_1$ and $f_2$ by
\begin{equation}
	f_1(x,y)^2 = L^2 \frac{y^2-x^2}{(x^2-1)(s^2-x^2)}~, \qquad f_2(x,y)^2 = L^2 \frac{y^2-x^2}{(y^2-1)(y^2 - s^2)}~.
\end{equation}
This form of the solution makes the $U(1) \times U(1)$ isometry manifest, but it is somewhat inconvenient to use it to compute the on-shell action since the coordinate ranges of $\Phi$ and $\Psi$ are dependent on one another.  To ameliorate this, we make the change of variables
\begin{equation}
	\Psi = L^2\left(\frac{s \phi_2 - \phi_1}{s^2-1}\right)~, \qquad \Phi = L^2 \left(\frac{s \phi_1 - \phi_2}{s(s^2-1)}\right)~,
\end{equation}
such that the solution takes the form
\begin{equation}\begin{aligned}
	ds^2 &= f_1(x,y)^2 dx^2 + f_2(x,y)^2 dy^2 + L^2\left(\frac{(x^2-1)(y^2-1)}{s^2-1}\right) d\phi_1^2 \\
	&\quad + L^2\left(\frac{(s^2-x^2)(y^2-s^2)}{s^2(s^2-1)}\right) d\phi_2^2~, \\
	W &= \frac{L}{x+y}\left((s^2+xy)d\phi_2 - (1+xy)d\phi_1\right)~.
\end{aligned}\end{equation}
The solution depends on a single real parameter $s \geq 1$. In the limit $s \rightarrow 1$ the gauge field vanishes and we recover the Euclidean AdS$_4$ solution. The coordinate ranges are:
\begin{equation}
	x \in [1,s]~, \quad y \in [s,\infty)~, \quad \phi_1 \in [0,2\pi)~, \quad \phi_2 \in [0,2\pi)~.
\end{equation}
The conformal boundary is approached as $y \to \infty$, and thus in order to evaluate the on-shell action we introduce a cut-off on the $y$ range of integration at a finite value $y_b$. Using the counterterms discussed above and evaluating explicitly this on-shell action we find
\begin{equation}\label{eq:FchiU1U1}
	\mathcal{F} = \frac{(s+1)^2}{4 s}~, \qquad \chi = 1~.
\end{equation}
To obtain the full result for the on-shell action these values should be plugged in the general formula \eqref{eq:ionshell}.

%%%%%%%%%%%%%%%%%%%%
\subsubsection{$SU(2) \times U(1)$ squashed sphere}
%%%%%%%%%%%%%%%%%%%%

There is another supersymmetric Euclidean solution with a squashed $S^3$ conformal boundary with $SU(2) \times U(1)$ isometry. It is holographically dual to a 3d $\mathcal{N}=2$ SCFT placed on the supersymmetric background studied in \cite{Imamura:2011wg}. The Euclidean supergravity solution can be found in~\cite{Martelli:2011fw}, see also \cite{AlonsoAlberca:2000cs} for its Lorentzian counterpart, and in our conventions reads
\begin{equation}\begin{aligned}
	ds^2 &= \frac{r^2 - s^2}{\Omega(r)} dr^2 + (r^2 - s^2) (\sigma_1^2 + \sigma_2^2) + \frac{4 s^2 \Omega(r)}{r^2 - s^2} \sigma_3^2~, \\
	W &= 2 s \left(\frac{r-s}{r+s}\right)\sqrt{\frac{4s^2}{L^2}-1}\,\sigma_3~,
\end{aligned}\end{equation}
where $\sigma_i$ denote the left-invariant Maurer-Cartan one-forms of $SU(2)$, given by
\begin{equation}\begin{aligned}
	\sigma_1 &= \sin\theta \cos\psi\,d\phi-\sin\psi\,d\theta~, \\
	\sigma_2 &= \sin\theta \sin\psi\,d\phi+\cos\psi\,d\theta~, \\
	\sigma_3 &= \cos\theta\,d\phi+d\psi~,
\end{aligned}
\end{equation}
and we have defined the function $\Omega(r)$ in the metric as
\begin{equation}
	\Omega(r) = (r-s)^2\left(1 + \frac{(r-s)(r+3s)}{L^2}\right)~.
\end{equation}
The parameter $s$ has units of length and is chosen such that $0 < s < \frac{L}{2}$. The coordinate $r$ spans the range $s < r <\infty$, with the conformal boundary at $r \to \infty$. In the limit where $s \to \frac{L}{2}$, the gauge field vanishes, the asymptotic $S^3$ boundary is not squashed and we recover the Euclidean AdS$_4$ solution.  The background described above represents a family of $\frac{1}{2}$-BPS solutions labelled by the free parameter $s$.  The on-shell action can be computed using the familiar counterterms and one finds
\begin{equation}
	\mathcal{F} = \frac{4 s^2}{L^2}~, \qquad \chi = 1~.
\end{equation}
As usual the full four-derivative on-shell action is obtained by plugging these quantities in~\eqref{eq:ionshell}.

%%%%%%%%%%%%%%%%%%%%
\subsubsection{AdS-Kerr-Newman black hole} 
\label{subsubsec:AdSKN}
%%%%%%%%%%%%%%%%%%%%

It is also instructive to study a black hole solution of the two-derivative action. We consider the 4d AdS-Kerr-Newman (AdS-KN) black hole solution. The metric in Euclidean signature is given by, see for example~\cite{Caldarelli:1998hg,Cassani:2019mms,Bobev:2019zmz},
\begin{equation}\label{eq:metKN}
ds^2 = \frac{\Delta_r}{V}\Bigl(d\tau + \frac{\alpha}{\Xi}\sin^2\theta d\phi\Bigr)^2 + V\Bigl(\frac{dr^2}{\Delta_r} + \frac{d\theta^2}{\Delta_\theta}\Bigr) + \frac{\Delta_\theta\sin^2\theta}{V}\Bigl(\alpha\,d\tau - \frac{\tilde{r}^2 - \alpha^2}{\Xi}d\phi\Bigr)^2 \, , 
\end{equation}
where
\begin{equation}\label{eq:fnsKN}
\begin{split}
\tilde{r} =&\; r + 2\,m\sinh^2\delta \, , \qquad \Xi = 1 + \frac{\alpha^2}{L^2} \, , \qquad V(r,\theta) = \tilde{r}^2 - \alpha^2\cos^2\theta \, , \\
\Delta_r(r) =&\; r^2 - \alpha^2 - 2\,m\,r + \frac{\tilde{r}^2}{L^2}\,(\tilde{r}^2 - \alpha^2) \, , \qquad \Delta_\theta(\theta) = 1 + \frac{\alpha^2}{L^2}\cos^2\theta \, .
\end{split}
\end{equation}
The gauge field is given by
\begin{equation}\label{eq:AKN}
W = 2i\,m\sinh(2\delta)\,\frac{\tilde{r}}{V}\Bigl(d\tau + \frac{\alpha}{\Xi}\sin^2\theta\,d\phi\Bigr) \, ,
\end{equation}
where the factor of~$i$ is due to the fact that we work in Euclidean signature. We note that this solution depends on three parameters $(m,\delta,\alpha)$ which are related to the mass, electric charge, and angular momentum.\footnote{One could also add a magnetic charge to the solution and study the more general dyonic Kerr-Newman black hole. We do not present explicitly this solution here, see \cite{Caldarelli:1998hg,Caldarelli:1999xj}.} To ensure that the background above is regular one has to impose that there is an outer horizon, i.e. a locus $r_{+}$ where the function $\Delta_r$ vanishes. To proceed further we define $s_\delta=\sinh\delta $ and~$c_\delta=\cosh\delta $, and a new radial coordinate~$R$ via~$r = R - 2\,m\,s_\delta{}^2$. The quadratic equation~$\Delta_r(R_+) = 0$, which determines the location of the outer horizon in the new coordinate, is solved for the following (complex) values of~$m$:
\begin{equation}
m = \pm\,i\,\frac{R_+^2 + 1 - (1 \pm i\,R_+)\coth(2\delta)}{2\,s_\delta\,c_\delta} \, .
\end{equation}
To ensure regularity of the $\tau$ circle at the location of the outer horizon we need to impose that $\tau$ is periodic with $\tau \sim \tau+\beta$ where
\begin{equation}\label{eq:betaKNdef}
\beta =\; 4\pi\,(\tilde{r}{\,}^2 - \alpha^2)\left(\frac{d\Delta_r(r)}{dr}\right)^{-1}\Bigg\vert_{r = r_+} =\; \frac{2\pi\,(R_+^2 - \alpha^2)}{2\,R_+^3 + R_+(1 - \alpha^2) - m\cosh(2\delta)} \, .
\end{equation}
Note that the parameter~$\delta$ controls the electric charge of the AdS-KN solution. The uncharged AdS-Kerr solution is obtained by setting~$\delta = 0$. 

The regularized two-derivative on-shell action for this solution, $I_{2\partial}$, can be computed using the counterterms above and \eqref{eq:I2deronshell}. After a lengthy calculation we find
\begin{equation}
\begin{split}
\mathcal{F} = -\frac{\beta}{\pi\,(L^2 + \alpha^2)}&\Bigg[\,\frac{r_+^2}{L^2} + 6\,m\,\frac{r_+^2}{L^2}\,s_\delta{}^2 - \frac{r_+^2}{L^2}\left(\alpha^2 - 12\,m^2 s_\delta{}^4\right) - m \\
& - 2\,m\,s_\delta{}^2\,\left(\frac{\alpha^2}{L^2} - \frac{4\,m}{L^2}\,s_\delta{}^4 + 1\right) - 4\,m^2\,\frac{c_\delta{}^2 s_\delta{}^2(2\,m\,s_\delta{}^2 + r_+)}{\alpha^2 - (2\,m\,s_\delta{}^2 + r_+)^2}\Bigg] \, .
\end{split}
\end{equation}
To obtain this result we integrated over the radial coordinate from the location of the outer horizon to asymptotic infinity and made use of the result for $\beta$ in \eqref{eq:betaKNdef} when integrating over the $\tau$ circle. To compute the regularized on-shell GB action we proceed in a similar fashion and after a tedious calculation, using the counterterms in~\eqref{eq:4derSct}, we find the simple result
\begin{equation}\label{eq:IGBKNgen}
I_{\text{GB}} = 64\,\pi^2 \,  \qquad \text{or} \qquad \chi=2\,.
\end{equation}

The three-parameter family of Euclidean AdS-KN solutions above admits a BPS limit obtained by imposing the relation
\begin{equation}\label{eq:susyrelKN}
\alpha = \frac{2i}{e^{4\delta} - 1} \, .
\end{equation}
As shown in~\cite{Cassani:2019mms}, see also \cite{Bobev:2019zmz}, the two-derivative regularized on-shell action for these BPS backgrounds can be written as
\begin{equation}
\mathcal{F}=\frac{(\omega + 1)^2}{2\,\omega}\,,
\end{equation}
where we have defined, 
\begin{equation}\label{eq:omUpsdef}
\omega = \frac{\cosh(2\delta) - 2\sinh(2\delta)}{\cosh(2\delta) + 2 i R_+ \sinh(2\delta)} \, .
\end{equation}
The on-shell GB action is not affected by the BPS limit \eqref{eq:susyrelKN} and is the same as in \eqref{eq:IGBKNgen}. Notice that there is a two-parameter family of supersymmetric Euclidean solutions labelled by $(\delta,m)$ but the on-shell action depends only the specific combination of these parameters given by $\omega$  in \eqref{eq:omUpsdef}. The interpretation of this in the dual 3d SCFT is similar to the one discussed below \eqref{eq:AdSRNFchidef}. The supersymmetric Euclidean KN solution is the gravity dual of the superconformal index of the 3d SCFT on $S^1\times S^2$ \cite{Bhattacharya:2008zy}. This index is different from the one discussed around \eqref{eq:ZTIdef}. Nevertheless, it is still expected on general grounds that the index does not depend on the size of the $S^1$ which on the gravity side is given by the combination of parameters that determine $\beta$ in \eqref{eq:betaKNdef}. It is reassuring that our explicit supergravity on-shell action calculations are in harmony with this expectation.

Combining the results above we obtain the following regularized HD Euclidean on-shell action of the supersymmetric AdS-KN solution,
\begin{equation}\label{eq:AdSKNI4d}
I_{\text{KN}} = \Bigg[\frac{\pi L^2}{2\,G_N} + 32\,\pi^2\,(c_2 - c_1)\Bigg]\,\frac{(\omega + 1)^2}{2\,\omega} + 64\,\pi^2\,c_1 \, .
\end{equation}
We discuss this result and its relation to the superconformal index of the dual SCFT further in Section~\ref{sec:susyloc}.

We end our discussion of the KN solution by noting that the Euclidean KN supersymmetric solution presented above can be analytically continued into a regular supersymmetric Lorentzian black hole solution by setting
\begin{equation}
\alpha = i \mathfrak{a} \,.
\end{equation}
This Lorentzian solution is smooth and free of CTCs only if one further relates the mass and rotation parameters as 
\begin{equation}\label{eq:maKNBH}
m=\mathfrak{a}(1+\mathfrak{a})\sqrt{2+\mathfrak{a}}\,,
\end{equation}
and restricts the rotation parameter to lie in the range $0\leq \mathfrak{a} < 1$. We therefore conclude that while there is a two-parameter family of regular Euclidean supersymmetric KN solutions there is only a one-parameter family of regular Lorentzian black holes that preserve supersymmetry.

%%%%%%%%%%%%%%%
\subsection{Comments on localization of the action}
%%%%%%%%%%%%%%%

After presenting the calculation for the regularized on-shell action for several non-trivial supergravity solutions we now pause to comment on the structure of the on-shell action for general supersymmetric backgrounds in the four-dimensional minimal supergravity. Every supersymmetric supergravity solution has a canonical Killing vector which can be obtained as a bilinear from the preserved Killing spinor. An elegant general formula for the on-shell action of supersymmetric solutions of the two-derivative minimal supergravity was recently derived in \cite{BenettiGenolini:2019jdz} in terms of the fixed loci of this Killing vector. These loci can be either points, referred to as NUTs, or two-dimensional submanifolds, called Bolts. The two-derivative on-shell action for a supersymmetric solution of the minimal supergravity theory can then be expressed as a sum over all these fixed loci with a prescribed contributions from each NUT and Bolt. For instance for a solution with several NUTs one finds
\begin{equation}
\label{eq:2derNUT}
	I_{\text{NUTs}} = \frac{\pi L^2}{2G_N}\,  \sum_{\text{NUTs}_{\mp}}\pm \frac{(b_1\pm b_2)^2}{4\, b_1b_2}\,.
\end{equation}
Here the real numbers $(b_1,b_2)$ determine the precise form of the canonical Killing vector in the neighborhood of a given NUT and the $\pm$ sign is fixed by the chirality of the Killing spinor at the NUT. Examples of solutions with NUT loci are given by the squashed sphere solutions and the supersymmetric Euclidean KN solution discussed above. Based on the explicit results for the four-derivative on-shell action for these solutions it is natural to conjecture a generalization of the results in \cite{BenettiGenolini:2019jdz} to our four-derivative supergravity model. For a solution with only NUT fixed loci we propose that \eqref{eq:2derNUT} generalizes to
\begin{equation}
\label{eq:4derNUT}
	I_{\text{NUTs}} = \left(\frac{\pi L^2}{2G_N} + 32\pi^2 c_2\right) \,  \sum_{\text{NUTs}_{\mp}}\pm \frac{(b_1\pm b_2)^2}{4\, b_1b_2}  -  32\pi^2 c_1\, \,  \sum_{\text{NUTs}_{\mp}}\pm \frac{(b_1\mp b_2)^2}{4\, b_1b_2} \, .
\end{equation}
The results above extend also to solutions with Bolts for which the analog of \eqref{eq:2derNUT} involves integrals over the first Chern class of the tangent and normal bundle to the Bolt, see \cite{BenettiGenolini:2019jdz}. These two-derivative results also extend to the four-derivative on-shell action. For instance for the on-shell action of the Bolt$_{\pm}$ solution discussed around \eqref{eq:BoltpmFchi} we find the four-derivative on-shell action
\begin{equation}
\label{eq:Bolt}
	I_{\text{Bolt}_\pm} =\left(\frac{\pi L^2}{2G_N} + 32\pi^2 c_2\right)\, \left( 1 - {\mathfrak g} \mp \frac{p}{4} \right) - 32\pi^2 c_1\, \left( 1 - {\mathfrak g} \pm \frac{p}{4} \right) \, .
\end{equation}
It would be desirable and very interesting to repeat the analysis of \cite{BenettiGenolini:2019jdz} for the most general supersymmetric solutions of the four-derivative action \eqref{eq:ihd} and establish the results above for the NUT and Bolt on-shell actions more rigorously.

It is tempting to speculate even further about what the results in \eqref{eq:4derNUT} and \eqref{eq:Bolt} may imply for the structure of higher-derivative supergravity more generally. It was suggested in \cite{Hosseini:2019iad} (see also \cite{Choi:2019dfu}), based on \cite{Beem:2012mb}, that the matter-coupled generalization of the two derivative result \eqref{eq:2derNUT} can be understood in terms of gluing basic gravitational blocks. At two derivatives one can recover the on-shell action of many different supergravity solutions by using a single universal {\it building block} and a set of {\it gluing rules} dictated by the background at hand. The two separate, albeit very similar, terms in the formulae \eqref{eq:4derNUT} and \eqref{eq:Bolt} suggest that at the four-derivative level there are {\it two distinct} gravitational building blocks. This result appears to be compatible with the general field theoretic gluing formula \cite{Beem:2012mb} that includes a summation over different ``Bethe vacua'' in the 3d SCFT supersymmetric partition function, see also \cite{Aharony:2021zkr} for a recent discussion. The reason we might be uncovering the need of a ``second building block'', generalizing the analysis in \cite{Hosseini:2019iad}, is that the four-derivative supergravity action has access to the first subleading correction to the SCFT partition function in the large $N$ limit. Clearly we need a much more detailed study of matter-coupled higher-derivative supergravity in order to understand this structure better. We return to this and similar questions in Section~\ref{sec:ExtGen} but leave a more comprehensive analysis for future work.

Another generalization of the results in \cite{BenettiGenolini:2019jdz} can be found by studying the effect on the on-shell action of the two- and four-derivative parity violating terms discussed in Section~\ref{sec:theta}. A full analysis along these lines is beyond the scope of this work and here we just note that at the two-derivative level the on-shell action will contain topological information about the instanton number of the Maxwell field of a given supergravity solution while the four-derivative on-shell action will contain the Pontryagin number. Both of these topological invariants should be calculated by adding appropriate boundary terms similar to the ones in  \eqref{eq:4derSct} needed to compute the Euler number $\chi$ in \eqref{eq:IGBonshell}.

%%%%%%%%%%%%%%%%%
\section{Spectrum and two-point functions}
\label{sec:univ}
%%%%%%%%%%%%%%%%%

After presenting the four-derivative supergravity theory and some of its solutions together with their on-shell action, we now change gears to study the masses and correlation functions of the fluctuations of the supergravity fields around the AdS$_4$ vacuum solution.

%%%%%%%%%%%
\subsection{Stress-energy tensor two-point function}
%%%%%%%%%%%

An important observable in any CFT is the two-point correlation function of the stress-energy tensor operator. For a $d$-dimensional CFT on $\mathbb{R}^d$ it is constrained by conformal invariance, see \cite{Osborn:1993cr}, to take the form\footnote{To make contact with the discussion in Section~\ref{sec:susyloc} we use the same normalization for $C_T$ as in \cite{Chester:2018aca}.}
\begin{equation}\label{eq:CTdef3dCFT}
\langle T_{\mu\nu} (x) T_{\rho\sigma}(0)\rangle = \frac{d-1}{d} \frac{C_T}{|x|^{2d}} \mathcal{I}_{\mu\nu\rho\sigma}(x)\,,
\end{equation}
where
\begin{equation}
\mathcal{I}_{\mu\nu\rho\sigma}(x) = \frac{1}{2} \left[\mathcal{I}_{\mu\rho}(x)\mathcal{I}_{\nu\sigma}(x)+\mathcal{I}_{\mu\sigma}(x)\mathcal{I}_{\nu\rho}(x)\right] - \frac{1}{d}\delta_{\mu\nu}\delta_{\rho\sigma}\,, \qquad \mathcal{I}_{\mu\nu}(x) = \delta_{\mu\nu}- 2\frac{x_{\mu}x_{\nu}}{|x|^2}\,.
\end{equation}
The coefficient $C_T$ is real and positive for a unitary CFT and depends on the dynamics of the theory. 

Holography provides a way to calculate $C_T$ in strongly coupled CFTs. One can use AdS/CFT to map the two-point function of the stress-energy tensor to the two-point function of the graviton propagating in the AdS vacuum and in this way extract the coefficient $C_T$. This is a standard calculation for a two-derivative gravitational theory which becomes more involved in the presence of higher-derivative corrections to the Einstein-Hilbert action. In~\cite{Sen:2014nfa} it was shown how to compute two- and three-point functions of the stress energy tensor in holographic CFTs for a general class of higher-derivative theories of gravity. Here we briefly summarize the results of~\cite{Sen:2014nfa} and apply them to our four-derivative supergravity action.

In~\cite{Sen:2014nfa}, the authors consider a Lorentzian bulk action of the form
\begin{equation}
	S_{\rm SS} = \int d^{4}x\,\sqrt{g}\, \bigg{[} b_0 + b_1 \Delta R + \frac{b_4}{2} \Delta R^2 + \frac{b_5}{2} \Delta R_{\mu\nu} \Delta R^{\mu\nu} + \frac{b_6}{2} \Delta R_{\mu\nu\rho\sigma} \Delta R^{\mu\nu\rho\sigma}\bigg{]}~,
\label{eq:sp1}
\end{equation}
where $b_i$ are real coefficients and we have ignored boundary terms.\footnote{We use $b_i$ to denote the coefficients in \eqref{eq:sp1}, instead of the $c_i$ used in~\cite{Sen:2014nfa}, to avoid confusion with the coefficients $c_{1,2}$ in the four-derivative supergravity action used in this work.} We set $d=3$ in the results of \cite{Sen:2014nfa} since we are interested in a theory with $d+1=4$ bulk dimensions. We also note that the coefficients $b_{0,1}$ are dimensionful, while $b_{4,5,6}$ are dimensionless. The tensor $\Delta R_{\mu\nu\rho\sigma}$ in \eqref{eq:sp1} is defined as
\begin{equation}
	\Delta R_{\mu\nu\rho\sigma} = R_{\mu\nu\rho\sigma} + \frac{1}{L^2}\left(g_{\mu\rho} g_{\nu\sigma} - g_{\mu\sigma} g_{\nu\rho}\right)~.
\end{equation}
Using this, we find the following relations:
\begin{equation}\begin{aligned}
	\Delta R_{\mu\nu\rho\sigma} \Delta R^{\mu\nu\rho\sigma} &= R_{\mu\nu\rho\sigma} R^{\mu\nu\rho\sigma} + \frac{4 R}{L^2} + \frac{24}{L^4}~, \\
	\Delta R_{\mu\nu} \Delta R^{\mu\nu} &= R_{\mu\nu} R^{\mu\nu} + \frac{6 R}{L^2} + \frac{36}{L^4}~, \\
	\Delta R^2 &= R^2 + \frac{24 R}{L^2} + \frac{144}{L^4}~.
\end{aligned}\end{equation}
We can apply these results to rewrite the action \eqref{eq:sp1} as
\begin{equation}\begin{aligned}
	S_{\rm SS} = \int d^4x\,\sqrt{g}\,\bigg{[}&\left(b_0 + \frac{12 b_1}{L^2} + \frac{72 b_4 + 18 b_5 + 12 b_6}{L^4}\right) \\
	& + \left(b_1 + \frac{12 b_4 + 3 b_5 + 2 b_6}{L^2}\right) R \\
	& + \frac{b_4}{2} R^2 + \frac{b_5}{2} R_{\mu\nu}R^{\mu\nu} + \frac{b_6}{2} R_{\mu\nu\rho\sigma} R^{\mu\nu\rho\sigma}\bigg{]}~.
\label{eq:sp2}
\end{aligned}\end{equation}
As shown in \cite{Sen:2014nfa} one can study the linearized fluctuations of the graviton around the AdS$_4$ solution of this general gravitational theory and find the following result for the two-point function coefficient $C_T$ 
\begin{equation}
	C_T^\text{SS} = \frac{48 L^2}{\pi^2}\left( b_1 + \frac{2 b_6}{L^2} \right)~.
\label{eq:ctt1}
\end{equation}
Note that we have reinstated a factor of $L^2$ in Equation (1.12) of \cite{Sen:2014nfa} for dimensional reasons. More specifically, we have that both $C_T$ and $b_6$ are dimensionless,  while $b_1$ has units of inverse length squared. 

To apply this result in our setup we need to map the coefficients $b_i$ used in \cite{Sen:2014nfa} to the constants appearing in the supergravity Lagrangian \eqref{eq:PHD} of interest here. Note that in \cite{Sen:2014nfa} the authors did not include any Maxwell fields in their action and therefore to compare with their results we need to study only the terms in \eqref{eq:PHD} that involve the metric and its derivatives. Converting \eqref{eq:PHD} to Lorentzian signature we find that it can be written as
\begin{equation}\begin{aligned}
	S_\text{4-deriv} &= \int d^4x\,\sqrt{g}\,\bigg{[} \frac{3}{8\pi GL^2} + \frac{1}{16\pi G}R - \frac{(c_1 + 2 c_2)}{3} R^2 \\
	&\quad+ 2 (c_1 + c_2) R_{\mu\nu}R^{\mu\nu}  - c_1 R_{\mu\nu\rho\sigma}R^{\mu\nu\rho\sigma} + \text{(field strength terms)}\bigg{]}~,
\label{eq:i4us}
\end{aligned}
\end{equation}
where we have omitted the terms involving the field strengths of the gauge field. Comparing the four-derivative action \eqref{eq:i4us} to the action of \cite{Sen:2014nfa} as given in \eqref{eq:sp2} above, we find the following relations:
\begin{equation}\begin{aligned}
	b_0 &= - \frac{3}{8\pi G L^2} - \frac{24 c_2}{L^4}~, \qquad b_1 = \frac{1}{16\pi G} + \frac{4 c_2}{L^2}~, \\
	b_4 &= - \frac{2(c_1 + 2 c_2)}{3}~, \qquad b_5 = 4 (c_1 + c_2)~, \qquad b_6 = -2 c_1~.
\end{aligned}\end{equation}
Combining this with \eqref{eq:ctt1} we find
\begin{equation}
	C_T^\text{SS} = \frac{48 L^2}{\pi^2}\left( b_1 + \frac{2 b_6}{L^2} \right) = \frac{ 3 L^2}{\pi^3 G} + \frac{192 (c_2 - c_1)}{\pi^2} ~.
\label{eq:ctt2}
\end{equation}
We note that the authors of \cite{Sen:2014nfa} used a different normalization for $C_T$ than the one employed above in \eqref{eq:CTdef3dCFT}. The relation between the two conventions is $3C_T = 32 \pi^2 C_T^\text{SS}$ which yields
\begin{equation}
\boxed{
	C_T = \frac{ 32 L^2}{\pi G} + 2048 (c_2 - c_1)~.}
\label{eq:ctt3}
\end{equation}
This result is an AdS/CFT prediction for $C_T$ for any 3d $\mathcal{N}=2$ SCFT holographically dual to the four-derivative supergravity theory presented in Section~\ref{sec:HD}. Note that for $c_1=c_2=0$ the result in \eqref{eq:ctt3} reduces to the well-known result for $C_T$ arising from the two-derivative Einstein-Hilbert action, see for instance \cite{Buchel:2009sk}. When $c_{1,2}\neq 0$ we find small corrections to this result due to the four-derivative terms in the supergravity action.

For 3d CFTs with $\mathcal{N}=2$ supersymmetry there is a Ward identity that relates the coefficient $C_T$ to the supersymmetric partition function of the theory on a squashed $S^3$, see for instance \cite{Closset:2012ru}. The Ward identity can be written as
\begin{equation}\label{eq:CTWI}
C_T = \frac{32}{\pi^2}\frac{\partial^2(\log Z_{S^3_b})}{\partial b^2}\bigg|_{b=1}\,,
\end{equation}
where $Z_{S^3_b}$ is the supersymmetric partition function of the SCFT on the $U(1) \times U(1)$ squashed sphere discussed in Section~\ref{subsubsec:U1U1sq} and the squashing parameters $s$ and $b$ are related by $s=b^2$. The relation in \eqref{eq:CTWI} provides a non-trivial consistency check of our results. We can use the on-shell action in \eqref{eq:ionshell} and \eqref{eq:FchiU1U1} and the relation $I^{\rm (HD)}_{\rm on-shell} = \log Z_{S^3_b}$ to compute $C_T$. After a short calculation we indeed find that \eqref{eq:CTWI} leads to the same result as in \eqref{eq:ctt3}. It is reassuring that we have arrived at the result \eqref{eq:ctt3} for $C_T$ by performing two non-trivial and distinct calculations in supergravity and holography.

Three-point functions of the stress-energy tensor in CFTs are also largely constrained by the conformal symmetry. In general number of space-time dimensions the three-point function is fully determined by three real constants one of which is proportional to the coefficient $C_T$ appearing in the two point function and the other two, typically denoted as $t_2$ and $t_4$, contain independent dynamical information about the CFT, see \cite{Erdmenger:1996yc,Hofman:2008ar,Buchel:2009sk}. For 3d CFTs the tensor structure associated with the  coefficient $t_2$ is absent and thus $C_T$ and $t_4$ fully determine the stress-energy three-point function. Given this, it is natural to ask whether one can calculate $t_4$ holographically for higher-derivative gravitational theories. In \cite{Sen:2014nfa} it was shown how to do this for a general class of such models. One finds that $t_4$ is proportional to the couplings of the six-derivatives term in the gravitational action and therefore in our setup we find that $t_4=0$ since we have restricted to four-derivative supergravity actions. We note that unitarity of the CFT imposes non-trivial constraints on the coefficients $C_T$ and $t_4$, see \cite{Hofman:2008ar,Buchel:2009sk}. In our four-derivative setup these constraints reduce to the the inequality $C_T>0$. Since $L^2/G_N$ is manifestly positive and is parametrically larger than the coefficients $c_{1,2}$ we conclude that this inequality is obeyed by the holographic result in \eqref{eq:ctt3} and does not lead to non-trivial constraints on the supergravity couplings. We note in passing that according to the results of \cite{Kats:2007mq} a four-derivative gravitational Lagrangian of the type we study here does not lead to any correction of the ratio between the shear viscosity and entropy density, $\eta/s=1/4\pi$, established for the two-derivative Einstein-Hilbert theory.
	
%%%%%%%%%%%%%	
\subsection{Linearized spectrum}
\label{subsec:linspec}
%%%%%%%%%%%%%

To gain further information about the supergravity model we study it is instructive to calculate the spectrum of linearized fluctuations of all supergravity fields around the maximally supersymmetric AdS$_4$ vacuum solution in the theory. In Appendix~\ref{app:spec} we show in some detail how to calculate the masses for all bosonic fluctuations of the supergravity fields. Similar spectrum calculations in AdS$_4$ for non-supersymmetric higher-derivative gravitational theories were done in \cite{Lu:2011zk,Smolic:2013gz}. It is important to note that in the presence of the four-derivative terms in the Lagrangian some of the auxiliary fields used in the conformal supergravity construction outlined in Section~\ref{sec:HD} acquire kinetic term and lead to dynamical excitations. Since the AdS$_4$ background is supersymmetric, the linearized bosonic fluctuations in our model should combine with the linearized fluctuations for the fermionic fields to form supergravity multiplets. While we have not explicitly computed the masses for all fermionic fluctuations, we have indeed confirmed that the bosonic fluctuations nicely fit into two supersymmetric multiplets. To discuss these results further it is convenient to use AdS/CFT and map all linearized supergravity modes around AdS$_4$ to operators in the dual 3d $\mathcal{N}=2$ SCFT. In the language of 3d $\mathcal{N}=2$ superconformal representation theory, see \cite{Cordova:2016emh} for the notation we use and a comprehensive review, we find that the supergravity modes organize into one short and one long multiplet. 

The short multiplet is denoted by $A_1\bar{A}_1[1]_{2}^{(0)}$ in \cite{Cordova:2016emh} and is simply the stress-energy multiplet of the SCFT, see Table~\ref{tab:EMmodes}. The supergravity modes in this multiplet are the massless metric and graviphoton excitations as well as the corresponding gravitino modes.
\begin{table}[ht]
	\begin{center}{\small
		\setlength{\tabcolsep}{7pt}
		\renewcommand{\arraystretch}{1.2}
		\begin{tabular}{ c || c | c | c|} \hline
	\multicolumn{1}{|c||}{$s$}	 & $2$ &  $\frac{3}{2}$ &{\color{red}$1$} \\ \hline
		\multicolumn{1}{|c||}{$\Delta$} & $3$ &  $\frac{5}{2}$ &{\color{red}$2$}  \\ \hline
		\multicolumn{1}{|c||}{$r$ } & $0$ & $ \pm1$  & {\color{red}$0$} \\ \hline
		\end{tabular}}
	\end{center}
	\caption{Spin $s$, conformal dimension $\Delta$, and R-charge $r$ of the operators in the stress-energy tensor multiplet $A_1\bar{A}_1[1]_{2}^{(0)}$. The superconformal primary operator is indicated in red.}
	\label{tab:EMmodes}
\end{table}

The long multiplet is denoted by $L\bar{L}[1]_{\delta}^{(0)}$ in \cite{Cordova:2016emh}, where $[j]$ indicates the Lorentz spin of the superconformal primary, the subscript is its conformal dimension and the superscript is its R-charge. The operators comprising this multiplet are summarized in Table~\ref{tab:longmodes}. The conformal dimensions of the operators in this multiplet are not determined in terms of their $r$ charge and are given in terms of the quantity
\begin{equation}\label{eq:deltadimdef}
\delta = \frac{1}{2}+\frac{1}{2}\sqrt{1+\frac{L^2}{8\pi G_N(c_1-c_2)}}\,.
\end{equation}

It is well-known that gravitational theories with higher-curvature corrections often have ghosts \cite{Stelle:1977ry}, see \cite{Asorey:1996hz} for a review and further references. These ghosts are manifested by the fact that the massive spin-2 mode arising from the metric has negative energy due to a wrong sign kinetic term. Our model is no exception to this general feature of four-derivative gravitational theories and also suffers from this ghost problem. As can be seen from~\eqref{eq:masspsi}, massive spin-2 modes with negative energy will be present for any value of the HD coefficients as long as~$c_1 \neq c_2$. In addition to this, our four-derivative supergravity model can suffer from instabilities associated to the violations of the BF bound by the massive scalar mode. Violations of the BF bound are associated with complex conformal dimensions in the dual CFT. Using the form of the conformal dimension in \eqref{eq:deltadimdef}  we find that the BF bound is violated when
\begin{equation}\label{eq:BFboundc1c2}
\frac{L^2}{8\pi G_N(c_1-c_2)} < -1\,.
\end{equation}
In a specific microscopic model given by the ABJM theory arising from $N$ M2-branes we find that in the large $N$ limit the 4d supergravity parameters scale as $L^2/G_N \sim N^{3/2}$ and $c_{1,2} \sim N^{1/2}$. Moreover we find that $c_1=3c_2 <0$, see the discussion around \eqref{eq:c1STU}, which indeed leads to a violation of the BF bound according to \eqref{eq:BFboundc1c2}. More generally, we expect that in any supergravity model with a consistent higher-derivative expansion $L^2/G_N$ is parametrically larger than $c_{1,2}$ and thus the BF bound will be violated for all models with~$c_1<c_2$. 

\begin{table}[ht]
	\begin{center}{\small
		\setlength{\tabcolsep}{4pt}
		\renewcommand{\arraystretch}{1.5}
		\begin{tabular}{ c || c | c | c| c | c | c| c | c | c |c | c|} \hline
	\multicolumn{1}{|c||}{$s$}	 & $2$ &  $\frac{3}{2}$ & $\frac{3}{2}$ & $1$ &  $1$ &  $1$ &  $1$ &{\color{red}$1$} & $\frac{1}{2}$ &  $\frac{1}{2}$ &0 \\ \hline
		\multicolumn{1}{|c||}{$\Delta$} & $\delta+1$ &  $\delta+\tfrac{3}{2}$ & $\delta+\tfrac{1}{2}$ & $\delta+2$ &  $\delta+1$ &$\delta+1$ & $\delta+1$ &{\color{red}$\delta$} & $\delta+\frac{3}{2}$ &  $\delta+\frac{1}{2}$ &$\delta+1$  \\ \hline
		\multicolumn{1}{|c||}{$r$ } & $0$ & $ \pm1$  & $ \pm1$ & $0$ &  $\pm 2$ &  $0$ &$0$ & {\color{red}$0$} & $\pm1$ &  $\pm1$ &0 \\ \hline
		\end{tabular}}
	\end{center}
	\caption{Spin $s$, conformal dimension $\Delta$, and R-charge $r$ of the operators in the long $L\bar{L}[1]_{\delta}^{(0)}$ multiplet. The superconformal primary operator is indicated in red.}
	\label{tab:longmodes}
\end{table}

Violations of the BF bound and the presence of propagating ghost modes will both generally lead to violations of unitarity in the theory.  In the case at hand it is possible that the presence of both phenomena simultaneously could somehow be compatible with unitarity, though concluding this would require a much more thorough exploration of the dynamics of the theory.   If unitarity is violated, it should be viewed as an artefact of the four-derivative approximation of the gravitational theory that we have used here; the UV completion of the 4d $\mathcal{N}=2$ supergravity theory by string or M-theory is expected to restore unitarity.

%%%%%%%%%%%%%%%%%
\section{Black hole thermodynamics}
\label{sec:BHTD}
%%%%%%%%%%%%%%%%%

We now turn our attention to studying black holes.  In particular, we study how the HD terms in our gravitational theory affect the thermodynamic properties of a wide class of AdS$_4$ black holes.  HD effects on black hole thermodynamics in anti-de Sitter spacetimes have been investigated previously in the literature, but this analysis can typically only be carried out perturbatively, since the black hole solutions themselves are generically modified by HD effects.  The theory we presented in Section~\ref{sec:HD}, by contrast, has the unique property that any solution to the two-derivative theory is also a solution to the four-derivative theory.  This guarantees that black hole solutions will not be modified by HD effects, thus allowing us to do an in-depth and complete analysis of HD modifications to the black hole thermodynamic quantities.

In order to look at proper black hole solutions, we first need to translate the actions presented in Section~\ref{sec:HD} from Euclidean into Lorentzian signature.  In particular, as discussed in Section~\ref{sec:Lorentzian}, this is accomplished simply by a Wick-rotation of one of the Euclidean tangent space coordinates into a Lorentzian time direction.  After implementing this, we can present the bulk Lorentzian gravitational action $S_\text{bulk}$ succinctly as follows:\footnote{Note that the overall minus sign on each term is due to the action picking up a sign when going from Euclidean to Lorentzian signature, see \eqref{eq:Wick-rotation}.}
\begin{equation}
	S_\text{bulk} = -S_{2\partial} + \left( c_2 - c_1\right) S_{\text{W}^2} - c_2 S_\text{GB}~,
\label{eq:sbulk}
\end{equation}
where $S_{2\partial}$, $S_{\text{W}^2}$, and $S_\text{GB}$, are defined in \eqref{eq:i2deriv}, \eqref{eq:iw2}, and \eqref{eq:igb}, respectively 

As discussed in Section~\ref{sec:holorenorm}, these bulk actions must be supplemented with particular boundary counterterm actions in order to regularize divergences.  We studied these boundary counterterms in detail for the Euclidean theory, and a similar analysis holds for the Lorentzian bulk theory~\eqref{eq:sbulk}.  The end result is that each of the three pieces of the action have their own associated counterterm, which read:
\begin{equation}\begin{aligned}
	S_{2\partial}^\text{CT} &= -\frac{1}{8\pi G_N} \int d^3x\,\sqrt{-h}\,\left( K - \frac{L}{2} \mathcal{R} - \frac{2}{L}\right)~, \\
	S_{\text{GB}}^\text{CT} &= 4 \int d^3x\,\sqrt{-h}\,\left( \mathcal{J} - 2 \mathcal{G}_{\mu\nu} K^{\mu\nu}\right)~, \\
	S_{\text{W}^2}^\text{CT} &= S_{\text{GB}}^\text{CT} + \frac{64 \pi G_N}{L^2} S_{2\partial}^\text{CT}~.
\end{aligned}\end{equation}
The full counterterm action is therefore given by
\begin{equation}\begin{aligned}
	S_\text{CT} &= -S_{2\partial}^\text{CT} + (c_2 - c_1) S_{\text{W}^2}^\text{CT} - c_2 S_{\text{GB}}^\text{CT} = \left(1 + \frac{64\pi G_N( c_2 - c_1)}{L^2}\right)  S_{2\partial}^\text{CT} - c_1 S_{\text{GB}}^\text{CT}~.
\end{aligned}\end{equation}
The full Lorentzian action for our higher-derivative theory, including both the bulk pieces and the boundary counterterm pieces, is given simply by the sum of the bulk and boundary actions:
\begin{equation}
	S_\text{HD} = S_\text{bulk} + S_\text{CT}~.
\label{eq:shd}
\end{equation}

This action $S_\text{HD}$ is our starting point for analyzing how four-derivative terms affect black hole thermodynamics in our supergravity theory.  We now consider an arbitrary stationary black hole solution to the original two-derivative theory, equipped with at least two Killing vectors: one time-like Killing vector associated with time-translations, and one space-like Killing vector associated with azimuthal rotations.  Correspondingly, the two-derivative black hole will have conserved total mass and azimuthal angular momentum, which we denote by $M_0$ and $J_0$, respectively.  Additionally, we allow the solution to be charged under the graviphoton gauge field with electric charge $Q_0$ and magnetic charge $P_0$.  We emphasize that the zero subscript indicates that these conserved quantities are all read off in the original two-derivative theory.  

In the following section, we show how these charges as well as other thermodynamic properties of the black hole are modified when we consider it as a solution to the four-derivative theory at hand.  We then go on to show how the higher-derivative corrections to thermodynamic quantities are compatible with the quantum statistical relation, before moving on to discuss the relations of these results to some recent proposals related to the weak gravity conjecture.

%%%%%%%%%%%%%%%%%%%%
\subsection{Black hole entropy}
\label{sec:bhentropy}
%%%%%%%%%%%%%%%%%%%%

We first analyze how the entropy of the black holes in consideration is modified in our theory.  We do so using the Wald formalism~\cite{Wald:1993nt}, which generalizes the Bekenstein-Hawking area law to incorporate the effects of higher-derivative terms in the action.\footnote{There are instances that arise in string theory where the Wald formalism does not fully account for the entropy of black holes, but these situations only arise when there are Chern-Simons terms in the action~\cite{Tachikawa:2006sz}.  In the 4d setting we study, there are no such terms, and so our action is amenable to simply using the Wald entropy without modification.}  The Wald entropy is given by
\begin{equation}
	S_\text{Wald} = -2 \pi \int_H d^2x\,\sqrt{\gamma}\, \frac{\delta \mathcal{L}}{\delta R_{\mu\nu\rho\sigma}} \epsilon_{\mu\nu} \epsilon_{\rho\sigma}~,
\end{equation}
where $\gamma$ denotes the determinant of the induced metric on the two-dimensional horizon $H$, and $\epsilon_{\mu\nu}$ denotes the unit binormal to the horizon, normalized such that $\epsilon_{\mu\nu}\epsilon^{\mu\nu} = - 2$.  Our goal now is to apply this formula to the three different pieces that go into the bulk action~\eqref{eq:sbulk}.

First, we consider the two-derivative action $S_{2\partial}$ and the Gauss-Bonnet action $S_{\text{GB}}$.  Varying their Lagrangian densities with respect to the Riemann tensor yields
\begin{equation}
	\frac{\delta \mathcal{L}_{2\partial}}{\delta R_{\mu\nu\rho\sigma}} = \frac{1}{16\pi G_N}g^{\mu\rho} g^{\nu\sigma}~, \quad \frac{\delta \mathcal{L}_\text{GB}} {\delta R_{\mu\nu\rho\sigma}} = 2 \left(R^{\mu\nu\rho\sigma} - 4 R^{\mu\rho} g^{\nu\sigma} + R g^{\mu\rho} g^{\nu\sigma}\right)~.
\label{eq:waldvar1}
\end{equation}
From here, it is straightforward to compute their corresponding contributions to the Wald entropy for general stationary black holes:
\begin{equation}
	S_\text{Wald}^{(2\partial)} = \frac{A_H}{4 G_N}~, \quad S_\text{Wald}^{(\text{GB})} = 32 \pi^2 \chi(H)~,
\label{eq:walds1}
\end{equation}
where $A_H$ is the area of the black hole horizon and $\chi(H)$ is the Euler characteristic of the horizon, defined by integrating the Ricci scalar $\mathcal{R}[\gamma]$ of the induced horizon metric over the entire horizon:
\begin{equation}
	\chi(H) \equiv \frac{1}{4\pi} \int_H d^2x\,\sqrt{\gamma}\, \mathcal{R}[\gamma]~.
\end{equation}
These results are well-established in the literature~\cite{Jacobson:1993xs,Iyer:1994ys}, and we repeat them here for completeness.  The W$^2$ action, on the other hand, is novel to our setup, and so we must determine its contribution to the Wald entropy carefully.  We first compute its variation with respect to the Riemann tensor:
\begin{equation}\begin{aligned}
	\frac{\delta \mathcal{L}_{\text{W}^2}}{\delta R_{\mu\nu\rho\sigma}} &= 2 R^{\mu\nu\rho\sigma} - 4 R^{\mu\rho} g^{\nu\sigma} + \frac{2}{3} R g^{\mu\rho}g^{\nu\sigma} - 2 F^{\mu\lambda}F\ind{^\rho_\lambda} g^{\nu\sigma} + \frac{1}{2} F_{\lambda \tau}F^{\lambda\tau} g^{\mu\rho}g^{\nu\sigma}~.
\label{eq:waldvar2}
\end{aligned}\end{equation}
This expression has dependence on both the Riemann tensor and the electromagnetic field strength tensor, and a priori this could result in a Wald entropy contribution that is heavily dependent on the details of the particular solution of interest.  However, by restricting ourselves to solutions to the original two-derivative equations of motion, we are free to use them to simplify this expression.  In particular, we find that we can trade all instances of the electromagnetic field strength $F_{\mu\nu}$ for geometric quantities and the AdS length scale $L$, resulting in
\begin{equation}\begin{aligned}
	\frac{\delta \mathcal{L}_{\text{W}^2}}{\delta R_{\mu\nu\rho\sigma}} &= 2\left(R^{\mu\nu\rho\sigma} - 4 R^{\mu\rho} g^{\nu\sigma} + R g^{\mu\rho}g^{\nu\sigma}\right) + \frac{4}{L^2} g^{\mu\rho} g^{\nu\sigma}~.
\label{eq:waldvar3}
\end{aligned}\end{equation}
Upon comparison of \eqref{eq:waldvar1} and \eqref{eq:waldvar3}, it is clear that we can express the variation of the $\text{W}^2$ Lagrangian as a linear combination of the variations of the two-derivative and Gauss-Bonnet Lagrangians:
\begin{equation}
	\frac{\delta \mathcal{L}_{\text{W}^2}}{\delta R_{\mu\nu\rho\sigma}} = \frac{64 \pi G_N}{L^2} \frac{\delta \mathcal{L}_{2\partial}}{\delta R_{\mu\nu\rho\sigma}} + \frac{\delta \mathcal{L}_\text{GB}} {\delta R_{\mu\nu\rho\sigma}}~.
\end{equation}
Correspondingly, the Wald entropy associated with the $\text{W}^2$ action will be a linear combination of the entropies in \eqref{eq:walds1}, and thus we find that
\begin{equation}
	S_\text{Wald}^{(\text{W}^2)} = \frac{16 \pi A_H}{L^2} + 32 \pi^2 \chi(H)~.
\end{equation}
Remarkably, this demonstrates that the Wald entropy associated with the $\text{W}^2$ action can in general be expressed solely in terms of geometric invariants of the horizon, with no explicit dependence on any details of the full black hole spacetime.  This simplicity is surprising, since one can expect that the presence of the Weyl tensor in the four-derivative action will generically introduce complicated dependence on the black hole parameters into the Wald entropy.

Putting together the Wald entropy results above, we find that the total entropy for any two-derivative black hole solution to our theory is given by
\begin{equation}
	\boxed{S = \left(1 + \frac{64 \pi G_N (c_2 - c_1)}{L^2}\right) \frac{A_H}{4 G_N} - 32 \pi^2 c_1 \chi(H)}~.
\label{eq:entropy_final}
\end{equation}
We therefore find that there are two modifications to the black hole entropy.  The first is a new topological term, proportional to the Euler characteristic $\chi(H)$ of the black hole horizon, with the constant of proportionality determined by the constant $c_1$ in the action.  The second is a rescaling of the usual Bekenstein-Hawking area law, the size of which is controlled by $c_2 - c_1$.  This result is universal for all stationary black holes, and it yields a relatively simple expression for the entropy of even the intricate AdS-Kerr-Newman black hole solution.

A few comments are in order.  We first want to stress that the Euler characteristic $\chi(H)$ is topological and thus insensitive to continuous variations of the black hole parameters.  In particular, this means that a BPS black hole retains the same topological entropy correction as any of its non-BPS counterparts with the same topology, and so the same correction will apply arbitrarily far away from extremality.  In fact, this same topological term is also present when higher-derivative corrections are added to ungauged supergravity~\cite{Charles:2016wjs,Charles:2017dbr}, lending further support to the hope of having precision microstate counting for non-supersymmetric black holes in string theory.

The last point we want to make is that the rescaling of the Bekenstein-Hawking area law in \eqref{eq:entropy_final} is most naturally interpreted as a redefinition of Newton's constant $G_N$.  Dropping the topological correction for now, we can write the entropy as
\begin{equation}
	S = \frac{A_H}{4 G_\text{eff}}~, \quad \frac{1}{G_\text{eff}} \equiv \frac{1}{G_N} + \frac{64 \pi (c_2 - c_1)}{L^2}~.
\label{eq:gren}
\end{equation}
In other words, the black hole entropy still takes the usual Bekensten-Hawking form, but with an effective gravitational coupling constant $G_\text{eff}$ that receives modifications from the four-derivative operators in the theory.  This effective rescaling of Newton's constant is interesting because the theory becomes strongly coupled in the UV and so $\frac{1}{G_N} \to 0$ as we go to higher and higher energies.  This in turn leads to quadratic divergences in the spectrum of the theory, despite the fact that all observables (and entropies) must be finite in a quantum theory of gravity.  A possible resolution of this tension is that $G_\text{eff}$ could remain finite in the UV even if $G_N$ blows up.  This is similar in spirit to the mechanisms proposed in~\cite{Susskind:1994sm,Larsen:1995ax} for taming ultraviolet divergences in string theory, and it would be interesting to investigate this further by studying the ultraviolet behavior of the higher-derivative terms in our supergravity theory.

%%%%%%%%%%
\subsection{Conserved charges and the quantum statistical relation}
\label{sec:qsr}
%%%%%%%%%%

We now move on to analyzing how the higher-derivative terms in the action \eqref{eq:shd} alter the conserved charges that a general stationary black hole solution is imbued with.  We first focus on the mass and angular momentum of the black hole before moving on to studying the electromagnetic charges. We end with a brief discussion of how our results are compatible with the quantum statistical relation.

%%%%%%%%%
\subsubsection{Mass and angular momentum}
%%%%%%%%%

Any stationary, asymptotically-AdS$_4$ black hole spacetime is equipped with at least two isometries, one associated with time translations and one associated with azimuthal rotations.  We denote the corresponding Killing vectors that generate these isometries by $\mathcal{K}_{(t)}$ and $\mathcal{K}_{(\phi)}$, respectively.  There are conserved quantities associated with these isometries, namely the mass $M$ and azimuthal angular momentum $J$, that play an important role in the thermodynamic properties of the black hole.  In order to compute these quantities, we follow the formalism developed in~\cite{Balasubramanian:1999re} for computing conserved charges in asymptotically-AdS spacetimes via a Komar integral of the boundary stress-tensor.  This method is the most natural in holographic settings, since it yields charges that are automatically free of divergences while also removing the ambiguities that can arise in other schemes, such as the Brown-York procedure~\cite{Brown:1992br}.

Let $\Sigma$ denote the conformal boundary of the spacetime at spatial infinity, where the unit normal to the boundary is $n^\mu$, the induced boundary metric is $h_{\mu\nu}$, the extrinsic curvature is $K_{\mu\nu}$ (with $K \equiv h^{\mu\nu} K_{\mu\nu}$), and the boundary Riemann tensor is $\mathcal{R}_{\mu\nu\rho\sigma}$.  We then denote by $\partial \Sigma$ a constant time slice of the boundary.  The conserved charge $\mathcal{Q}$ associated with a Killing vector $\mathcal{K}$ is computed by the Komar integral
\begin{equation}
	\mathcal{Q}[\mathcal{K}] = \int_{\partial \Sigma} d^2x\,\sqrt{\gamma}\,u^\mu \mathcal{K}^\nu \tau_{\mu\nu}~,
\label{eq:komar}
\end{equation}
where $\gamma$ is the induced metric on $\partial \Sigma$, $u$ is the unit normal to $\partial \Sigma$, and  $\tau_{\mu\nu}$ denotes the boundary stress-tensor, defined by
\begin{equation}
	\tau_{\mu\nu} \equiv \frac{2}{\sqrt{-h}} \frac{\delta \mathcal{L}}{\delta h^{\mu\nu}}~.
\end{equation}
Importantly, the boundary stress-tensor is computed by varying both the bulk action as well as the boundary counterterm action, since the counterterm action is required for a well-posed variational principle that ensures no derivatives of the metric fluctuation appear.  

For the two-derivative minimal supergravity action and the Gauss-Bonnet action, the boundary stress-tensors are well-known in the literature (see e.g. \cite{Emparan:1999pm,Davis:2002gn}), and we go through their computation extensively in Appendix~\ref{app:bdry}.  The results are:
\begin{equation}
	\tau_{\mu\nu}^{(2\partial)} = \frac{1}{8\pi G_N}\left( K_{\mu\nu} - K h_{\mu\nu} - L \mathcal{G}_{\mu\nu} + \frac{2}{L} h_{\mu\nu}\right)~,
\label{eq:tau2d}
\end{equation}
and
\begin{equation}
	\tau_{\mu\nu}^{(\text{GB})} = 12 \mathcal{J}_{\mu\nu} - 4 \mathcal{J} h_{\mu\nu} - 8 \mathcal{P}_{\mu\rho\nu\sigma} K^{\rho\sigma}~,
\label{eq:taugb}
\end{equation}
where $\mathcal{G}_{\mu\nu} \equiv \mathcal{R}_{\mu\nu} - \frac{1}{2} h_{\mu\nu}\mathcal{R}$ is the boundary Einstein tensor, the boundary tensor $\mathcal{J}_{\mu\nu}$ is defined in \eqref{eq:jtens}, and $\mathcal{P}_{\mu\nu\rho\sigma}$ is the divergence-free part of the boundary Riemann tensor:
\begin{equation}
	\mathcal{P}_{\mu\nu\rho\sigma} \equiv \mathcal{R}_{\mu\nu\rho\sigma} - 2 \mathcal{R}\ind{_{\mu[\rho}} h\ind{_{\sigma]\nu}} + 2 \mathcal{R}\ind{_{\nu[\rho}} h\ind{_{\sigma]\mu}} + \mathcal{R} h\ind{_{\mu[\rho}} h\ind{_{\sigma]\nu}}~.
\end{equation}
With these boundary stress-tensors, we can now compute the mass $M_0$ and angular momentum $J_0$ for any black hole in the two-derivative theory simply by utilizing the Komar integral~\eqref{eq:komar} with the corresponding Killing vectors and the two-derivative boundary stress-tensor:
\begin{equation}\begin{aligned}
	M_0 &= \int_{\partial \Sigma} d^2x\,\sqrt{\gamma}\, u^\mu \mathcal{K}_{(t)}^\nu \tau_{\mu\nu}^{(2\partial)}~, \\
	J_0 &= \int_{\partial \Sigma} d^2x\,\sqrt{\gamma}\, u^\mu \mathcal{K}_{(\phi)}^\nu \tau_{\mu\nu}^{(2\partial)}~.
\end{aligned}\end{equation}
For the Gauss-Bonnet boundary stress-tensor, on the other hand, the corresponding Komar integrals vanish and thus the Gauss-Bonnet term yields no contribution to the mass and angular momentum.  The particular sum of boundary tensors that show up in~\eqref{eq:taugb} conspire such that $\tau_{\mu\nu}^{(\text{GB})}$ becomes very subleading in the radial coordinate, and so in the limit where the radial cutoff is sent to infinity the Komar integral~\eqref{eq:komar} dies out:
\begin{equation}
	\int_{\partial \Sigma} d^2x\,\sqrt{\gamma}\,u^\mu \mathcal{K}_{(t)}^\nu \tau_{\mu\nu}^{(\text{GB})} = \int_{\partial \Sigma} d^2x\,\sqrt{\gamma}\,u^\mu \mathcal{K}_{(\phi)}^\nu \tau_{\mu\nu}^{(\text{GB})} = 0~.
\end{equation}
Physically, this is simply a manifestation of the topological nature of the Gauss-Bonnet invariant in four dimensions, which forbids the Gauss-Bonnet term from affecting conserved charges~\cite{Myers:1988ze}.  In higher dimensions this is no longer the case, and the Gauss-Bonnet term can yield non-trivial corrections to thermodynamic properties of black holes.

We now have to determine the boundary stress-tensor for the $\text{W}^2$ component of the full HD action~\eqref{eq:shd}.  This is in general a hard problem to tackle, as the variation of $S_{\text{W}^2}$ with respect to the boundary metric is fairly complicated.  However, since we are only interested in considering black holes that are solutions to the original two-derivative equations of motion, we can apply these equations to drastically simplify the variation.  Additionally, the W$^2$ boundary counterterm $S_{\text{W}^2}^\text{CT}$ is simply a linear combination of the two-derivative and Gauss-Bonnet counterterms $S_{2\partial}^\text{CT}$ and $S_{\text{GB}}^\text{CT}$, and so their variations with respect to the boundary metric will be similarly related.  Putting all of this together, we find that the W$^2$ boundary stress-tensor is simply given by a linear combination of $\tau_{\mu\nu}^{(2\partial)}$ and $\tau_{\mu\nu}^{(\text{GB})}$, i.e.
\begin{equation}
	\tau_{\mu\nu}^{(\text{W}^2)} = \frac{64 \pi G_N}{L^2} \tau_{\mu\nu}^{(2\partial)} + \tau_{\mu\nu}^{(\text{GB})}~.
\label{eq:tauw2}
\end{equation}
Again, we stress that this boundary stress-tensor will take a more complicated form when considering the four-derivative equations of motion as a whole, but when we restrict to only those solutions that satisfy the two-derivative equations of motion, it is forced to take the simpler form in \eqref{eq:tauw2}.

Putting all this together, we find that the full boundary stress-tensor for our four-derivative supergravity theory is given by
\begin{equation}
	\tau_{\mu\nu} = \left(1 + \frac{64 \pi G_N(c_2 - c_1)}{L^2}\right) \tau_{\mu\nu}^{(2\partial)} - c_1 \tau_{\mu\nu}^{(\text{GB})}~.
\label{eq:tau4}
\end{equation}
The mass $M$ and angular momentum $J$ in the four-derivative theory are then computed by inserting the boundary stress-tensor \eqref{eq:tau4} into the Komar integral \eqref{eq:komar}.  Since the Gauss-Bonnet stress-tensor yields no contribution to the charges and the two-derivative stress-tensor in \eqref{eq:tau4} is modified only by an overall rescaling, we find that
\begin{equation}
	\boxed{M = \left(1 + \frac{64\pi G_N (c_2 - c_1)}{L^2}\right) M_0~, \quad J = \left(1 + \frac{64 \pi G_N (c_2 - c_1)}{L^2}\right) J_0}~.
\label{eq:4dermass}
\end{equation}
That is, the mass $M$ and angular momentum $J$, as computed in the four-derivative theory, are related to the original mass $M_0$ and angular momentum $J_0$ in the two-derivative theory by a constant rescaling.  This relation holds for general stationary black hole solutions, so once the mass and angular momentum in the two-derivative theory are computed, we can immediately find their values in the four-derivative theory by simply using \eqref{eq:4dermass}.

The last point we would like to stress is that the W$^2$ counterterm is absolutely essential for deriving the results~\eqref{eq:4dermass}.  Indeed, even though it evaluates to zero on-shell for all explicit solutions we have studied, its variation with respect to the metric is non-zero, which allows us to cancel bulk terms in the variation of $S_{\text{W}^2}$ and end up with a well-defined boundary stress-tensor.  The presence of this novel boundary term also explains the discrepancy between our results and those of~\cite{Cremonini:2019wdk}, where it was argued that certain bulk four-derivative terms cannot affect the mass of the black hole.  Our results therefore serve as an important reminder of the significance of finite boundary counterterms in holographic settings.

%%%%%%%%%
\subsubsection{Electromagnetic charges}
%%%%%%%%%

We now turn to studying higher-derivative corrections to the electromagnetic charges of the black hole solution.  In order to do so, we first recall how these charges are computed in the first place.  In the two-derivative Einstein-Maxwell theory, the Maxwell equation of motion is simply given by 
\begin{equation}
	\nabla_\mu F^{\mu\nu} = 0~.
\end{equation}
Additionally, since the theory is free of any sources, the Maxwell fields must also satisfy the Bianchi identity $\nabla_{[\mu} F_{\nu\rho]} = 0$, or equivalently 
\begin{equation}
	\nabla_\mu \star F^{\mu\nu} = 0~,
\end{equation}
where $\star F$ denotes the two-form Hodge dual of $F$.  Thus, the combined Maxwell-Bianchi equations tell us that both $F^{\mu\nu}$ and its Hodge dual $\star F^{\mu\nu}$ are conserved.  This in turn means that integrating $\star F$ and $F$ over a two-dimensional constant time slice $\partial \Sigma$ of the conformal boundary will result in conserved charges that are independent of our choice of slice.  These charges are (up to some constants of proportionality that are unimportant for our analysis) the electric charge $Q_0$ and magnetic charge $P_0$, respectively:
\begin{equation}\begin{aligned}
	Q_0 = \int_{\partial \Sigma} \star F~, \quad P_0 = \int_{\partial \Sigma} F~.
\label{eq:2dercharges}
\end{aligned}\end{equation}
Again, we use the zero subscript to stress that these are the charges as measured in the two-derivative theory.

We now move on to the full four-derivative theory.  Since the action \eqref{eq:shd} is free of any electromagnetic sources, the Bianchi identity is untouched, and so the Maxwell field must still satisfy $\nabla_\mu \star F^{\mu\nu} = 0$.  The Maxwell equations, however, are altered, and are no longer simply given by $\nabla_\mu F^{\mu\nu} = 0$.  Instead, if we define a new tensor $G_{\mu\nu}$ via
\begin{equation}
	\star G_{\mu\nu} \equiv -32 \pi G_N \frac{\delta \mathcal{L}_{\text{HD}}}{\delta F^{\mu\nu}}~,
\end{equation}
then the four-derivative Maxwell equation can be presented succinctly as $\nabla_\mu \star G^{\mu\nu} = 0$.  Thus, the combined Maxwell-Bianchi equations are given by
\begin{equation}
	\nabla_\mu \star G^{\mu\nu} = \nabla_\mu \star F^{\mu\nu} = 0~.
\end{equation}
In form language, this can equivalently be written as $d G = d F = 0$, and so we can obtain conserved charges by integrating the two-forms $G$ and $F$ over $\partial \Sigma$, by analogy with the two-derivative case:
\begin{equation}
	Q = \int_{\partial \Sigma} G~, \quad P = \int_{\partial \Sigma} F~.
\label{eq:newcharges}
\end{equation}
From this expression, it is clear that the magnetic charge $P$ will take the same value $P_0$ as in the two-derivative theory, as expected since our theory does not modify the Bianchi identity.  The electric charge, on the other hand, is modified, due to the four-derivative dependence found inside of $G_{\mu\nu}$.

From the full four-derivative bulk action in \eqref{eq:sbulk} we can explicitly compute that the dual of the modified field strength $G_{\mu\nu}$ is given by
\begin{equation}\begin{aligned}
	\star G_{\mu\nu} &= F_{\mu\nu} + 32 \pi G_N (c_1 - c_2)\left[ \left(R-\frac{2}{L^2} -\frac{1}{2} F_{\rho\sigma} F^{\rho\sigma}\right)F_{\mu\nu} + \left(2 F_{\mu\rho}F^{\sigma\rho} - 4 R\ind{_\mu^\sigma}\right)F_{\sigma \nu}\right]~.
\end{aligned}\end{equation}
Upon implementing the two-derivative Einstein equation, this expression simplifies drastically and we are left simply with
\begin{equation}
	\star G_{\mu\nu} = \left(1 + \frac{64 \pi G_N (c_2 - c_1)}{L^2}\right) F_{\mu\nu}~.
\end{equation}
That is, the four-derivative terms in the action leave $\star G \propto F$, albeit with a modified constant of proportionality.  The electric charge $Q$, defined in \eqref{eq:newcharges} by integrating $G$ over the closed surface $\partial \Sigma$, will therefore be given by
\begin{equation}
	Q = \int_{\partial \Sigma} G = \left(1 + \frac{64 \pi G_N (c_2 - c_1)}{L^2}\right) \int_{\partial \Sigma} \star F~.
\end{equation}
Putting all of these results together and comparing them to the original two-derivative electric and magnetic charges in \eqref{eq:2dercharges}, we find that the charges in our four-derivative theory are related to the original ones via
\begin{equation}
	\boxed{Q = \left(1 + \frac{64 \pi G_N (c_2 - c_1)}{L^2}\right) Q_0~, \quad P = P_0}~.
\label{eq:4dercharges}
\end{equation}
The magnetic charge is unaffected by the four-derivative terms in the theory, while the electric charge is simply rescaled by a constant.  Importantly, this is universally true for general stationary black hole solutions; we have assumed nothing further about the details of the solution itself.  In fact, the four-derivative charge relation \eqref{eq:4dercharges} will be valid for any solution imbued with electromagnetic charges and not just black holes; our derivation relies solely on integrating the Maxwell-Bianchi equations over the boundary, with no reference to the details of the solution itself.

%%%%%%%%%%%%%%%%
\subsubsection{Quantum statistical relation}
%%%%%%%%%%%%%%%%

Armed with the entropy and charges for any stationary black hole in our theory, we can perform a consistency check of our results by confirming that they obey the  laws of black hole thermodynamics. In particular, we can check the first law of thermodynamics, or equivalently the so-called quantum statistical relation~\cite{Gibbons:1976ue,Gibbons:2004ai}, which posits that the thermodynamic properties of black holes are related as follows:
\begin{equation}
	I = \beta\left(M - TS - \Phi Q - \omega J\right)~.
	\label{eq:qsr}
\end{equation}
This is a relation between the Eucliean on-shell action $I$, the temperature $T=\beta^{-1}$, the mass $M$, the entropy $S$, the electric charge $Q$, the electrostatic potential $\Phi$, the angular momentum $J$, and the angular velocity $\omega$.  The extensive quantities in this expression (i.e. the ones that scale with the size of the system) are $I$, $S$, $M$, $Q$, and $J$, all of which receive modifications from higher-derivative terms in the supergravity theory, as detailed above.  The intensive quantities (i.e. the ones that do not scale with the system size) are $T$, $\Phi$, and $\omega$, all three of which can be read off directly from the solution itself.  Crucially, since the black hole solutions we study are not modified by the four-derivative terms in the supergravity Lagrangian, the intensive quantities will not be modified either.

To be more explicit, we consider each of the three intensive quantities.  The temperature $T$ can be computed via the usual method, where we Wick-rotate the black hole to Euclidean signature via a time coordinate rotation of the form $t \to - i \tau$, and then demand that $\tau$ is periodic with some periodicity $\beta$ required to avoid conical singularities where the spacetime caps off in the bulk.  Then, by identifying $\beta$ with $T^{-1}$, we obtain the temperature of the black hole.  The electrostatic potential $\Phi$ is given by
\begin{equation}
	\Phi = \left(\mathcal{K}_{(t)}^\mu W_\mu \right)_{r = r_H} - \left(\mathcal{K}_{(t)}^\mu W_\mu \right)_{r \to \infty}~,
\end{equation} 
where the contraction of the time-like Killing vector $\mathcal{K}_{(t)}^\mu$ of the spacetime with the gauge field $W_\mu$ is evaluated both at the horizon $r_H$ and at the conformal boundary and we take the difference between the two.  The Killing vector $\mathcal{K}_{(t)}^\mu$ is defined purely from the black hole geometry, while the gauge field is part of the specified solution.  Finally, the angular velocity $\omega$ can be read off by defining the linear combination of Killing vectors
\begin{equation}
	\xi \equiv \mathcal{K}_{(t)} + \omega \mathcal{K}_{(\phi)}~,
\end{equation}
and then tuning $\omega$ such that $\xi$ is a null vector at the black hole event  horizon.  Therefore for all three intensive quantities we have explicit expressions that depend only on the background two-derivative solution, which we know to be preserved by the four-derivative operators in our theory.

Now that we have defined all relevant quantities in the quantum statistical relation \eqref{eq:qsr}, we can analyze it for the theory at hand.  At the two-derivative level, using the notation defined in \eqref{eq:i2deriv} and throughout this section, it takes the form
\begin{equation}
	\frac{\pi L^2}{2 G_N} \mathcal{F}(\mathbb{S}) = \beta\left(M_0 - \frac{T A_H}{4 G_N} - \Phi Q_0 - \omega J_0\right)~.
\label{eq:qsr_2d}
\end{equation}
The two-derivative quantum statistical relation \eqref{eq:qsr_2d} holds for all known asymptotically AdS$_4$ black hole solutions in Einstein-Maxwell theory.  At the four-derivative level, if we combine the expressions for the on-shell action \eqref{eq:ihd}, the entropy \eqref{eq:entropy_final}, the mass and angular momentum \eqref{eq:4dermass}, and the electric charge \eqref{eq:4dercharges}, the quantum statistical relation takes the following form:
\begin{equation}\begin{aligned}
	&\left(1 + \frac{64 \pi G_N (c_2 - c_1)}{L^2}\right) \frac{\pi L^2}{2 G_N} \mathcal{F}(\mathbb{S}) + 32 \pi^2 c_1 \chi(\mathbb{S}) \\
	&\quad\quad= \beta \left(1 + \frac{64 \pi G_N (c_2 - c_1)}{L^2}\right)\left(M_0 -  \frac{T A_H}{4 G_N} - \Phi Q_0 - \omega J_0\right) + 32 \pi^2 c_1 \chi(H)~.
\label{eq:qsr_4d}
\end{aligned}\end{equation}
If we compare the four-derivative relation \eqref{eq:qsr_4d} to the two-derivative relation \eqref{eq:qsr_2d}, we can see that most terms cancel automatically, due to the way in which the higher-derivative terms in the theory conspire to rescale certain thermodynamic quantities.  After eliminating these terms, we are left with only the topological terms.  We therefore find that the quantum statistical relation holds at the four-derivative level if and only if
\begin{equation}
	\chi(\mathbb{S}) = \chi(H)~,
\label{eq:chivschi}
\end{equation}
i.e. if the Euler characteristic of the full spacetime is equal to the Euler characteristic of the horizon. The above equality follows from applying the Atiyah-Singer index theorem to the de Rahm complex on the space-time manifold described by~$\mathbb{S}$, which relates~$\chi(\mathbb{S})$ to the fixed points of isometries~\cite{Gibbons:1979xm}. For black holes, such fixed points are either isolated (so-called NUTs) or form a two-dimensional surface corresponding to their horizon (so-called Bolts), and in both cases~\eqref{eq:chivschi} follows.  
%
%We have verified that \eqref{eq:chivschi} holds for all known asymptotically AdS$_4$ black hole solutions, but we are not aware of any general proof of this statement.\footnote{The non-compact nature of asymptotically AdS spacetimes also makes this statement somewhat subtle to prove in general.}  So, while it would be interesting to establish \eqref{eq:chivschi} via a formal proof, we nonetheless have verified that it holds in all known cases, 
%
%thus
This establishes that our four-derivative supergravity theory satisfies the quantum statistical relation.

%%%%%%%%%%%%%%%
\subsection{Implications for the weak gravity conjecture}
%%%%%%%%%%%%%%%

The weak gravity conjecture (WGC), in its simplest incarnation, posits that in order for a gravitational theory with a $U(1)$ gauge symmetry to admit a UV completion, there must exist some state in the Hilbert space of the theory with $U(1)$ charge $Q$ and mass $M$ that is \emph{superextremal}, i.e. its charge-to-mass ratio (in appropriate units) exceeds the black hole extremality bound:
\begin{equation}
	\frac{Q}{M} > \frac{Q}{M} \bigg{|}_\text{ext.}~.
\label{eq:wgc1}
\end{equation}
To be more precise, given a state with mass $M$ and charge $Q$, we consider an electrically charged black hole with the same mass $M$ but whose charge is tuned such that the black hole is extremal. If the charge-to-mass ratio of the state exceeds that of its extremal black hole counterpart, it is considered to be superextremal.

One of the primary motivations for the WGC is to ensure that there is a decay channel for extremal black holes in order to avoid having a large number of black hole remnants littering the universe.  Since its inception, though, the WGC has been shown to be intricately linked to a whole host of other swampland conjectures (see e.g.~\cite{Palti:2019pca} for a review), and so its motivation goes far beyond simply avoiding the remnant problem.

One mechanism proposed in~\cite{ArkaniHamed:2006dz,Kats:2006xp} for realizing the WGC is to incorporate the effects of higher-derivative operators in the theory.  These higher-derivative operators can modify the extremality bound for black holes and potentially allow for black holes themselves to be superextremal, in the sense that their charge-to-mass ratio exceeds that of an extremal black hole in the original two-derivative theory.  This opens up a scenario where charged black holes can decay into smaller charged black holes, thus avoiding the remnant problem in a universal way that makes no reference to the field content of the theory beyond gravity and the $U(1)$ gauge field.  It has been argued in a number of bottom-up approaches that this black hole version of the WGC will hold if certain assumptions are placed on the UV-completion of the theory, including positivity of black hole entropy corrections~\cite{Cheung:2018cwt,Cheung:2019cwi,Goon:2019faz,Cremonini:2019wdk}, unitarity and causality~\cite{Cheung:2014ega,Hamada:2018dde,Loges:2019jzs}, and  scattering amplitude positivity~\cite{Bellazzini:2019xts,Loges:2020trf}.  Most of these analyses are for asymptotically flat black holes, therefore it is worthwhile to further our understanding of the black hole WGC in asymptotically-AdS spacetimes.  Moreover, as we have shown in Section~\ref{sec:superconf}, supersymmetry leads to powerful constraints on higher derivative terms in the theory, and it is our goal now to analyze the interplay of these constraints and the WGC.

For the particular supergravity theory described by the action in \eqref{eq:shd} we have shown that the higher-derivative terms conspire to rescale the charge $Q$ and mass $M$ of the black hole by a constant prefactor, as shown in \eqref{eq:4dermass} and \eqref{eq:4dercharges}.  Since both quantities are modified by the same factor the charge-to-mass ratio $Q/M$ is unaltered as compared to the charge-to-mass ratio in the  two-derivative theory:
\begin{equation}
	\frac{Q}{M} = \frac{\left(1+ \frac{64\pi G_N (c_2 - c_1)}{L^2}\right)Q_0}{\left(1+ \frac{64\pi G_N (c_2 - c_1)}{L^2}\right)M_0} = \frac{Q_0}{M_0}~.
\end{equation}
Since there is no way to violate the extremality bound in the original two-derivative theory without introducing naked singularities, we are left with no possible way to satisfy the WGC \eqref{eq:wgc1} by black holes alone in this four-derivative theory.  Moreover, our results hold for general black hole solutions, not just the ones that admit a BPS limit, and so we can conclude that supersymmetry at the level of the four-derivative theory obstructs the black hole WGC.  This is corroborated by the results of~\cite{Charles:2019qqt}, where it was shown that the charge-to-mass ratio of general black holes in ungauged supergravity are unaffected by higher-derivative effects as well.

An important aspect to consider in AdS spacetimes is that the WGC presented in~\eqref{eq:wgc1} is not necessarily sufficient to allow extremal black holes to decay; the charge-to-mass ratio of the superextremal state must satisfy an even more stringent bound, in order to overcome the more attractive nature of gravity in spacetimes with a negative cosmological constant~\cite{Nakayama:2015hga}.  The precise details of how the WGC \eqref{eq:wgc1} must be modified so that AdS black holes can decay depend on the details of the black hole of consideration.    However, for all versions of the WGC based on the kinematics of black hole decay, the end result is that satisfying the WGC requires a charge-to-mass ratio that at least satisfies \eqref{eq:wgc1}, and so we can view \eqref{eq:wgc1} as the most conservative version of the WGC in AdS spacetimes~\cite{Montero:2018fns}.  Since the four-derivative corrections in our theory are such that black holes cannot satisfy this mildest form of the WGC, they clearly do not satisfy any of the stronger versions discussed in~\cite{Nakayama:2015hga,Montero:2018fns}.\footnote{There are also versions of the WGC in asymptotically-AdS spacetimes that posit the existence of operators in the dual CFT whose conformal dimension $\Delta$ and charge $q$ under the $U(1)$ current in the CFT satisfy certain inequalities.  These are interesting in their own right, but they are beyond the scope of our discussion.}

Another proposed version of the WGC for black holes posits that the same combination of higher-derivative operators control corrections to the charge-to-mass ratio as well as the entropy of the black hole.  This has been demonstrated explicitly in a number of different examples~\cite{Cheung:2018cwt,Cheung:2019cwi,Goon:2019faz,Cremonini:2019wdk}, all of which take the schematic form
\begin{equation}
	\Delta M_\text{ext.}(Q) \propto  -\Delta S(M,Q)\bigg{|}_{M \to M_\text{ext.}(Q)}~.
\label{eq:wgc2}
\end{equation}
Here, the left-hand side $\Delta M_\text{ext.}(Q)$ denotes the shift in the extremal mass due to higher-derivative operators when comparing the two-derivative black hole solution to the four-derivative solution at the same fixed charge $Q$.  The right-hand side $\Delta S(M,Q)$ denotes the change in entropy of the black hole due to higher-derivative corrections at fixed mass and charge.  If the left-hand and right-hand sides of \eqref{eq:wgc2} are proportional and have the same sign, then demanding that the black hole is superextremal (i.e. the mass shift at fixed charge is negative) will correspond to positive entropy corrections.  This has led to a new version of the black hole WGC in~\cite{Cheung:2018cwt,Cheung:2019cwi,Cremonini:2019wdk} that posits that the entropy corrections due to higher-derivative effects in a UV-complete theory of gravity must be positive:
\begin{equation}
	\Delta S > 0~.
\label{eq:wgc3}	
\end{equation}

For the supergravity theory of consideration in this work we find that both of the proposals~\eqref{eq:wgc2} and \eqref{eq:wgc3} for entropic versions of the WGC are not satisfied in general.  It is easy to see that the mass shift relation \eqref{eq:wgc2} cannot hold in general, because the Gauss-Bonnet term in the action yields a topological contribution to the entropy in \eqref{eq:entropy_final}, proportional to the Euler characteristic $\chi(H)$ of the horizon, that has no analogue in the mass $M$ computed in \eqref{eq:4dermass}.  Additionally, it is well-known that 4d AdS-Reissner-Nordstr\"om black holes can have horizons with the topology of a Riemann surface $\Sigma_{\mathfrak{g}}$ with arbitrary genus $\mathfrak{g}$.  The Euler characteristic of such a Riemann surface horizon is given by
\begin{equation}
	\chi(H) = 2(1 - \mathfrak{g})~.
\end{equation}
For spherical horizons with $\mathfrak{g} = 0$, we find that $\chi(H) > 0$, while for higher-genus surfaces with $\mathfrak{g} > 1$ we find that $\chi(H) < 0$.  The topological correction to the black hole entropy therefore cannot take any definite sign; it instead depends on the solution of interest.  There are therefore no constraints that we can set on the coefficients $c_1, c_2$ in our action that force $\Delta S > 0$ for all black holes, and so we also violate \eqref{eq:wgc3}.  Similar violations of these entropic versions of the WGC can also be found in ungauged supergravity~\cite{Charles:2016wjs} and in heterotic string theory~\cite{Cano:2019oma}, due again to the presence of a Gauss-Bonnet term in the four-dimensional low-energy theory.  Since the Gauss-Bonnet term in the action is compatible with supersymmetry, as discussed in Section~\ref{sec:superconf}, there is no way to rule out its existence in generic 4d string and M-theory compactifications or consistent truncations.  We therefore conclude that the proposed relations \eqref{eq:wgc2} and \eqref{eq:wgc3} cannot be used directly as criteria for delineating the landscape of string theory and should be modified appropriately.

Importantly, even if we go beyond the four-derivative truncation used in our theory and consider six- and higher-derivative operators, the WGC proposals~\eqref{eq:wgc2} and \eqref{eq:wgc3} will still be violated in our theory.  The mismatch between the mass shift and the entropy shift in \eqref{eq:wgc2} arises entirely because of the topological nature of the Gauss-Bonnet term in the four-derivative Lagrangian, and the same mismatch will be present even if higher-derivative operators are incorporated into our theory.  For the WGC proposal in~\eqref{eq:wgc3}, as long as we consider a sufficiently large black hole, the contributions of any six-derivative or higher terms in the action to the Wald entropy will be suppressed by powers of the size of the black hole~\cite{Sen:2012dw,Cheung:2018cwt,Cheung:2019cwi}.  This is in contrast to the topological Euler characteristic term in the Wald entropy in equation~\eqref{eq:entropy_final}, which is constant and does not scale with the size of the black hole.  Thus, if we consider a sufficiently large black hole for which this topological term in the entropy causes the entropy shift to be negative in our four-derivative theory, going beyond four-derivative order for this particular black hole will not ameliorate the situation, and so the proposal \eqref{eq:wgc3} will still not hold for general black holes.

The only connection between the charge-to-mass ratio corrections and the entropy corrections that black holes in our theory satisfy in general is the universal relation proved in~\cite{Goon:2019faz}, which for the theory at hand takes the form
\begin{equation}\begin{aligned}
	\frac{\partial M_\text{ext}(Q)}{\partial c_1} &= \lim_{M \to M_\text{ext}(Q)}\left(- T(M,Q)\, \frac{\partial S(M,Q)}{\partial c_1}\right)~, \\
	\frac{\partial M_\text{ext}(Q)}{\partial c_2} &= \lim_{M \to M_\text{ext}(Q)}\left(- T(M,Q)\, \frac{\partial S(M,Q)}{\partial c_2}\right)~,
\label{eq:goon}
\end{aligned}\end{equation}
where $M_\text{ext}(Q)$ is the mass of the black hole in the extremal limit at a fixed charge $Q$, $T(M,Q)$ and $S(M,Q)$ are the temperature and entropy of the black hole expressed in terms of the mass and charge, and $c_1$, $c_2$ are the coefficients of the four-derivative terms in the action.  These relations essentially follow from assuming that the laws of thermodynamics hold even when higher-derivative terms are present, which is a fairly mild assumption.  Unfortunately, these relations are not especially useful in the case at hand because, despite their appearance, they do not actually lead to useful relations between the entropy and mass of the black hole.  The topological correction to the entropy coming from the Gauss-Bonnet term is constant in temperature, and so taking the extremal limit on the right-hand side of \eqref{eq:goon} implies that the topological correction vanishes as $T \to 0$.  So, as also discussed previously in~\cite{Cano:2019oma}, these general relations cannot be leveraged to constrain the higher-derivative coupling constants when the Gauss-Bonnet term is present in the action.

%%%%%%%%%%%%%%%%%
\section{Supersymmetric localization and holography}
\label{sec:susyloc}
%%%%%%%%%%%%%%%%%

Our discussion so far has been restricted mostly to 4d supergravity. The connection of the results in Sections~\ref{sec:onshell}, \ref{sec:univ}, and \ref{sec:BHTD} to holography was made without specifying a particular embedding of the supergravity theory in string or M-theory, or put differently, without specifying the dual 3d $\mathcal{N}=2$ SCFT. Here we change gears and show how the unknown coefficients in the supergravity HD action can be determined using the holographic results above in conjunction with supersymmetric localization results for specific classes of 3d $\mathcal{N}=2$ SCFTs. Our focus is on 3d SCFTs arising on the world-volume of $N$ coincident M2-branes in M-theory, i.e. parity invariant rank $N$ Chern-Simons matter theories. In the large $N$ limit, these models admit a holographic description in terms AdS$_4\times X^7$ Freund-Rubin solutions of 11d supergravity, where $X^7$ is a 7d Sasaki-Einstein manifold threaded by $N$ units of flux. As shown in \cite{Gauntlett:2007ma} 11d supergravity on such a background admits a consistent truncation to the minimal 4d $\mathcal{N}=2$ supergravity we study in this work. This consistent truncation has been established at the two-derivative level of the 11d and 4d supergravity theories. We will assume here that the consistent truncation also exists in the presence of higher-derivative corrections in 4d and 11d. This assumption is justified also by the field theory universality of partition functions discussed in \cite{Benini:2015bwz,Azzurli:2017kxo,Bobev:2017uzs,Bobev:2019zmz}, see also \cite{Hosseini:2020mut}. It will be very interesting to generalize the consistent truncation results of \cite{Gauntlett:2007ma} by studying the eight-derivative corrections to 11d supergravity and reducing the resulting Lagrangian on general Sasaki-Einstein manifolds. This is a technically involved calculation that we will not pursue here. The results presented below pass several highly non-trivial consistency checks which serve as strong additional justification for our assumptions. We note that we focus on AdS$_4$/CFT$_3$ dual pairs with embedding in M-theory for two main reasons. First, in AdS$_4$ Freund-Rubin backgrounds of 11d supergravity and M-theory there is a single dimensionless parameter given by the ratio of the AdS$_4$ length scale and the 11d Planck length, which is related to a power of $N$. This in turn simplifies the map between the dimensionless parameters $(L^2/G_N,c_1,c_2)$ of the 4d HD theory and this single 11d parameter.\footnote{For AdS$_4$ backgrounds in type II string theory we have an additional dimensionless quantity given by the string coupling constant $g_s$ which in general may complicate the map between 4d and 10d parameters.} Second, for precisely this class of 3d $\mathcal{N}=2$ SCFTs there are powerful supersymmetric localization techniques that have been explored in detail in the large $N$ limit and there is a collection of explicit results that facilitate our analysis.

The two-derivative consistent truncation combined with the flux quantization in M-theory and the standard holographic dictionary lead to the following leading order scaling of the dimensionless ratio between the AdS$_4$ scale and the 4d Newton constant
\begin{equation}\label{eq:L2GNN32}
\frac{L^2}{2G_N} =  A\,N^{\frac{3}{2}}\,.
\end{equation}
The real constant $A$ does not scale with $N$ and is determined by the volume of the internal Sasaki-Einstein manifold. More specifically, one finds $A = \sqrt{\frac{2\pi^4}{27 {\rm vol}(X^7)}}$.\footnote{When the Sasaki-Einstein manifold is $S^7$ we take it to have unit radius, i.e. the volume of the sphere is ${\rm vol}(S^7) = \frac{\pi^4}{3}$.} This result is also corroborated by supersymmetric localization in the dual CFT, see for instance \cite{Marino:2011nm} for a review.

It is expected that the higher-derivative corrections to M-theory will modify the leading order relation in \eqref{eq:L2GNN32} by a term proportional to $N^{\frac{1}{2}}$. In addition, the  coefficients $c_{1,2}$ in the four-derivative supergravity Lagrangian \eqref{eq:PHD} are also expected to scale as $N^{\frac{1}{2}}$. This can be summarized as follows
\begin{equation}
\label{eq:q-holo-dict}
	\frac{L^2}{2G_N} =  A\,N^{\frac{3}{2}} + a\,N^{\frac{1}{2}}\,, \qquad c_i =  v_i \frac{N^{\frac{1}{2}}}{32\pi} \, .
\end{equation}
Here the constants $(A,a,v_1,v_2)$ are all real numbers of order $1$, i.e. they do not scale with $N$. We emphasize that the identities in \eqref{eq:q-holo-dict} should be viewed as valid to leading and subleading order in the large $N$ expansion, i.e. they are expected to receive additional corrections that scale with smaller powers of $N$.

We can now use this information from the semiclassical limit of the UV complete quantum gravity theory, i.e. M-theory, in the 4d on-shell action in \eqref{eq:ionshell} to find
\begin{equation}
\label{eq:IAv1v2}
	I_{\text{HD}} = \pi\,\mathcal{F}\,\left[ A\,N^{\frac{3}{2}} + (a+v_2)\,N^{\frac{1}{2}}\right] - \pi\,(\mathcal{F} - \chi)\,v_1\,N^{\frac{1}{2}} \, .
\end{equation}
This expression can now be viewed as a supergravity holographic prediction for the free energy, or logarithm of the partition function, of the dual 3d $\mathcal{N}=2$ SCFT on a given compact Euclidean 3-manifold which captures the leading term and the first subleading correction in the large $N$ limit. The 3-manifold is determined by the conformal boundary of the asymptotically locally Euclidean AdS$_4$ solution with a given $\mathcal{F}$ and $\chi$.

Similarly, we find that in the large $N$ limit the two-point function coefficient $C_T$ in \eqref{eq:ctt3} of the SCFTs at hand should take the form
\begin{equation}\label{eq:CTsusyloccomp}
C_T = \frac{64}{\pi}\left[ A N^{\frac{3}{2}} +(a+v_2-v_1)N^{\frac{1}{2}} \right]\,.
\end{equation}
The round $S^3$ is a compact Euclidean background which preserves the full 3d $\mathcal{N}=2$ superconformal symmetry and thus plays a special role. In this case we have $\mathcal{F}=\chi=1$ and using $\eqref{eq:IAv1v2}$ we find the following result for the SCFT free energy on the round $S^3$
\begin{equation}\label{eq:FS3susyloccomp}
F_{S^3} = \pi\,\left[ A\,N^{\frac{3}{2}} + (a+v_2)\,N^{\frac{1}{2}}\right]  \, .
\end{equation}
As we discuss below there are supersymmetric localization results in the literature for two classes of SCFTs which lead to expressions for $C_T$ and $F_{S^3}$ of the form in \eqref{eq:CTsusyloccomp} and \eqref{eq:FS3susyloccomp}.\footnote{See \cite{Pestun:2016zxk} for a review on supersymmetric localization and further references.}. This allows us to uniquely fix the constants $A$, $a+v_2$, and $v_1$ for this class of models.

We have fixed the dependence of the 4d dimensionless constants on the large parameter $N$ using a microscopic embedding in M-theory. However, one can in principle use more bottom-up arguments, like the ones in \cite{Camanho:2014apa}, to argue that the four-derivative coefficients $c_1$ and $c_2$ are parametrically smaller than the  dimensionless parameter, $L^2/G_N$, that controls the two-derivative supergravity action. By the same token, the coefficients of six- and higher-derivative terms in the 4d supergravity Lagrangian should be parametrically smaller than the four-derivative coefficients $c_1$ and $c_2$. In general the large parameter that controls this higher-derivative expansion is the dimension of the lightest higher-spin single-trace operator in the CFT. This was denoted by $\Delta_{\rm gap}$ in \cite{Camanho:2014apa} and it will be interesting to revisit their discussion in the context of the higher-derivative supergravity results discussed here. We discuss this further in Section~\ref{sec:ExtGen}.

%%%%%%%%%%%%%%%%%%%%%
\subsection{Chern-Simons matter theories and supersymmetric localization}
%%%%%%%%%%%%%%%%%%%%%

To implement the idea outlined above we focus on two classes of SCFTs arising from M2-branes. The first is the ABJM theory, see \cite{Aharony:2008ug}. This is an $\mathcal{N}=6$ $U(N)_{-k}\times U(N)_k$ gauge theory coupled to matter where the integer $k$ specifies the Chern-Simons levels for both gauge groups. This model arises from M2-branes probing $\mathbb{C}^4/\mathbb{Z}_k$ where the $\mathbb{Z}_k$ orbifold acts on the four complex coordinates as
\begin{equation}\label{eq:ZkorbC4}
(z_1,z_2,z_3,z_4) \to e^{2\pi i/k}(z_1,z_2,z_3,z_4)\,.
\end{equation}
For $k=1,2$ there is supersymmetry enhancement to $\mathcal{N}=8$, see \cite{Bashkirov:2010kz}. The other model is 3d $\mathcal{N}=4$ SYM coupled to 1 adjoint hypermultiplet and $N_f$ hypermultiplets in the fundamental representation. This theory is realized on the worldvolume of M2-branes probing $\mathbb{C}^4/\mathbb{Z}_{N_f}$ where the $\mathbb{Z}_{N_f}$ orbifold action is
\begin{equation}\label{eq:ZNforbC4}
(z_1,z_2,z_3,z_4) \to (z_1,z_2,e^{2\pi i/N_f}z_3,e^{2\pi i/N_f}z_4)\,.
\end{equation}
See \cite{Mezei:2013gqa} for more details of this model and its embedding in M-theory. As shown in \cite{Bashkirov:2010kz} for $N_f=1$ the SYM theory preserves $\mathcal{N}=8$ supersymmetry and is actually dual to the ABJM theory at $k=1$. We will focus on the large $N$ limit of these two models with the parameters $k$ and $N_f$ held fixed. In this limit there is a holographically dual description in terms of 11d supergravity on AdS$_4\times S^7/\mathbb{Z}_k$ and AdS$_4\times S^7/\mathbb{Z}_{N_f}$, respectively. The precise form of the orbifold action on $S^7$ can be determined from \eqref{eq:ZkorbC4} and \eqref{eq:ZNforbC4} by embedding $S^7$ in $\mathbb{C}^4$.

%\NPB{Summary of susy localization results for the round $S^3$ partition function and $C_T$ for two classes of models - ABJM and $\mathcal{N}=4$ SYM with one adjoint $N_f$ flavors.}

The round $S^3$ free energy of ABJM at large $N$ for general fixed $k$ has the following leading and subleading terms, see \cite{Fuji:2011km,Marino:2011eh},
\begin{equation}\label{eq:ABJMround}
F_{S^3}^{\rm ABJM} = \frac{\sqrt{2k}\pi}{3}  N^{\frac{3}{2}} - \frac{\sqrt{2k}\pi}{6} \left(\frac{k}{8}+\frac{1}{k}\right) N^{\frac{1}{2}} \,.
\end{equation}
The round $S^3$ free energy of the SYM theory has also been computed by supersymmetric localization in the large $N$, see \cite{Mezei:2013gqa}. For the leading and subleading terms one finds
\begin{equation}\label{eq:SYMround}
F_{S^3}^{N_f} = \frac{\sqrt{2N_f} \pi}{3} N^{\frac{3}{2}} - \frac{\sqrt{2N_f}\pi}{4}\left(\frac{1}{N_f}-\frac{N_f}{4}\right)  N^{\frac{1}{2}} \,.
\end{equation}

For both families of SCFTs it is possible to compute $C_T$ to leading and subleading order in the large $N$ limit, see \cite{Chester:2020jay} as well as the earlier work in \cite{Agmon:2017xes,Chester:2018aca}. For the ABJM theory an important ingredient in deriving this result is a supersymmetric Ward identity that relates derivatives with respect to the real masses of the $S^3$ partition function of the theory to $C_T$. The dependence of the $S^3$ partition function of the ABJM theory on real masses was studied in \cite{Nosaka:2015iiw}. 

The expression for $C_T$ for the ABJM theory at fixed $k$ in the large $N$ limit can be found in \cite{Chester:2018aca}, see also \cite{Agmon:2017xes} for the expressions for $k=1,2$, and reads\footnote{Note that there is a typo in Equation (2.11) of \cite{Chester:2018aca}. We are grateful to Shai Chester and Silviu Pufu for a useful communication on this.}
\begin{equation}\label{eq:CTABJMCPY}
C_T^{\rm ABJM} = \frac{64 \sqrt{2k}}{3\pi} N^{\frac{3}{2}} + \frac{4(16-k^2)\sqrt{2}}{3\pi\sqrt{k}} N^{\frac{1}{2}}\,.
\end{equation}
For the SYM theory at fixed $N_f$ the leading terms in the large $N$ limit were found in \cite{Chester:2020jay} and take the form
\begin{equation}\label{eq:CTNfShai}
C_T^{N_f} = \frac{64 \sqrt{2N_f}}{3\pi} N^{\frac{3}{2}} + \frac{4\sqrt{2}}{3\pi} \frac{8+7N_f^2}{\sqrt{N_f}} N^{\frac{1}{2}}\,.
\end{equation}

We can now combine the supersymmetric localization results in \eqref{eq:ABJMround}, \eqref{eq:SYMround}, \eqref{eq:CTABJMCPY}, and \eqref{eq:CTNfShai} with the supergravity calculation in \eqref{eq:CTsusyloccomp} and \eqref{eq:FS3susyloccomp} to find the constants appearing in the higher-derivative supergravity on-shell action \eqref{eq:IAv1v2}. For the ABJM theory we find
\begin{equation}\label{eq:Av1v2ABJM}
A = \frac{\sqrt{2 k}}{3}\,, \qquad a+v_2 = -\frac{k^2 + 8}{24 \sqrt{2 k}}\,, \qquad v_1=-\frac{1}{\sqrt{2 k}}\,.
\end{equation}
For the $\mathcal{N}=4$ SYM theory with 1 adjoint and $N_f$ fundamental hypers we have
\begin{equation}\label{eq:Av1v2SYM}
A = \frac{\sqrt{2 N_f}}{3}\,, \qquad a+v_2 = \frac{N_f^2 - 4}{8 \sqrt{2 N_f}}\,, \qquad v_1=-\frac{N_f^2 + 5}{6 \sqrt{2 N_f}}\,.
\end{equation}

The results for the round sphere free energy and $C_T$ above are sufficient to fix the unknown coefficients in the supergravity on-shell action \eqref{eq:IAv1v2}. We can now use this on-shell action result to calculate the logarithm of the partition function of the ABJM theory or the $\mathcal{N}=4$ SYM on any compact three-dimensional manifold to leading and subleading order in the large $N$ expansion. Before we move on to present a few explicit results that illustrate the utility of this approach it is worthwhile to demonstrate that our calculations pass several non-trivial consistency checks.

In \cite{Hatsuda:2016uqa} it was shown that the large $N$ limit of the squashed $S^3$ partition function with $U(1)\times U(1)$ invariance of the $\mathcal{N}=4$ SYM theory can be evaluated for the special value of the squashing parameter $b^2=3$ by exploiting a relation to matrix models arising from topological string theory on a non-compact CY manifold. More specifically it was found that for $b^2=3$ the leading and subleading term in the large $N$ limit are given by
\begin{equation}\label{eq:FS3b23}
F_{S^3_{b^2=3}}^{N_f} = \frac{4\sqrt{2N_f}}{9} \pi N^{\frac{3}{2}} -  \frac{2\sqrt{2N_f}}{3}\left(\frac{7}{24N_f}-\frac{N_f}{6}\right) \pi N^{\frac{1}{2}} \,.
\end{equation}
The result above for $N_f=1$ is the same as for the ABJM theory with $k=1$.\footnote{We are not aware of an extension of the calculation of \cite{Hatsuda:2016uqa} applicable to the ABJM theory at general level $k$.} This result can be compared to our holographic calculation. For the squashed sphere partition function we found that $\mathcal{F} = \frac{1}{4}(b+b^{-1})^2$ and $\chi=1$, see \eqref{eq:FchiU1U1}.\footnote{The supergravity squashing parameter $s$ is related to the parameter $b$ more commonly used in the supersymmetric localization literature via $s=b^2$.} We can combine this with the results for $(A,v_1,a+v_2)$ in \eqref{eq:Av1v2SYM} and the on-shell action result in \eqref{eq:IAv1v2} to see that indeed for $b^2=3$, the result for $F_{S^3_{b^2=3}}^{N_f}$ in \eqref{eq:FS3b23} agrees with the holographic prediction. Very recently, prompted by our results in \cite{Bobev:2020egg}, the subleading terms in the large $N$ expansion of the ABJM squashed $S^3$ partition function for $k=1,2$ were studied in \cite{Chester:2021gdw}. The authors of \cite{Chester:2021gdw} studied the 4th and 5th derivative of the free energy with respect to $b$ evaluated at $b=1$ and found perfect agreement with our supergravity results for the on-shell action. We note that the results of \cite{Chester:2021gdw} can be viewed as computing the leading non-trivial terms in an expansion of the squashed sphere free energy around $b=1$ while our supergravity on-shell action results in \eqref{eq:IAv1v2}, see also \eqref{eq:FS3bABJM} and \eqref{eq:FS3bSN4SYM} below, are valid for a general squashing parameter $b$. 

Another supersymmetric partition function for this class of SCFTs that has been studied to subleading order in the large $N$ expansion is the ABJM topologically twisted index on $S^1\times S^2$. This index was introduced in \cite{Benini:2015noa} and was used in the large $N$ limit in \cite{Benini:2015eyy} for a microscopic derivation of the Bekenstein-Hawking entropy for supersymmetric AdS$_4$ black holes. The subleading terms in the large $N$ expansion of the ABJM topologically twisted index on $S^1\times S^2$ were studied numerically in \cite{Liu:2017vll}. To a good numerical accuracy it was found that for the so-called universal twist, see \cite{Azzurli:2017kxo}, the index reads
\begin{equation}\label{eq:S1S2TTI}
\log Z_{S^1\times S^2}^{\rm ABJM} =  -\frac{\sqrt{2k}\pi}{3} N^{\frac{3}{2}} + \frac{\sqrt{2k}\pi}{3} \left(\frac{k}{16}+\frac{2}{k}\right)N^{\frac{1}{2}}\,.
\end{equation}
This supersymmetric partition function should be compared with the higher-derivative on-shell action of the $\mathfrak{g}=0$ Euclidean Romans solution for which we found $\mathcal{F}= 1-\mathfrak{g}$ and $\chi = 2(1-\mathfrak{g})$, see \eqref{eq:AdSRNFchidef}. Using this, together with \eqref{eq:Av1v2ABJM} and the on-shell action in \eqref{eq:IAv1v2}, we indeed find that the supersymmetric localization result for the topologically twisted index in \eqref{eq:S1S2TTI} agrees with our holographic calculation.

%%%%%%%%%%%%%%%%%%%%%
\subsection{Holographic predictions at order $N^{\frac{1}{2}}$}
%%%%%%%%%%%%%%%%%%%%%

These non-trivial consistency checks of our holographic results increase our confidence that one can use the results for the constants $(A,v_1,a+v_2)$ in \eqref{eq:Av1v2ABJM} and \eqref{eq:Av1v2SYM} in conjunction with the on-shell action in \eqref{eq:IAv1v2} to derive supergravity predictions for the leading and subleading behavior in the large $N$ limit of other SCFT partition functions. Below we collect some explicit results in this spirit by focusing on the partition functions of the two classes of 3d SCFTs discussed above placed on compact Euclidean manifolds that arise at the asymptotic boundary of the supergravity solutions we described in Section~\ref{subsec:examples}.

We start with the $U(1)\times U(1)$ squashed sphere partition function with a squashing parameter $b$ which can be determined from the supergravity solution in Section~\ref{subsubsec:U1U1sq} with $s=b^2$. For the ABJM theory we find that the result for the free energy to order $N^{1/2}$ is
\begin{equation}\label{eq:FS3bABJM}
\begin{split}
F_{S^3_b}^{\rm ABJM} = \tfrac{\pi\sqrt{2k}}{12}\left[\left(b+\tfrac{1}{b}\right)^2\left(N^{\frac{3}{2}}+\tfrac{16-k^2}{16k}N^{\frac{1}{2}}\right)   - \tfrac{6}{k}N^{\frac{1}{2}}\right]\,. 
\end{split}
\end{equation}
Similarly for the squashed sphere free energy of the $\mathcal{N}=4$ SYM theory we find
\begin{equation}\label{eq:FS3bSN4SYM}
\begin{split}
F_{S^3_b}^{N_f} = \tfrac{\pi\sqrt{2N_f}}{12}\left[\left(b+\tfrac{1}{b}\right)^2\left(N^{\frac{3}{2}}+\left(\tfrac{1}{2N_f}+\tfrac{7N_f}{16}\right)N^{\frac{1}{2}}\right)   - \tfrac{N_f^2+5}{N_f}N^{\frac{1}{2}}\right]\,. 
\end{split}
\end{equation}

The topologically twisted index for the universal twist and general smooth Riemann surface of genus $\mathfrak{g}$ can be found by using the Euclidean Romans solutions presented in Section~\ref{subsubsec:EucRomans}. For the ABJM theory we find the result
\begin{equation}\label{eq:TTIABJM}
\begin{split}
\log Z_{S^1\times \Sigma_{\mathfrak{g}}}^{\rm ABJM} = (\mathfrak{g}-1)\tfrac{\pi\sqrt{2k}}{3}\left(N^{\frac{3}{2}} -\tfrac{32+k^2}{16k} N^{\frac{1}{2}}\right)\,.
\end{split}
\end{equation}
For the partition function of more general Seifert manifolds determined by an $S^1$ fibration over $\Sigma_{\mathfrak{g}}$ we can employ the AdS-Taub-Bolt solution described around \eqref{eq:BoltpmFchi} to find
\begin{equation}\label{eq:SeifABJM}
\begin{split}
\log Z_{\mathcal{M}_{\mathfrak{g},p}}^{\rm ABJM} = \tfrac{\pi\sqrt{2k}}{3}\left((\mathfrak{g}-1 \pm \tfrac{p}{4})\left(N^{\frac{3}{2}} +\tfrac{16-k^2}{16k} N^{\frac{1}{2}}\right)-\tfrac{3}{k}(\mathfrak{g}-1)N^{\frac{1}{2}}\right)\,.
\end{split}
\end{equation}
We can proceed similarly for the $\mathcal{N}=4$ SYM theory and find that the topologically twisted index takes the form
\begin{equation}\label{eq:TTIN4SYM}
\begin{split}
\log Z_{S^1\times \Sigma_{\mathfrak{g}}}^{N_f} = (\mathfrak{g}-1)\tfrac{\pi\sqrt{2N_f}}{3}\left(N^{\frac{3}{2}} -\tfrac{32+N_f^2}{16N_f} N^{\frac{1}{2}}\right)\,,
\end{split}
\end{equation}
while for more general Seifert manifolds we have
\begin{equation}\label{eq:SeifN4SYM}
\begin{split}
\log Z_{\mathcal{M}_{\mathfrak{g},p}}^{N_f} = \tfrac{\pi\sqrt{2N_f}}{3}\left((\mathfrak{g}-1 \pm \tfrac{p}{4})\left(N^{\frac{3}{2}} +\left(\tfrac{1}{2N_f}+\tfrac{7N_f}{16}\right)N^{\frac{1}{2}}\right)- \tfrac{N_f^2+5}{2N_f}(\mathfrak{g}-1)N^{\frac{1}{2}}\right)\,.
\end{split}
\end{equation}
The Euclidean Romans solutions with $\mathfrak{g}>1$ admit an analytic continuation to a regular supersymmetric Lorentzian black hole. The Bekenstein-Hawking entropy of this universal black hole is captured by the leading $N^{\frac{3}{2}}$ term in the topologically twisted index in \eqref{eq:TTIABJM} and \eqref{eq:TTIN4SYM}, see \cite{Benini:2015eyy,Azzurli:2017kxo}. The results in \eqref{eq:TTIABJM} and \eqref{eq:TTIN4SYM} lead to a precise prediction for the first subleading correction to the black hole entropy which we discuss further below.

The superconformal index of a 3d SCFT is another important physical observable that contains detailed information about the spectrum of BPS operators. By using the supersymmetric AdS-Kerr-Newman solution discussed in Section~\ref{subsubsec:AdSKN} we can find the leading and subleading terms in the large $N$ limit of the superconformal index. For the ABJM theory we find
\begin{equation}\label{eq:FS1S2ABJM}
\begin{split}
-\log Z_{S^1\times_{\omega} S^2}^{\rm ABJM} = \tfrac{\pi\sqrt{2k}}{3}\left[\tfrac{(\omega+1)^2}{2\omega}\left(N^{\frac{3}{2}}+\tfrac{16-k^2}{16k}N^{\frac{1}{2}}\right)   - \tfrac{3}{k}N^{\frac{1}{2}}\right]\,. 
\end{split}
\end{equation}
We note that this result applies to the superconformal index with vanishing flavor fugacities and $\omega$ is the fugacity for the angular momentum on $S^2$. For the superconformal index of the $\mathcal{N}=4$ SYM theory we have the expression
\begin{equation}\label{eq:FS1S2N4SYM}
\begin{split}
-\log Z_{S^1\times_{\omega} S^2}^{N_f} = \tfrac{\pi\sqrt{2N_f}}{3}\left[\tfrac{(\omega+1)^2}{2\omega}\left(N^{\frac{3}{2}}+\left(\tfrac{1}{2N_f}+\tfrac{7N_f}{16}\right)N^{\frac{1}{2}}\right)   - \tfrac{N_f^2+5}{2N_f}N^{\frac{1}{2}}\right]\,. 
\end{split}
\end{equation}
The superconformal index should account for the entropy of the supersymmetric Kerr-Newman black hole presented in Section~\ref{subsubsec:AdSKN} and our results above lead to a prediction for the leading and subleading terms in this entropy as we show explicitly below.

We note in passing that there is a curious relation between the squashed $S^3$ free energy and the superconformal index. Namely we find that $2F_{S^3_b}=-\log Z_{S^1\times_{\omega} S^2}$ after setting $\omega=b^2$. This relation is valid for the leading and subleading terms in the large $N$ limit of the SCFT and follows from the specific form of the on-shell action in \eqref{eq:IAv1v2} and the particular values of $\mathcal{F}$ and $\chi$ for these curved manifolds. It will be interesting to understand whether this relation between the two supersymmetric partition functions can be extended to more subleading terms in the large $N$ expansion.

%%%%%%%%%%%%%%%%%%%%%
\subsection{Corrections to the Bekenstein-Hawking entropy}
%%%%%%%%%%%%%%%%%%%%%

As shown in Section~\ref{sec:BHTD} the Bekenstein-Hawking entropy of a black hole in the two-derivative Einstein-Maxwell theory is corrected by the four-derivative terms and takes the simple form in \eqref{eq:entropy_final}. Using the same logic we applied above to the on-shell action we can convert this expression into a prediction for the large $N$ limit of the entropy of the M2-brane system which comprises a given black hole when it is realized as an M-theory background. Using \eqref{eq:q-holo-dict} we find that the black hole entropy takes the form
\begin{equation}\label{eq:SBHcorrAv12}
S = \left(A\,N^{\frac{3}{2}} +(v_2+a-v_1)N^{\frac{3}{2}}\right) \frac{A_H}{2L^2} - \pi v_1 \chi(H) N^{\frac{1}{2}}\,.
\end{equation}
This is a general expression that can be applied to any black hole solution of the 4d Einstein-Maxwell equations of motion which in turn can be embedded as a solution of 11d supergravity using the consistent truncation results in \cite{Gauntlett:2007ma}. Below we present explicit results for the entropy of three different black hole solutions that should describe the coarse grained behavior of states in the ABJM and the $\mathcal{N}=4$ SYM SCFTs.

The Euclidean Romans solution presented in Section~\ref{subsubsec:EucRomans} with $Q=0$ and $\mathfrak{g}>1$ can be analytically continued to Lorentzian signature where it can be viewed as a supersymmetric magnetic AdS-Reissner-Nordstr\"om black hole, see \cite{Caldarelli:1998hg} and also \cite{Azzurli:2017kxo} where the universal nature of this black hole solution and its holographic interpretation were studied in more detail. The area and the Euler number of the horizon of this black hole are easy to calculate and read
\begin{equation}
A_H = 2\pi (\mathfrak{g}-1) L^2\,, \qquad \chi(H)=2(1-\mathfrak{g})\,.
\end{equation}
We can use this in \eqref{eq:SBHcorrAv12} together with the results for $(A,v_1,a+v_2)$ in \eqref{eq:Av1v2ABJM} to find that the entropy of this black hole, when interpreted as arising from microscopic states in the ABJM theory, reads
\begin{equation}
S_{\rm RN}^{\rm ABJM} = \pi (\mathfrak{g}-1)\frac{\sqrt{2 k}}{3}\,\left(N^{\frac{3}{2}} +\frac{16-k^2 }{16k}N^{\frac{1}{2}}\right)  -2\pi (\mathfrak{g}-1) \frac{1}{\sqrt{2 k}}  N^{\frac{1}{2}}\,.
\end{equation}
Similarly we can use \eqref{eq:Av1v2SYM} to find that the entropy for the case of the $\mathcal{N}=4$ SYM theory is
\begin{equation}
S_{\rm RN}^{N_f} = \pi (\mathfrak{g}-1)\frac{\sqrt{2 N_f}}{3}\,\left(N^{\frac{3}{2}} +\frac{7N_f^2 +8}{16 N_f}N^{\frac{1}{2}}\right)  - 2\pi (\mathfrak{g}-1) \frac{N_f^2 + 5}{6 \sqrt{2 N_f}}  N^{\frac{1}{2}}\,.
\end{equation}

The AdS-Kerr-Newman solution in Section~\ref{subsubsec:AdSKN} admits a supersymmetric limit in which it can be interpreted as a black hole with a regular horizon. The thermodynamic properties of the general non-supersymmetric AdS-Kerr-Newman black hole were studied in \cite{Caldarelli:1999xj}. In the supersymmetric limit the entropy of the black hole simplifies significantly, see for instance \cite{Cassani:2019mms,Bobev:2019zmz}, and one finds that the area and the Euler number of the horizon are
\begin{equation}
A_H = 2\pi L^2\left[\sqrt{1+4G_N^2Q^2}-1\right]\,, \qquad \chi(H)=2\,.
\end{equation}
The parameter $Q$ is the electric charge of the black hole solution and is given by
\begin{equation}
Q = \frac{m}{G_N\Xi^2}\sinh(2\delta)\,,
\end{equation}
where in the BPS black hole limit the parameters $(m,\alpha,\delta)$ in the general AdS-Kerr-Newman solution in \eqref{eq:metKN}--\eqref{eq:AKN} obey the relations in \eqref{eq:susyrelKN} and \eqref{eq:maKNBH}. Using these results we find that the supersymmetric AdS-Kerr-Newman black hole entropy with a microscopic implementation in the ABJM theory is
\begin{equation}
S_{\rm KN}^{\rm ABJM} = \pi \left[\sqrt{1+4G_N^2Q^2}-1\right]\frac{\sqrt{2 k}}{3}\,\left(N^{\frac{3}{2}} +\frac{16-k^2 }{16k}N^{\frac{1}{2}}\right)  +2\pi  \frac{1}{\sqrt{2 k}}  N^{\frac{1}{2}}\,.
\end{equation}
Similarly, for the $\mathcal{N}=4$ SYM the supersymmetric AdS-Kerr-Newman entropy we find
\begin{equation}
S_{\rm KN}^{N_f} = \pi \left[\sqrt{1+4G_N^2Q^2}-1\right]\frac{\sqrt{2 N_f}}{3}\,\left(N^{\frac{3}{2}} +\frac{7N_f^2 +8}{16 N_f}N^{\frac{1}{2}}\right)  + 2\pi  \frac{N_f^2 + 5}{6 \sqrt{2 N_f}}  N^{\frac{1}{2}}\,.
\end{equation}

The final example we consider is the AdS-Schwarzschild metric which describes a non-supersymmetric black hole. The metric is given by
\begin{equation}
ds^2 = \left(\frac{r^2}{L^2} + 1 - \frac{m}{r}\right)\,d\tau^2 + \left(\frac{r^2}{L^2} + 1 - \frac{m}{r}\right)^{-1}\,dr^2 + r^2 d\Omega_2^2 \, ,
\end{equation}
where $d\Omega_2^2$ is the metric on the round $S^2$. The location of the outer horizon~$r_+$ is related to the mass parameter $m$ as
\begin{equation}
m = \frac{r_+^3}{L^2} + r_{+} \, .
\end{equation}
The area and Euler number of the horizon are given by
\begin{equation}
A_H = 4\pi\,r_+^2\,, \qquad\qquad \chi(H) = 2\,.
\end{equation}
When the AdS-Schwarzschild solution is interpreted holographically as describing a thermal state in the ABJM theory the leading and subleading terms in the large $N$ expansion of its entropy can be determined using the results above and take the form
\begin{equation}
S_{\rm Sch}^{\rm ABJM} = \frac{2\pi r_{+}^2}{L^2}\frac{\sqrt{2 k}}{3}\,\left(N^{\frac{3}{2}} +\frac{16-k^2 }{16k}N^{\frac{1}{2}}\right)  +2\pi  \frac{1}{\sqrt{2 k}}  N^{\frac{1}{2}}\,.
\end{equation}
Similarly, for the $\mathcal{N}=4$ SYM AdS-Schwarzschild entropy we find
\begin{equation}
S_{\rm Sch}^{N_f} = \frac{2\pi r_{+}^2}{L^2}\frac{\sqrt{2 N_f}}{3}\,\left(N^{\frac{3}{2}} +\frac{7N_f^2 +8}{16 N_f}N^{\frac{1}{2}}\right)  + 2\pi  \frac{N_f^2 + 5}{6 \sqrt{2 N_f}}  N^{\frac{1}{2}}\,.
\end{equation}
It will be most interesting to understand how to reproduce these expressions for the thermal entropy in the dual SCFTs. Given that supersymmetry is broken by the finite temperature this is a hard problem to address in general.

%%%%%%%%%%%%%%%%%
\section{Extensions and generalizations}
\label{sec:ExtGen}
%%%%%%%%%%%%%%%%%

We can generalize the four-derivative supergravity construction in Section~\ref{sec:HD} in two important ways while still making use of the two HD invariants~$\mathrm{W}^2$ and~$\mathbb{T}$. The first extension is to include general higher-derivative terms of order six or higher. This is allowed by $\mathcal{N}=2$ supersymmetry and in general there could be an infinite series of such HD corrections. The other extension is to couple the minimal supergravity theory to physical vector multiplets and discuss the effect of the four-derivative corrections in this more general setting.\footnote{One could in general also couple the supergravity theory to arbitrary numbers of physical tensor multiplets or hypermultiplets but we will not discuss this possibility here.} We study both of these generalizations below.

Keeping only a single auxiliary hypermultiplet but an arbitrary number of vector multiplets labelled by $I = 0, 1, \ldots , n_V$, we can write a superconformally invariant Lagrangian density using the chiral density formula~\eqref{eq:chiral-superspace}, where~$\mathscr{L}_\pm$ is now built as homogeneous functions of the vector multiplets~$\mathcal{X}^I_\pm$ and of the chiral multiplets~$\mathcal{W}_\pm^2$ and~$\mathbb{T}_\pm$ presented in Section~\ref{sec:HD}. Such functions are called the prepotentials~$F^\pm(X^I_\pm,\mathcal{A}_\pm)$, and the bosonic Lagrangian resulting from the construction is~\cite{deWit:2017cle}
\begin{equation}
\label{eq:gen-action}
\begin{split}
e^{-1}\mathcal{L} =&\; e^{-K}\bigl(\tfrac16\,R - D\bigr) - N_{IJ}\,\mathcal{D}_\mu X^I_+\mathcal{D}^\mu X^J_- + \tfrac18\,F_I^+\,\widehat{F}_{ab}^{+I} T^{ab+} + \tfrac18\,F_I^-\,\widehat{F}_{ab}^{-I} T^{ab-} \\
&\;-\tfrac14\,F_{IJ}^+\,\widehat{F}_{ab}^{-I}\widehat{F}^{ab-J} - \tfrac14\,F_{IJ}^-\,\widehat{F}_{ab}^{+I}\widehat{F}^{ab+J} \\
&\;+\tfrac18\,N_{IJ}\,Y_{ij}^I Y^{ijJ} + \tfrac{1}{32}\,F^+\bigl(T_{ab}^+)^2 + \tfrac1{32}\,F^-\bigl(T_{ab}^-\bigr)^2 \\
&\;+\chi_\mathrm{H}\bigl(\tfrac16\,R + \tfrac12\,D\bigr) - \tfrac12\,\varepsilon^{ij}\Omega_{\alpha\beta}\,\mathcal{D}_\mu A_i{}^\alpha \mathcal{D}^\mu A_j{}^\beta \\
&\;+2\,g^2\,\varepsilon^{ij}\Omega_{\alpha\beta}\,A_i{}^\alpha\,(\xi_I X^I_+)(\xi_J X^J_-)\,t^\beta{}_\gamma t^\gamma{}_\delta\,A_j{}^\delta - \tfrac12\,g\,\Omega_{\alpha\beta}\,A_i{}^\alpha\,(\xi_I Y^{ijI})\,t^\beta{}_\gamma\,A_j{}^\gamma \\
&\;- \tfrac12\,F^+_{\mathcal{A}}\,\mathcal{C}_+ - \tfrac12\,F^-_{\mathcal{A}}\,\mathcal{C}_- - \tfrac14\,F^+_{\mathcal{A}I}\,\mathcal{B}^{ij}_+\,Y_{ij}^I - \tfrac14\,F^-_{\mathcal{A}I}\,\mathcal{B}^{ij}_-\,Y_{ij}^I \\
&\;+ \tfrac12\,F^+_{\mathcal{A}I}\,\mathcal{F}_{ab}^-\,\widehat{F}^{ab-I} + \tfrac12\,F^-_{\mathcal{A}I}\,\mathcal{F}_{ab}^+\,\widehat{F}^{ab+I} + \tfrac18\,F^+_{\mathcal{A}\mathcal{A}}\,\mathcal{B}_{ij+}\mathcal{B}^{ij}_+ + \tfrac18\,F^-_{\mathcal{A}\mathcal{A}}\,\mathcal{B}_{ij-}\mathcal{B}^{ij}_- \\
&\;+ \tfrac14\,F^+_{\mathcal{A}\mathcal{A}}\,\mathcal{F}_{ab}^-\mathcal{F}^{ab -} + \tfrac14\,F^-_{\mathcal{A}\mathcal{A}}\,\mathcal{F}_{ab}^+\mathcal{F}^{ab+} \, .
\end{split}
\end{equation}  
We have a free gauging parameter $\xi_I$ for each vector field, parametrizing the gauging of the auxiliary hypermultiplet isometry (which is the R-symmetry from an on-shell point of view). The parameters $\xi_I$ are called Fayet-Iliopoulos (FI) parameters. They are the natural generalization of the gauge coupling $g$ in minimal supergravity. The fields~$(\mathcal{A}_\pm,\,\mathcal{B}^{ij}_\pm,\,\mathcal{F}^{\mp}_{ab},\,\mathcal{C}_\pm)$ denote the components of the (anti-)chiral multiplets whose lowest components are a linear combination of the lowest components of the~$\mathcal{W}_\pm^2$ and~$\mathbb{T}_\pm$ multiplets,
\begin{equation}
\label{eq:fullA}
\mathcal{A}_\pm \equiv \frac1{32}\,c_1\,A_\pm\big\vert_{\mathcal{W}^2_\pm} + c_2\,A_\pm\big\vert_{\mathbb{T}_\pm} \, .
\end{equation}
With this choice, supersymmetry uniquely fixes the form of the remaining bosonic fields $(\mathcal{B}^{ij}_\pm,\,\mathcal{F}^{\mp}_{ab},\,\mathcal{C}_\pm)$ that feature prominently in the above Lagrangian density. We have also defined
\begin{equation}
e^{-K} \equiv X^I_+F_I^- + X^I_- F_I^+ \, , \qquad \text{and} \qquad N_{IJ} \equiv F_{IJ}^+ + F_{IJ}^- \, ,
\end{equation}
where $F^\pm_{\mathcal{A}},\,F^\pm_{I}$, etc\ldots~denote derivatives of the functions $F^\pm$ with respect to the fields $\mathcal{A}_\pm,\,X^I_\pm$, and so on. The Lagrangian density~\eqref{eq:gen-action} is fully specified by the choice of gauging as well as by the choice of prepotentials. The latter are homogeneous functions of degree two in the scalars $X^I_\pm$ (each of Weyl weight~$w=1$) and~$\mathcal{A}_\pm$ (of Weyl weight~$w=2$). Schematically, every additional power of $\mathcal{A}_\pm$ in $F^\pm$ leads to new terms with two additional derivatives in the Lagrangian. Thus, the derivative expansion can be written as an expansion of the prepotentials in powers of $\mathcal{A}_\pm$, while the coupling to physical vector multiplets leads to additional freedom in the functional form of $F^\pm(X^I_\pm)$ at each power of $\mathcal{A}_\pm$.

The two possible generalizations of the discussion in Section~\ref{sec:HD}, i.e. including more than four derivatives and coupling the minimal theory to physical vector multiplets, are in principle independent of each other. We could tackle both at once and ask the question of how to deal with arbitrary higher-derivative terms in matter-coupled gauged supergravity, which would amount to using the most general prepotentials compatible with the given field content in the Lagrangian density~\eqref{eq:gen-action}. However, since each of the two generalizations poses its own specific technical challenges, we opt for splitting the discussion in two different subsections. We begin by analyzing the effect of arbitrary higher-derivative terms on full-BPS solutions, and in the process make some observations on the possibility of extending these results to less supersymmetric configurations. The detailed analysis of the BPS conditions is relegated to Appendix~\ref{app:susy-vars}. We then discuss the addition of physical vector multiplets, and focus more specifically on four-derivative terms in the so-called $STU$ model. There, we are able to relate our supergravity results to subleading terms in the large $N$ expansion of the Airy function that controls the supersymmetric partition function of the ABJM theory on the round $S^3$. This in turn prompts us to make a natural conjecture for the form of the four-derivative prepotential for the $STU$ model.  

%%%%%%
\subsection{Infinite derivative expansion}
\label{subsec:minHDinf}
%%%%%%

Let us consider the case of minimal supergravity already discussed in Section~\ref{sec:HD}. The index $I = 0$ in the density \eqref{eq:gen-action} becomes obsolete and we can think of it as only running over the auxiliary scalars $X_\pm$, with $\xi_0 = 1$. It is important to note that the $\mathrm{W}^2$ and $\mathbb{T}$ invariants in minimal supergravity can already lead to an infinite order of higher-derivatives {\it on-shell}. This was formally discussed in detail already around \eqref{eq:C-phi-prime}. Very concretely, we can take the following prepotentials:
\begin{equation}
\label{eq:prepot}
F^\pm(X_\pm,\mathcal{A}_\pm) = (X_\pm)^2 + \mathcal{A}_\pm + \sum_{n=1}^{\infty}\,d_n\,\frac{(\mathcal{A}_\pm)^{n+1}}{(X_\pm)^{2n}} \, ,
\end{equation}
where $d_n$ are real constants and we separated the first order in $\mathcal{A}_\pm$ to make it manifest that it precisely leads to the four-derivative terms we discussed extensively in Section~\ref{sec:HD}. We now turn to analyzing the pure AdS$_4$ solution of this HD supergravity theory and its on-shell action. We then discuss the near-horizon supersymmetric AdS$_2 \times \Sigma_{\mathfrak{g} > 0}$ solution and comment on the technical challenges arising from the presence of the HD terms.

%%%%%%
\subsubsection*{The AdS$_4$ action at all orders}
%%%%%%

Let us first look at the simplest case, the maximally (super)symmetric space AdS$_4$. As a first step, we set~$d_n = 0$, $\forall\;n \geq 1$. In this simple case, we have
\begin{equation}
F^\pm_I = 2\,X_\pm \, , \quad F^\pm_{IJ} = 2 \, , \quad F_{\mathcal{A}}^\pm = 1 \, , \quad F^\pm_{\mathcal{A}\mathcal{A}} = F^\pm_{\mathcal{A}I} = 0 \, .
\end{equation}
Furthermore, the conformal supergravity formalism allows us to analyze supersymmetry conditions \emph{off-shell}. We can thus derive the consequences of demanding that a field configuration be full-BPS independently of the precise form of the prepotentials~$F^\pm$. In Appendix~\ref{app:susy-vars}, we show that maximal supersymmetry highly constrains the value of the superconformal fields. The conditions we obtain after imposing the K-, D- and V-gauges are
\begin{equation}
F_{ab}^\pm = T_{ab}^\pm = R(A)_{ab}^\pm = R(\mathcal{V})^\pm_{ab}{}^i{}_j = D = C_{abcd} = 0 \, ,
\end{equation}
together with
\begin{equation}
\label{eq:fullBPScond}
Y_{ij} = 4\,g\,X_+\,X_-\,\varepsilon_{ik}\,t^k{}_j \, , \quad \text{and} \quad R = 48\,g^2\,X_+\,X_- \, ,
\end{equation}
where~$X_\pm$ are constant. If we now fix the A-gauge $X_+ = X_- \equiv X$, the resulting configuration has vanishing graviphoton and constant negative (after implementing the redefinitions in Footnote~\ref{foot:redef}) scalar curvature. It is therefore pure Euclidean AdS$_4$. On this configuration, we obtain simple expressions for the components of the (anti-)chiral multiplets beginning with~$\mathcal{A}_\pm$ given in~\eqref{eq:fullA},
\begin{equation}
\label{eq:full-BPS-chiral-bckgd}
\begin{split}
\mathcal{A}_\pm =&\; 2\,c_2\,(4\,g^2\,X^2 - D) \, , \qquad \mathcal{B}_{\pm ij} = -16\,c_2\,g\,X\,(4\,g^2\,X^2 + D)\,\varepsilon_{ik}\,t^k{}_j \, , \\
\mathcal{F}_{ab}^\pm =&\; 0 \, , \qquad\qquad\qquad\qquad\qquad \mathcal{C}_\pm = -2\,c_2\,\bigl(192\,g^4\,X^4 - 3\,D^2 - 2\,\square D\bigr) \, .
\end{split}
\end{equation}
Note that we have refrained from imposing the full-BPS~$D=0$ condition at this stage. The reason is that we must keep at least the terms linear in $D$ in the Lagrangian density in order to impose the corresponding equation of motion when going on-shell. On this full-BPS configuration,~\eqref{eq:gen-action} evaluates to
\begin{equation}
e^{-1}\mathcal{L} = 48\,g^2\,X^4 + D\,(\kappa^{-2} - 4\,X^2) + 384\,c_2\,g^4\,X^4 - 6\,c_2\,D^2 - 4\,c_2\,\square D \, .
\end{equation}
From this, we find that the $D$ equation of motion imposes
\begin{equation}
X = \frac{1}{2\,\kappa} \, , 
\end{equation}
after using the full-BPS condition~$D=0$. Having fixed the value of the scalar~$X$, we arrive at the following simple result for the Lagrangian density:
\begin{equation}
\label{eq:4der-action}
e^{-1}\mathcal{L}\big\vert_{\mathrm{EAdS}_4} = \bigl(1 + \alpha\bigr)\,\frac{3g^2}{\kappa^4} \, , \qquad \text{with} \qquad \alpha \equiv 8\,c_2\,g^2 \, .
\end{equation}
This agrees with the result derived in Section~\ref{sec:onshell}, i.e. we have established that putting all $d_n = 0$ in \eqref{eq:prepot} brings us back to the four-derivative result.

Let us now consider the full prepotential~\eqref{eq:prepot} and repeat the previous calculation. The BPS conditions \eqref{eq:fullBPScond}-\eqref{eq:full-BPS-chiral-bckgd}, being derived {\it off-shell}, remain the same. The full Lagrangian however depends on the quantities
\begin{equation}
\begin{split}
F_I^\pm =&\; 2\,X_\pm - \sum_{n=1}^{\infty}2n\,d_n\,\frac{(\mathcal{A}_\pm)^{n+1}}{(X_\pm)^{2n+1}} \, , \quad F_{IJ}^\pm = 2 + \sum_{n=1}^{\infty}2n(2n+1)\,\frac{(\mathcal{A}_\pm)^{n+1}}{(X_\pm)^{2n+2}} \, , \\
F^\pm_{\mathcal{A}} =&\; 1 + \sum_{n=1}^{\infty}(n+1)\,d_n\,\frac{(\mathcal{A}_\pm)^{n}}{(X_\pm)^{2n}} \, , \qquad F^\pm_{\mathcal{A}\mathcal{A}} = \sum_{n=1}^{\infty}n\,(n+1)\,d_n\,\frac{(\mathcal{A}_\pm)^{n-1}}{(X_\pm)^{2n}} \, , \\
F^\pm_{\mathcal{A}I} =&\; -\sum_{n=1}^{\infty}2\,n\,(n+1)\,d_n\,\frac{(\mathcal{A}_\pm)^{n}}{(X_\pm)^{2n+1}} \, .
\end{split}
\end{equation}
Plugging in the full-BPS configuration~\eqref{eq:full-BPS-chiral-bckgd}, we can work at first order in the field~$D$ to write down the~$D$ equation of motion, since~$D$ vanishes on a full-BPS configuration. This gives
\begin{equation}
\begin{split}
F_I^\pm =&\; 2\,X - \frac{4\,c_2}{X}\,\sum_{n=1}^{\infty}\,n\,d_n\,\alpha^n\,\bigl(4\,g^2\,X^2 - (n+1)\,D\bigr) + \mathcal{O}(D^2) \, , \\
F_{IJ}^\pm =&\; 2 + \frac{4\,c_2}{X^2}\,\sum_{n=1}^{\infty}\,n(2n+1)\,d_n\,\alpha^n\,\bigl(4\,g^2\,X^2 - (n+1)\,D\bigr) + \mathcal{O}(D^2) \, , \\
F^\pm_{\mathcal{A}} =&\; 1 + \frac{1}{4\,g^2\,X^2}\,\sum_{n=1}^{\infty}\,(n+1)\,d_n\,\alpha^n\,\bigl(4\,g^2\,X^2 - n\,D\bigr) + \mathcal{O}(D^2) \, , \\
F^\pm_{\mathcal{A}\mathcal{A}} =&\; \frac{1}{4\,g^2\,X^4}\,\sum_{n=1}^{\infty}\,n(n+1)\,d_n\,\alpha^{n-1}\,\bigl(4\,g^2\,X^2 - (n-1)\,D\bigr) + \mathcal{O}(D^2) \, , \\
F^\pm_{\mathcal{A}I} =&\; -\frac{1}{2\,g^2\,X^3}\,\sum_{n=1}^{\infty}\,n(n+1)\,d_n\,\alpha^n\,\bigl(4\,g^2\,X^2 - n\,D\bigr) + \mathcal{O}(D^2) \, .
\end{split}
\end{equation}
Armed with this, we can collect the linear terms in the Lagrangian~\eqref{eq:gen-action} that contribute to the~$D$ equation of motion. They are
\begin{align}
e^{-1}\mathcal{L} \ni D\,\Bigl[&\kappa^{-2} - 4\,X^2 + 4\,X^2\,\sum_{n=1}^{\infty}\bigl(2n(n+1) + n - n(n+1)(2n+1)\bigr)\,d_n\,\alpha^{n+1} \\
&\!\!\!- 4\,X^2\,\sum_{n=1}^{\infty}\,\bigl(3n(n+1) - 4n(n-1)(n+1) + 2n(n-3)(n+1)\bigr)\,d_n\,\alpha^{n+1}\Bigr] \, . \nonumber
\end{align}
The bracket must vanish on-shell, and we find that it simplifies to a rather compact expression. As a result, the~$D$ equation of motion simply implies
\begin{equation}
\kappa^{-2} - 4\,X^2\Bigl(1 - \sum_{n=1}^{\infty}\,n\,d_n\,\alpha^{n+1}\Bigr) = 0 \, .
\end{equation}
This fixes the value of the scalar~$X$,
\begin{equation}
X = \frac{1}{2\,\kappa\,\beta} \, , \qquad \text{with} \qquad \beta \equiv \Bigl(1 - \sum_{n=0}^{\infty}\,n\,d_n\,\alpha^{n+1}\Bigr)^{\frac{1}{2}} \, .
\end{equation}
We can now compute the infinite derivative on-shell action for the maximally supersymmetric EAdS$_4$ configuration whose conformal boundary is the round three-sphere. First, the Lagrangian density~\eqref{eq:gen-action} with prepotential~\eqref{eq:prepot} evaluates to
\begin{equation}
e^{-1}\mathcal{L}\big\vert_{\mathrm{EAdS}_4} = \frac{1}{\beta^{4}}\Bigl[1 + \sum_{n=0}^{\infty}\,d_n\,\alpha^{n+1}\Bigr]\,\frac{3g^2}{\kappa^4} \, ,
\end{equation}
where we have set~$d_0 \equiv 1$ in accordance with the terms linear in $\mathcal{A}_{\pm}$ in \eqref{eq:prepot}. Clearly, when all~$d_{n>0}$ vanish, we recover the expression in~\eqref{eq:4der-action}. As in Section~\ref{sec:HD}, the EAdS$_4$ scale is related to the gauge coupling of the supergravity theory via $L = g^{-1}\kappa$. Now, in order to find a finite on-shell action that has a direct holographic meaning we use the standard holographic renormalization counterterms\footnote{Here we assume that there are no additional subtleties involving counterterms arising from the full tower of higher-derivative terms.} discussed in Section~\ref{sec:onshell}. After taking into account a factor of $4 \pi^2 L^4 / 3$ from the integration over the four-dimensional manifold, we obtain the following result for the regularized on-shell action of Euclidean AdS$_4$ in the presence of the HD terms in~\eqref{eq:4der-action}
\begin{equation}
\label{eq:S3pfinftyder}
	I_{\mathrm{EAdS}_4} =  \frac{1}{\beta^{4}}\Bigl[1 + \sum_{n=0}^{\infty}\,d_n\,\alpha^{n+1}\Bigr]\,\frac{\pi L^2}{2  {G}_N} \, .
\end{equation}
Note that this is independent of the constant~$c_1$, meaning that the $\mathrm{W}^2$ invariant does not contribute to the on-shell action of EAdS$_4$. This can be understood intuitively from the fact that EAdS$_4$ is conformally flat with vanishing Weyl tensor. Note also that the full-BPS conditions imply {\it all} equations of motion {\it on-shell} \cite{Hristov:2009uj}, and instead in the {\it off-shell} formalism we needed {\it only} the addition of the $D$ equation of motion. 

The fact that the field $D$ is forced to vanish for maximally supersymmetric solutions presents a major technical simplification, since we only needed to keep the linear terms without any derivatives on $D$. For solutions that preserve only partial supersymmetry, or none at all, one immediately faces at least two types of complications: there are a number of additional equations of motions to be verified and one needs to keep all orders of $D$ since it does not generically vanish. An illuminating example of these complications is the half-BPS configuration AdS$_2 \times \Sigma_{\mathfrak{g}}$, that has been worked out in \cite{Hristov:2016vbm} in the presence of the $\mathrm{W}^2$ invariant only. Indeed one finds a non-vanishing {\it off-shell} value for $D$ and the need to additionally verify the Maxwell equations and the equations of motion for the auxiliary fields $Y^{ij}$ on top of the BPS analysis. In the presence of the $\mathbb{T}$ invariant this introduces an infinite tower of additional terms one needs to keep in the Lagrangian. Analyzing this in full detail is beyond the scope of this work and is left for the future.

%%%%%%%%
\subsection{Coupling to matter}
%%%%%%%%

We now study another generalization of the HD supergravity construction in Section~\ref{sec:HD}. We restrict to four-derivative supergravity actions, but couple the theory to a number $n_V$ of Abelian vector multiplets described again by the general Lagrangian in \eqref{eq:gen-action}. To this end we take the following prepotentials:
\begin{equation}
\label{eq:matter-4der}
F^\pm(X^I_\pm,\mathcal{A}_\pm) = F^{(2)\pm}(X^I_\pm) + F^{(0)\pm} (X^I_\pm)\,\mathcal{A}_\pm\, , \qquad I \in \{0, 1,\ldots n_V \}\, ,
\end{equation}
with $\mathcal{A}_\pm$ defined in \eqref{eq:fullA},  and~$F^{(2)\pm}$,~$F^{(0)\pm}$ real homogeneous functions of degree two and zero, respectively. We proceed by analyzing the full-BPS background and its on-shell action. We focus on the most prominent example of a matter-coupled 4d $\mathcal{N}=2$ gauged supergravity theory known as the $STU$ model. At the two-derivative level, this model admits an embedding in maximal 4d supergravity and can be obtained as a consistent truncation from 11d supergravity on $S^7$ \cite{deWit:1986oxb,Cvetic:1999xp,Nicolai:2011cy}. Part of our motivation to perform this analysis is to relate our supergravity results to the calculation of the subleading $N^{\frac{1}{2}}$ corrections to the $S^3$ free energy of the ABJM theory in the presence of real masses studied in \cite{Nosaka:2015iiw}, which we discuss in detail below. 

%%%%%%%%
\subsubsection*{The AdS$_4$ action}
%%%%%%%%

To evaluate the supergravity action on the full-BPS EAdS$_4$ background, we first need to slightly generalize the BPS constraints derived in Appendix~\ref{app:full-BPS} to include the additional constraints coming from matter fermions. In the matter-coupled theory, the full-BPS conditions we obtain after imposing the K-, D- and V-gauges are  
\begin{equation}
\mathcal{V}_\mu{}^i{}_j = -2\,g\,\xi_IW_\mu^I\,t^i{}_j = 0\, , \qquad Y_{ij}^I = 4\,g\,\varepsilon_{ik} t^k{}_j\,\Xi_\mp X^I_\pm \, ,
\end{equation}
where~$\Xi_\pm \equiv \xi_I X^I_\pm$, together with 
\begin{equation}
F_{ab}^\pm = T_{ab}^\pm = R(A)^\pm_{ab} = R(\mathcal{V})^\pm_{ab}{}^i{}_j = D = \mathcal{R}(M)_{abcd} = 0 \, .
\end{equation}
In addition, the scalars~$X^I_\pm$ must all be constant and such that~$\Xi_- X^I_+ = \Xi_+ X^I_-$. We now fix the A-gauge~$\Xi_+ = \Xi_- \equiv \Xi$, which also implies that~$X_+^I = X_-^I \equiv X^I$ for all~$I$. This in turn implies that, just as in the minimal supergravity case, the $\mathrm{W}^2$ invariant gives a vanishing contribution to the on-shell action for any full-BPS configuration.

The contribution of the~$\mathbb{T}$ invariant is obtained by evaluating the components of the multiplet starting with~$\mathcal{A}_\pm$ in~\eqref{eq:fullA}. Recall from Section~\ref{sec:HD} that this invariant is built starting from an (anti-)chiral multiplet~$\Phi_\mp$ of non-zero Weyl weight. In a supergravity theory with physical vector multiplets, it seems that we have a priori different choices for this multiplet. However, since the bosonic components of~$\Phi_\mp$ are also constrained by the amount of supersymmetry that the background preserves, this ambiguity is in fact lifted for the full-BPS configuration. We find that the Lagrangian density~\eqref{eq:gen-action} with the prepotentials~\eqref{eq:matter-4der} evaluates on the above full-BPS configuration to
\begin{align}
\label{eq:matter-full-1}
e^{-1}\mathcal{L} =&\; \tfrac16\,R\,\bigl(e^{-K} + 2\kappa^{-2}\bigr) + 4\,g^2\,\Xi^2\,N_{IJ}\, X^I X^J - 16\,g^2\,\kappa^{-2}\,\Xi^2 \\
&\!\!\!\!- 2\,c_2 \bigl(F^{(0)+} + F^{(0)-}\bigr)\bigl[R^{ab}R_{ab} - \tfrac13\,R^2 - \tfrac13\,\mathcal{D}_a\mathcal{D}^a R\,\bigr] - \tfrac43\,g^2\,\Xi^2\,c_2\,R\,\bigl(F^{(0)+}_I + F^{(0)-}_I\bigr)X^I \, , \nonumber
\end{align}
where we have used that~$F^{(0)\pm}$ are homogeneous of degree zero. Because of our gauge-fixing choice, we have~$F^{(0)+} = F^{(0)-}$ and we simply write this quantity as~$F^{(0)}$. The same is of course true for~$F^{(2)\pm}$, and for the various derivatives with respect to~$X^I$. In the density~\eqref{eq:matter-full-1}, the quantities~$e^{-K}$ and~$N_{IJ}$ are built from the full prepotentials~\eqref{eq:matter-4der}. We can express them in terms of the two-derivative quantities derived from~$F^{(2)}$ as follows,
\begin{equation}
e^{-K} = e^{-K^{(2)}} - 16\,g^2\,c_2\,\Xi^2\, F^{(0)}_I X^I\, , \qquad N_{IJ} = N_{IJ}^{(2)} - 16\,g^2\,c_2\,\Xi^2\, F^{(0)}_{IJ}\, ,
\end{equation}
where we made use of the full-BPS constraints. Using this, we rewrite the Lagrangian evaluated on the full-BPS configuration as
\begin{equation}
\begin{split}
\label{eq:fullnonminads}
e^{-1}\mathcal{L} =&\; \tfrac16\,R\,\bigl(e^{-K^{(2)}} + 2\,\kappa^{-2}\bigr) + 4\,g^2\,\Xi^2\,\bigl(N^{(2)}_{IJ}\,X^I\,X^J - 4\,\kappa^{-2}\bigr) \\
&\; - 4\,c_2\, F^{(0)}\, \bigl[R^{ab}R_{ab} - \tfrac13\,R^2 - \tfrac13\,\mathcal{D}_a\mathcal{D}^a R\,\bigr] \\
&\;- \tfrac{16}3\,g^2\,c_2\,R\;\Xi^2\, F^{(0)}_I X^I - 64\,g^4\,c_2\,\Xi^2\, F^{(0)}_{IJ}\,X^I\,X^J \, .
\end{split}
\end{equation}
We stress that this form of the Lagrangian is valid for arbitrary functions~$F^{(2)}$ and~$F^{(0)}$ since we only made use of homogeneity properties and off-shell BPS constraints. \\

At this stage, let us first focus on the two-derivative case and set $c_2 = 0$. Using the last remaining full-BPS condition~$R = 48\,g^2\,\Xi^2$, the first line of \eqref{eq:fullnonminads} reads
\begin{equation}
\label{eq:2der-fullBPS}
e^{-1}\mathcal{L}\big\vert_{\mathrm{EAdS}_4} = 48\,g^2\,\Xi^2\, F^{(2)} \, .
\end{equation}
For the EAdS$_4$ geometry of scale~$L = g^{-1}\kappa$, the Ricci scalar is given by~$R = 12\,L^{-2}$, which fixes
\begin{equation}
\label{eq:constraint}
\xi_I X^I = \frac1{2\,\kappa} \, .
\end{equation}
To determine the values of the individual scalar fields we can now extremize~\eqref{eq:2der-fullBPS} subject to the above constraint. This is an alternative route to fixing all the scalars instead of explicitly solving the $D$ equation of motion and using the additional BPS constraints coming from the variations of the physical gaugini. The reason we take this route is so that we can present a partially off-shell expression for the AdS$_4$ on-shell action that we find illuminating for our purposes. The motivation for employing this approach is to relate the on-shell action result to the field theory calculation of the ABJM $S^3$ partition function with general real mass parameters.
To make the analysis explicit consider the two-derivative $STU$ model with $n_V = 3$ vector multiplets defined by the following prepotential and FI parameters
\begin{equation}\label{eq:F2STUdef}
	F^{(2)}_{STU} = \sqrt{X^0\,X^1\,X^2\,X^3} \, , \qquad \xi_I = \Bigl(\frac14,\,\frac14,\,\frac14,\,\frac14\Bigr) \, .
\end{equation}
To evaluate the EAdS$_4$ on-shell action we again take the conformal boundary to be $S^3$ and we employ holographic renormalization.\footnote{Note that we are not discussing possible subtle issues with holographic renormlization in this section. In particular, in the presence of extra matter fields there could be finite counterterms needed when performing holographic renormalization for supersymmetric solutions of a given supergravity theory, see for example \cite{Freedman:2013oja,Bobev:2013cja}. Since we are considering the maximally symmetric AdS$_4$ vacuum solution of the theory we do not expect such subtleties to arise.} Using the above full-BPS conditions, the finite EAdS$_4$ on-shell action takes the form
\begin{equation}
\label{eq:off-shell2der}
	I_{\mathrm{EAdS}_4}^{STU} (z) =   \frac{2 \pi\,L^{2}}{\mathcal{G}_N}\, \sqrt{z^0\,z^1\,z^2\,z^3} \, , \qquad \text{with} \qquad \sum_{I=0}^3 z^I = 2 \, .
\end{equation}
Here, we introduced the dimensionless variables 
\begin{equation}
z^I \equiv \kappa\, X^I\, .
\end{equation} 

It is worth clarifying the precise meaning of the ``on-shell'' action in \eqref{eq:off-shell2der}. It is computed by setting all the fields of the supergravity theory to their on-shell value, with the exception of the individual scalar fields~$X^I$. An efficient shortcut to find the correct on-shell values for the scalars is then to extremize $I_{\mathrm{EAdS}_4}^{STU} (z)$ as a function of $z^I$ subject to the linear constraint in \eqref{eq:off-shell2der}. Performing this extremization we find
\begin{equation}
z^0 = z^1 = z^2 = z^3 = \frac12\, ,
\end{equation}
and evaluating~\eqref{eq:2der-fullBPS} at this extremum yields
\begin{equation}\label{eq:I2derveconshell}
	I_{\mathrm{EAdS}_4}^{STU}\big\vert_* = \frac{\pi\,L^{2}}{2 \mathcal{G}_N} \,.
\end{equation}
This of course agrees with the two-derivative on-shell action in the absence of vector multiplets given in \eqref{eq:S3pfinftyder}. The main reason to present both the true on-shell value of the AdS$_4$ action in \eqref{eq:I2derveconshell} as well as the ``partially on-shell'' expression in \eqref{eq:off-shell2der} is that the result for $I_{\mathrm{EAdS}_4}^{STU}(z)$ in \eqref{eq:off-shell2der} has a clear holographic meaning. Namely, \eqref{eq:off-shell2der} can be thought of as the leading order in the large $N$ supersymmetric $S^3$ partition function of the $k=1$ ABJM theory in the presence of the three independent real mass parameters \cite{Jafferis:2011zi}. The extremization procedure to go from $I_{\mathrm{EAdS}_4}^{STU} (z)$ to the on-shell value in \eqref{eq:I2derveconshell} is the supergravity analogue of F-maximization \cite{Jafferis:2010un}. We emphasize that while the supergravity result for $I_{\mathrm{EAdS}_4}^{STU} (z)$ agrees with the ABJM $S^3$ free energy in the presence of real masses it is obtained in a somewhat ad-hoc way, see \cite{Hosseini:2016tor} for further discussion. The rigorous method to derive the gravitational dual of the ABJM theory on $S^3$ in the presence of general real masses is to construct new Euclidean supergravity solutions asymptotic to AdS$_4$ with an $S^3$ boundary that have non-trivial profiles for the supergravity scalar fields as in \cite{Freedman:2013oja}. In the discussion below we will continue to employ the short-hand procedure leading to the partially on-shell answer in \eqref{eq:off-shell2der} even in the presence of the HD corrections to the supergravity action. It is an interesting open problem to understand how to generalize the solutions of \cite{Freedman:2013oja} in the HD context. \\

We now return to the full-BPS Lagrangian \eqref{eq:fullnonminads} and study the effects of the HD terms. Since~$F^{(0)}$ is of degree zero, when~$c_2 \neq 0$ we find the simple result
\begin{equation}
\label{eq:4der-fullBPS}
e^{-1}\mathcal{L}\big\vert_{\mathrm{EAdS}_4} = 12\,L^{-2}\, F^{(2)} + 48\,c_2\,L^{-4}\, F^{(0)}\, ,
\end{equation}
which, following the short-hand procedure described above, should again be extremized under the condition \eqref{eq:constraint}. It is simple to show that the saddle point evaluation of the scalars $X^I$ remains exactly the same as in the two-derivative case, again due to the fact that $F^{(0)}$ is of degree zero. Therefore, we can take the four-derivative $STU$ model with a completely arbitrary function  $F^{(0)}_{STU}$ and still find the extremum
\begin{equation}\label{eq:zextrSTU4der}
z^0 = z^1 = z^2 = z^3 = \frac12\, .
\end{equation}
The four-derivative ``partially on-shell'' action for EAdS$_4$ with an~$S^3$ boundary is then
\begin{equation}
\label{eq:off-shell4der}
	I_{\mathrm{EAdS}_4}^{STU} (z) =   \frac{2 \pi\,L^{2}}{\mathcal{G}_N}\, \sqrt{z^0\,z^1\,z^2\,z^3}\, +  64 \pi^2 c_2\, F^{(0)}_{STU} (z^I)\,.
\end{equation}
The on-shell value can be obtained using the extremized $z^{I}$ in \eqref{eq:zextrSTU4der} and reads
\begin{equation}
	I_{\mathrm{EAdS}_4}^{STU}\big\vert_* = \frac{\pi\,L^{2}}{2 \mathcal{G}_N}\, +  64 \pi^2 c_2\, F^{(0)}_{STU} (z^I = \tfrac12) \, .
\end{equation}
These results hold for arbitrary functions~$F^{(0)}_{STU}(z^I)$ that are homogeneous of degree zero. 

%%%%%%%%%%
\subsubsection*{ABJM theory and the Airy function}
%%%%%%%%%%

To find the function $F^{(0)}_{STU} (z^I)$ that determines the four-derivative corrections to the $STU$ model we can take advantage of the supersymmetric localization calculations of the $S^3$ partition function of the ABJM theory, see \cite{Marino:2011eh,Fuji:2011km} and references therein. As discussed above, the two-derivative version of the $STU$ model can be obtained from a consistent truncation of 11d supergravity on $S^7$, which means that it is holographically dual to the ABJM theory at Chern-Simons level $k=1$. In the brief field theory discussion below we will keep $k$ arbitrary and finite since the supersymmetric localization results apply for general values of $k$. In particular, we will make use of the results of Nosaka~\cite{Nosaka:2015iiw} who determined the $S^3$ partition function of the ABJM theory away from the superconformal point, i.e. allowing some freedom in the R-charge assignment of the bi-fundamental multiplets, or equivalently studying the theory with non-vanishing real mass parameters.

Let us very briefly recall the basic features of the supersymmetric $S^3$ partition function of a general 3d ${\cal N} = 2$ Chern-Simons-matter theory arising from M2-branes. It can be written as a function of the gauge group rank $N$ via an inverse transformation of the so-called grand potential $J$,
\begin{equation}
	Z (N) = \int \frac{{\rm d} \mu}{2\pi i}\, e^{J(\mu) - \mu N}\, ,
\end{equation}
where $\mu$ is the chemical potential dual to $N$. The grand potential can be determined from an alternative formulation of the supersymmetric localization matrix model in terms of a Fermi gas. For the ABJM theory (and other similar models enjoying additional supersymmetry), the Fermi gas can be treated in the ideal gas approximation. One can then derive the following expressions for the perturbative part of the grand potential\footnote{One can also determine the non-perturbative part, which is related to instanton corrections and is beyond the reach of HD supergravity.}
\begin{equation}
\label{eq:grandpot}
	J^\text{pert} (\mu) = \frac{\tt C}{3}\, \mu^3 + {\tt B}\, \mu + {\tt A}\, ,
\end{equation}
for some constant parameters ${\tt A}, {\tt B}, {\tt C}$ that are specific to the theory in question and do not depend on $N$. The result for the perturbative part of $Z$ is then given in terms of the Airy function and reads
\begin{equation}\label{eq:Airydef}
	Z^\text{pert} (N) = e^{{\tt A}}\, {\tt C}^{-\frac{1}{3}}\, \text{Ai}[{\tt C}^{-\frac{1}{3}} (N-{\tt B})]\,.
\end{equation}
The Airy function admits a simple large $N$ expansion given by
\begin{equation}\label{eq:AirylargeN}
	- \log Z (N \rightarrow \infty) = \frac{2}{3 \sqrt{{\tt C}}}\, N^{\frac{3}{2}} - \frac{{\tt B}}{\sqrt{{\tt C}}}\, N^{\frac{1}{2}} + \frac14\, \log N + {\cal O} (N^0)\ .
\end{equation}
It is clear that the leading and first subleading term in the above large $N$ expansion are determined by the constants $\tt B$ and ${\tt C}$. One can use this fact, together with the AdS/CFT correspondence and the bulk on-shell results derived previously, to fix the functions~$F^{(2)}$ and~$F^{(0)}$ entering the supergravity prepotentials.

The ABJM theory admits a deformation by three real mass parameters that break the conformal symmetry of the theory but preserve $\mathcal{N}=2$ supersymmetry. These deformations can be organized in terms of four parameters $\Delta_i$ which obey the constraint (see for instance Section 5 in \cite{Jafferis:2011zi})
\begin{equation}
\label{eq:standardconstr}
	\sum_{i = 1}^4 \Delta_i = 2\ .
\end{equation}
In \cite{Nosaka:2015iiw} Nosaka calculated the $S^3$ partition function of the ABJM theory in the presence of two non-vanishing real masses. Translating these results in our notation leads to the following values of the parameters $\Delta_i$
\begin{equation}
	\Delta_{1,2} = \frac{1 + \zeta_{1,2}}{2}\ , \qquad \Delta_{3,4}  = \frac{1 - \zeta_{1,2}}{2}\,.
\end{equation}
We stress that the results of \cite{Nosaka:2015iiw} amount to turning on only two out of the three possible real mass parameters which is manifested in the following relations obeyed by the $\Delta_i$ 
\begin{equation}
\label{eq:redundantconstr}
	\Delta_{i+2} = 1 - \Delta_i\ .
\end{equation}
Having fixed our conventions, we can read off the coefficients ${\tt B}$ and ${\tt C}$ entering the Airy function~\eqref{eq:Airydef} from Equation (1.12) in \cite{Nosaka:2015iiw},
\begin{equation}\label{eq:BCABJMgenDelta}
	{\tt C}_\text{ABJM} = \frac{1}{8 \pi^2 k \Delta_1 \Delta_2 \Delta_3 \Delta_4}\ , \quad {\tt B}_\text{ABJM} = \frac{1}{48 k \Delta_1 \Delta_2 \Delta_3 \Delta_4} \left(2 k^2 \Delta_1 \Delta_2 \Delta_3 \Delta_4 + \sum_i \Delta_i^2 \right)\ .
\end{equation}
We note that the result in \cite{Nosaka:2015iiw} is derived for $\Delta_i$ obeying both the supersymmetry relation in \eqref{eq:standardconstr} and the extra condition in \eqref{eq:redundantconstr}. We however believe that the expressions \eqref{eq:BCABJMgenDelta} are valid for general $\Delta_i$ obeying only the constraint in \eqref{eq:standardconstr}, i.e. for the ABJM theory in the presence of three generic mass parameters. While we have no rigorous derivation of this statement it is strongly supported by the permutation symmetry between the four $\Delta_i$ and by the fact that at $k=1,2$ the ABJM theory enjoys $\mathcal{N}=8$ symmetry and the corresponding $SO(8)$ R-symmetry puts the three real mass parameters on equal footing.

Using these results, we arrive at the following large $N$ expansion for the partition function of the ABJM theory in the presence of three real mass parameters, 
\begin{equation}\label{eq:AiryNosakalargeN}
	- \log Z_\text{ABJM} = \frac{\pi ( 16\, N^{\frac{3}{2}} -k N^{\frac{1}{2}})}{12}\, \sqrt{2 k \Delta_1 \Delta_2 \Delta_3 \Delta_4}\,   - \frac{\pi\, N^{\frac{1}{2}} \sum_i \Delta_i^2}{12 \sqrt{2 k \Delta_1 \Delta_2 \Delta_3 \Delta_4}} + \frac14\log N\, .
\end{equation}
Note that to emphasize the similarity between this result and the HD supergravity action in \eqref{eq:off-shell4der} we have grouped the two terms above in order to make manifest that they are homogeneous of degree two and zero as a function of the parameters $\Delta_i$. We note that for~$k=1,2$, Equation~\eqref{eq:AiryNosakalargeN} is consistent with the relations derived in~\cite{Chester:2021gdw} for 3$d$~$\mathcal{N}=2$ SCFT partition functions of mass-deformed theories on the squashed 3-sphere. In what follows, we will work under the assumption that~\eqref{eq:AiryNosakalargeN} holds at general level~$k$ and at strong coupling. It would be interesting to gather further evidence for this.

%%%%%%%%
\subsubsection*{A conjecture for $F^{(0)}_{STU}$}
%%%%%%%%

We can combine the supersymmetric localization results~\eqref{eq:AiryNosakalargeN} with the ``partially on-shell'' supergravity action~\eqref{eq:off-shell4der} to determine the unknown four-derivative prepotential function $F^{(0)}_{STU} (z^I)$. As discussed below \eqref{eq:F2STUdef} and \eqref{eq:off-shell2der}, and elaborated upon in \cite{Hosseini:2016tor}, at the two derivative level we have the following relation between the $S^3$ partition function and the prepotential,
\begin{equation}
	- \log Z(\Delta_i) = I_{\mathrm{EAdS}_4}^{2\partial} = \frac{2 \pi\,L^{2}}{\mathcal{G}_N}\, F^{(2)} (z^I)\, \qquad \text{with} \qquad \Delta_i = z^I\, ,
\end{equation}
which holds for a large class of holographically dual pairs. For the ABJM theory and the $STU$ model at hand we can use \eqref{eq:F2STUdef} and \eqref{eq:off-shell2der} and the standard two-derivative holographic relation, see \eqref{eq:L2GNN32} and \eqref{eq:Av1v2ABJM} above,
\begin{equation}
	\frac{L^{2}}{2 \mathcal{G}_N} = \frac{\sqrt{2 k}}{3}\,  N^{\frac{3}{2}}\ .
\end{equation}
to reproduce the $N^{\frac{3}{2}}$ term in the ABJM partition function in \eqref{eq:AiryNosakalargeN}. It is natural to conjecture that the relation between the supergravity $STU$ model prepotential and the ABJM partition function in the presence of real masses continues to hold to subleading order in the large $N$ expansion and in the presence of higher-derivative terms. In particular, we conjecture that the result for the ABJM $S^3$ partition function in \eqref{eq:AiryNosakalargeN} should be related to the four-derivative partially on-shell action in \eqref{eq:off-shell4der} as
\begin{equation}
	- \log Z_\text{ABJM} (\Delta_i) =  I_{\mathrm{EAdS}_4}^{STU} (z^I) \, \qquad \text{with} \qquad \Delta_i = z^I\, , 
\end{equation}
This in turn allows us to uniquely determine the function $F^{(0)}_{STU}$ and find
\begin{empheq}[box=\fbox]{equation}
\label{eq:conjecture}
	F^{(0)}_{STU} (z) =\frac{(z^0)^2 + (z^1)^2 + (z^2)^2 + (z^3)^2}{8 \sqrt{z^0 z^1 z^2 z^3}} \, ,
\end{empheq}
with the following holographic relations between the constants in the HD supergravity action and the field theory parameters
\begin{empheq}[box=\fbox]{equation}
\label{eq:c2STU}
	\frac{L^{2}}{2 \mathcal{G}_N} = \frac{\sqrt{2 k}}{48}\, (16\,  N^{\frac{3}{2}} -k\, N^{\frac{1}{2}} )\, , \qquad 32 \pi c_2 = - \frac{N^{\frac{1}{2}}}{3 \sqrt{2 k}}  \ .
\end{empheq}
This result is compatible with the minimal supergravity calculation in \eqref{eq:Av1v2ABJM} and actually allows us to determine both the constants $a$ and $v_2$ in \eqref{eq:Av1v2ABJM} independently. Given that $c_1$ is already determined in \eqref{eq:Av1v2ABJM} we see that the $STU$ model results above combined with the supersymmetric localization calculation in \eqref{eq:AiryNosakalargeN} allow us to fully determine the HD supergravity Lagrangian at the four-derivative order. We note that the relation between the four-derivative supergravity couplings $c_{1,2}$ is
\begin{empheq}[box=\fbox]{equation}
\label{eq:c1STU}
	c_1 = 3\,c_2 < 0  \quad \Longrightarrow \quad c_1 < c_2 \ .
\end{empheq}
In particular this determines the sign of some of the corrections to the black hole entropy in \eqref{eq:entropy_final}.

To the best of our knowledge the four-derivative modification of the prepotential for the $STU$ model of gauged supergravity in \eqref{eq:conjecture} has not appeared in the literature. A simple consistency check for its validity is that after setting $\Delta_i = \frac12 \; \forall\,i$, i.e. for ABJM theory at the conformal point, we reproduce the minimal supergravity results discussed in Section~\ref{sec:susyloc}.  It would certainly be very interesting to devise more consistency checks to establish the validity of \eqref{eq:conjecture} or to derive it directly from the HD corrections to 11d supergravity.

%%%%%%%%
\subsubsection*{Further speculations}
%%%%%%%%

There are a number of natural questions and open problems that arise from the discussion above. Below we discuss some of them.

\begin{itemize}

	\item The infinite-derivative Lagrangian for the $\mathrm{W}^2$ and $\mathbb{T}$ invariants is governed by the formal expansion of the prepotential \eqref{eq:prepot}, which admits a natural generalization to matter-coupled supergravity. This raises the question of how such an infinite order expansion can arise from a string theory compactification. The analogous question in ungauged supergravity was addressed by the OSV conjecture \cite{Ooguri:2004zv}, where the prepotential expansion is related to the topological string partition function on Calabi-Yau manifolds. It is tempting to speculate that a similar result applies to compactifications of 11d supergravity on Sasaki-Einstein (SE$_7$) manifolds, all of which lead to minimal ${\cal N} = 2$ supergravity at two derivatives \cite{Gauntlett:2007ma}. We hope that the general form of the infinite derivative AdS$_4$ action in minimal supergravity, \eqref{eq:S3pfinftyder}, will help to uncover the underlying structure of SE$_7$ compactifications and its relation to string or M-theory.
		
	\item It is also very interesting to notice that subleading terms in the Airy function, in particular the coefficient $\tt B$ in \eqref{eq:grandpot}, directly relate to subleading terms in the prepotential. This means that the proper understanding of the string theoretic origin of the prepotential directly relates to the understanding of the exact expression for the Airy function, and vice versa. It would be interesting if one could explore this for more general Chern-Simons-matter theories with less supersymmetry than the ABJM model. For instance, note that the coefficient $\tt B$ for the ABJM theory only has two distinct terms of degree 2 and degree 0, which map to the respective supergravity prepotential at two and four derivatives. The lack of further subleading terms in $\tt B$ therefore suggests the absence of further HD corrections in the 4d consistent truncation of 11d supergravity on $S^7$. It will be very interesting to prove this conjecture explicitly.

		\item As noted above the $\mathrm{W}^2$ invariant completely drops out of the pure EAdS$_4$ on-shell action, at any order of derivatives and with arbitrary matter coupling. This was a major simplification of our analysis, but also means that we cannot leverage the Airy function to learn about another infinite tower of possible HD corrections in supergravity. On the other hand, the $\mathrm{W}^2$ invariant leads to a non-vanishing contribution at all orders in a derivative expansion and with arbitrary matter coupling for the near-horizon geometry of static black holes in AdS$_4$, as shown in \cite{Hristov:2016vbm}. It would be very interesting to generalize this analysis to include the $\mathbb{T}$ invariant and then compare with subleading corrections to the holographically dual topologically twisted index.

	\item Above we have mostly focused on the AdS$_4$ solution but it is desirable to have calculational control over the numerous other BPS and non-BPS supergravity backgrounds of holographic interest. We understood many such solutions in detail in minimal four-derivative supergravity, but  generalizing these results to infinitely many derivatives looks very cumbersome at present. For example we could consider the Wald entropy after adding the full expansion for the prepotential in the $\mathbb{T}$-invariant, but we need to first deal with the resulting infinite expansion of the Lagrangian in powers of the field $D$. It has been suggested before, see \cite{Mohaupt:2000mj} and references therein, that one could deal with such equations of motion order by order. The validity of this perturbative approach is yet to be carefully explored in gauged supergravity and for solutions of holographic interest.

	\item We have focused here on the $\mathrm{W}^2$ and $\mathbb{T}$ invariants which were sufficient for treating HD corrections in minimal gauged supergravity. Note however, that there could be other HD invariants that are relevant for the 4d matter coupled gauged supergravity theories that can be embedded in string/M-theory. We can again make a comparison with ungauged supergravity, where for example the infinite tower of $D$-term HD invariants were shown to completely vanish on BPS black holes \cite{deWit:2010za}. It would be very useful to extend such non-renormalization theorems to backgrounds in gauged supergravity, but this is certainly more technically challenging.
	
	\item As we discussed above, the 2-derivative $STU$ model can be embedded in maximal 4d gauged supergravity, and it is therefore natural to expect this to hold for the conjectured four-derivative prepotential \eqref{eq:conjecture} and coupling constant relation \eqref{eq:c1STU} as well. It would be very interesting to use the constraints of ${\cal N}=8$ supersymmetry to check whether the four-derivative $STU$ model Lagrangian we presented above is indeed consistent and unique.
	
\end{itemize}

%%%%%%%%%%%%%%%%%
\section{Discussion}
\label{sec:discussions}
%%%%%%%%%%%%%%%%%

In this paper we have used the interplay between HD gauged supergravity, holography and supersymmetric localization to determine the constants controlling the four-derivative minimal supergravity action. This in turn allowed us to make detailed quantitative predictions about the subleading corrections to the entropy of asymptotically AdS$_4$ black holes and to find the $N^{\frac{1}{2}}$ corrections to the partition functions of 3d Chern-Simons matter  SCFTs on compact manifolds. Our results can serve as a blueprint for future research on this topic and can be extended in numerous different directions. We discuss several of these possible generalizations below.

As explained in detail in Section~\ref{sec:HD} and summarized in Figure~\ref{fig:sol-subset}, all solutions of the two-derivative equations of motion in the minimal supergravity theory also solve the four-derivative ones. It is important to understand whether there are interesting new solutions of the four-derivative equations of motion and what is their holographic interpretation. It will be also interesting to uncover the structure of the supergravity solution space in the presence of the six- and higher-derivative corrections discussed in Section~\ref{sec:ExtGen} as well as in the presence of additional matter multiplets.

In Section~\ref{sec:susyloc} we employed holography and supersymmetric localization to determine the two constants appearing in the four-derivative minimal supergravity. It will be most interesting to establish that four-derivative minimal supergravity, like the two-derivative theory, arises as a consistent truncation from the HD corrections to 11d supergravity and to perform an explicit reduction from 11d to 4d in order to determine the full form of the 4d supergravity Lagrangian and check the results in Section~\ref{sec:susyloc}. It will also be interesting to understand whether our 4d supergravity approach can be used in conjunction with the ideas pursued in \cite{Chester:2018aca,Binder:2018yvd} to relate supersymmetric localization calculations to the HD corrections to 11d supergravity.

Here we have focused on 4d $\mathcal{N}=2$ supergravity but it should be possible to apply the same approach to supergravity theories with more supersymmetry. A natural extension of our work is to consider 4d $\mathcal{N}=4$ conformal supergravity where the structure of the HD terms is more constrained \cite{deRoo:1984zyh,deRoo:1985jh,Butter:2016mtk}. It is natural to expect that when the 4d $\mathcal{N}=2$ minimal supergravity is embedded in 4d $\mathcal{N}=4$ supergravity the additional supersymmetry will lead to a relation between the constants $c_1$ and $c_2$ controlling the four-derivative interactions. We are planning to investigate this question in the near future. The conformal supergravity approach cannot be applied for 4d supergravity theories with $\mathcal{N} >4$ supersymmetry due to the absence of a corresponding superconformal algebra. Therefore, it is desirable to develop tools to study the HD corrections to 4d $\mathcal{N}=8$ (or $\mathcal{N}=6$)  supergravity which will be of relevance for compactifications of 11d supergravity on $S^7$ (or $S^7/\mathbb{Z}_k$) and the holographically dual ABJM theory.

In Section~\ref{sec:susyloc} we used supersymmetric localization results for two families of 3d SCFTs arising from M2-branes in order to determine the unknown constants in the four-derivative minimal supergravity action. This in turn leads to a number of new results for the $N^{\frac{1}{2}}$ term in the large $N$ expansion of supersymmetric partition functions of these SCFTs. It is important to revisit supersymmetric localization calculations in the large $N$ limit of these models in order to verify our holographic predictions.\footnote{After some of our results were announced in \cite{Bobev:2020egg} such analysis was undertaken for the squashed $S^3$ partition function of the ABJM theory in \cite{Chester:2021gdw}.} Our methods can be applied for any 3d $\mathcal{N}=2$ SCFT with a weakly coupled holographic dual. As emphasized in \cite{Bobev:2017uzs}, such theories should admit a universal sector in the large $N$ limit that is captured by the 4d $\mathcal{N}=2$ minimal supergravity and its HD extensions we studied here. In addition to the two classes of M2-brane SCFTs discussed in Section~\ref{sec:susyloc} we also studied this for class $\mathcal{R}$ SCFTs arising from wrapped M5-branes \cite{Bobev:2020zov}. It will be most interesting to extend this analysis for other families of 3d $\mathcal{N}=2$ SCFTs arising in string and M-theory. This includes Chern-Simons matter theories that control the low energy dynamics of M2-branes on Sasaki-Einstein manifolds as well as similar theories arising from D2-D8 branes in massive IIA supergravity \cite{Guarino:2015jca,Fluder:2015eoa}. In addition, one could attempt to extend our results to more exotic 3d $\mathcal{N}=2$ SCFTs with an AdS$_4$ holographic duals in string theory, like the S-fold SCFTs discussed in \cite{Assel:2018vtq,Bobev:2021yya} or the theories obtained from wrapped D4-D8 branes studied in \cite{Hosseini:2018uzp,Crichigno:2018adf} as well as similar models arising from wrapped $(p,q)$ five-branes in type IIB supergravity. The supergravity result for the four-derivative on-shell action of the minimal supergravity theory we derived in \eqref{eq:ionshell} has a universal nature and it is tempting to speculate that the corresponding supersymmetric partition functions in these different classes of holographic SCFTs will have a similar universal structure in the large $N$ limit. We believe that this universality deserves further exploration by field theory methods.

We have limited our holographic analysis in Section~\ref{sec:susyloc} to the leading $N^{\frac{1}{2}}$ correction in the large $N$ expansion of physical  observables in the ABJM theory and the 3d $\mathcal{N}=4$ SYM theory. In general one should expect further corrections of order $N^{-\frac{n+1}{2}}$ for $n \geq -1$ as well as a $\log (N)$ term. This expectation is explicitly confirmed by the large $N$ expansion of the Airy function that controls the $S^3$ partition function of these theories, see \eqref{eq:Airydef} and \eqref{eq:AirylargeN}. It will be most interesting to calculate these corrections explicitly in supergravity. The $\log(N)$ correction is expected to arise from one-loop contributions from all supergravity modes and can be computed by applying the methods used in \cite{Bhattacharyya:2012ye,Liu:2017vbl}. For general asymptotically AdS$_4$ solutions of minimal supergravity this one-loop calculation has not been performed and it will be desirable to do so. We expect that the $N^{-\frac{n+1}{2}}$ corrections in the supersymmetric partition functions will be controlled by the six- and higher derivative coefficients $d_n$ in \eqref{eq:S3pfinftyder}. It will be very interesting, and probably technically challenging, to calculate these coefficients either from a first principle derivation using HD terms in 11d supergravity or by using supersymmetric localization. We note in passing that the Airy function that determines the round $S^3$ partition function of the ABJM theory (with vanishing real masses) was derived from AdS$_4$ supergravity by using supergravity supersymmetric localization in \cite{Dabholkar:2014wpa}. This result of course includes both the $\log(N)$ and $N^{-\frac{n+1}{2}}$ and it will be interesting to revisit the analysis of \cite{Dabholkar:2014wpa} in light of the HD on-shell action calculations we performed in this work.

While many of the details of our calculations were specific to 4d supergravity and 3d SCFTs it should be possible to apply our HD supergravity approach to theories in higher dimensions. The natural first target for this is 5d $\mathcal{N}=2$ gauged supergravity which admits a conformal supergravity treatment and the leading HD terms have been studied in \cite{Hanaki:2006pj,Ozkan:2013nwa}, see also \cite{Baggio:2014hua,Melo:2020amq} for a holographic application. It will be interesting to repeat our analysis above for the special case of minimal 5d $\mathcal{N}=2$ gauged supergravity which captures the universal stress-energy tensor dynamics of planar 4d $\mathcal{N}=1$ holographic SCFTs and we plan to study this in future work. Studying HD corrections to 6d and 7d supergravity will also be interesting and should be possible to do using conformal supergravity techniques for the half-maximal gauged supergravity theories. These models have twice the number of supercharges of the 4d $\mathcal{N}=2$ supergravity we treated in this work and their structure will be similar to the 4d $\mathcal{N}=4$ theory discussed above. It is clear that many of these generalizations should be accessible with currently available techniques and given the importance of studying HD corrections in top-down holography it will be most interesting to pursue them in the future.

%%%%%%%%%%%%%%%%%%%%%%
\section*{Acknowledgments}
We are grateful to Pietro Benetti Genolini, Guillaume Bossard, Daniel Butter, Shai Chester, Marcos Crichigno, Dongmin Gang, Fri\dh rik Freyr Gautason, Garrett Goon, Callum Jones, Stefanos Katmadas, Brian McPeak, Vincent Min, Miguel Paulos, Silviu Pufu, Chiara Toldo, Jesse van Muiden, Toine Van Proeyen, and Balt van Rees  for valuable discussions.  NB is supported in part by an Odysseus grant G0F9516N from the FWO. KH is supported in part by the Bulgarian NSF grants DN08/3, N28/5, and KP-06-N 38/11. NB, AMC, and VR are also supported by the KU Leuven C1 grant ZKD1118 C16/16/005. We would also like to acknowledge the work of Steve Wilhite and his team, whose development of the GIF format facilitated enjoyable online discussions during the social distancing necessitated by the pandemic.

%%%%%%%%%%%%%%%%%%%%%%

%%%%%%%%%%%%%%%%%%
\appendix
%%%%%%%%%%%%%%%%%%

%%%%%%%%%%%%%%%%%%%%%%%%%%%%%%%%%%%%%
\section{Conventions and conformal supergravity details}
\label{app:sugra}
%%%%%%%%%%%%%%%%%%%%%%%%%%%%%%%%%%%%%

In this appendix we review some details of the conformal supergravity formalism for four-dimensional~$\mathcal{N}=2$ theories in Euclidean signature, following~\cite{deWit:2017cle}. The Lorentzian counterpart is presented in many papers and reviews, see the recent book \cite{Lauria:2020rhc} for a comprehensive exposition. Conformal supergravity is based on the gauging of the superconformal algebra~$\mathrm{SU}^*(4|2)$. This algebra comprises the generators of the spatial diffeomorphisms, local tangent space rotations, dilatation, special conformal, chiral~$\mathrm{SU}(2)$ and~$\mathrm{SO}(1,1)$ R-symmetry, supersymmetry (Q) and special supersymmetry (S) transformations. The gauge fields associated with general coordinate transformations ($e_\mu{}^a$), dilatations ($b_\mu$), chiral R-symmetry ($\mathcal{V}_\mu{}^i{}_j$ and $A_\mu$) and Q-supersymmetry ($\psi_\mu{}^i$) are independent fields. The remaining gauge fields associated with the Lorentz ($\omega_\mu{}^{ab}$), special conformal ($f_\mu{}^a$) and S-supersymmetry transformations ($\phi_\mu{}^i$) are composite fields. The corresponding supercovariant curvatures and covariant fields are contained in a tensor chiral multiplet called the Weyl multiplet, which comprises $24 \oplus 24$ off-shell degrees of freedom. In addition to the independent superconformal gauge fields, it contains three other fields: a symplectic-Majorana spinor doublet~$\chi^i$, a real scalar~$D$, and a Lorentz antisymmetric tensor~$T_{ab}$, which decomposes into a self-dual and an anti-self-dual field. The Weyl and chiral weights for the fields in this multiplet are presented in Table~\ref{table:weyl}.

%%%%%%%%%%%%%%%%%%%%%%%%%%%%%%%%%%%%%%%%%%%%%%%%%%%%%%%
\begin{table}[ht]
\centering
\begin{tabular*}{14.8cm}{@{\extracolsep{\fill}}
    |c||cccccccc|ccc||ccc| }
\hline
 & &\multicolumn{9}{c}{Weyl multiplet} & &
 \multicolumn{2}{c}{parameters} & \\[1mm]  \hline \hline
 field & $e_\mu{}^{a}$ & $\psi_\mu{\!}^i$ & $b_\mu$ & $A_\mu$ &
 $\mathcal{V}_\mu{}^i{}_j$ & $T_{ab}^\pm $ &
 $ \chi^i $ & $D$ & $\omega_\mu^{\,ab}$ & $f_\mu{}^a$ & $\phi_\mu{\!}^i$ &
 $\epsilon^i$ & $\eta^i$
 & \\[.5mm] \hline
$w$  & $-1$ & $-\tfrac12 $ & 0 &  0 & 0 & 1 & $\tfrac{3}{2}$ & 2 & 0 &
1 & $\tfrac12 $ & $ -\tfrac12 $  & $ \tfrac12  $ & \\[.5mm] \hline
$c$  & $0$ & $\mp\tfrac12 $ & 0 &  0 & 0 & $\pm1$ & $\mp\tfrac{1}{2}$ & 0 &
0 & 0 & $\pm\tfrac12 $ & $ \mp\tfrac12 $  & $ \pm\tfrac12  $ & \\[.5mm] \hline
 $\tilde\gamma_5$   &  & $\pm$ &   &  $\pm$  &   &  $\pm$ & $\pm$ &  &  $\pm$&
 & $\pm$ & $\pm$  & $\pm $ & \\ \hline
\end{tabular*}
\vskip 2mm
\renewcommand{\baselinestretch}{1}
\parbox[c]{14.8cm}{\caption{\label{table:weyl}{\footnotesize
      Weyl weights $w$, chiral $\mathrm{SO}(1,1)$ weights $c$, and 
      chirality/duality $\tilde\gamma_5$ of the fields and spinors 
      for the four-dimensional Euclidean Weyl multiplet.}}}
\end{table}
%%%%%%%%%%%%%%%%%%%%%%%%%%%%%%%%%%%%%%%%%%%%%%%%%%%%%%%%%%

The composite gauge fields~$\omega_\mu{\!}^{ab}$,~$\phi_\mu{\!}^i$ and~$f_\mu{\!}^a$, associated with~$\mathrm{SO}(4)$ tangent space rotations, S-supersymmetry, and conformal boosts, respectively, are expressed in terms of the independent fields through the so-called conventional constraints. Their explicit expressions are
\begin{align}
  \label{eq:dependent-gf}
  \omega_\mu{\!}^{ab} =&\, -2\,e^{\nu[a} \,(\partial_{[\mu} + b_{[\mu}) e_{\nu]}{\!}^{b]}
     -e^{\nu[a}e^{b]\rho}\,e_{\mu c}\,(\partial_\rho +b_\rho) e_\nu{\!}^c
      -\tfrac{1}{4}\bigl(2\,\bar{\psi}_{\mu i}\gamma^5\gamma^{[a}\psi^{b]i}
     + \bar{\psi}^a_i\gamma^5\gamma_\mu\psi^{b\,i}\bigr) \, , \nonumber\\ 
     %%%
     \phi_{\mu}{\!}^i =& \, -\tfrac12\mathrm{i}\left( \gamma^{\rho
         \sigma} \gamma_\mu - \tfrac{1}{3} \gamma_\mu \gamma^{\rho
         \sigma} \right) \left(\mathcal{D}_\rho \psi_{\sigma}^i +
       \tfrac{1}{32}\mathrm{i}\,T_{ab}\,\gamma^{ab}
       \gamma_\rho\psi_{\sigma}^i + \tfrac{1}{4} \gamma_{\rho
         \sigma} \chi^i \right) \, ,\\ 
    %%%%%%%%
   f_\mu{\!}^{a} =& \, \tfrac12\,R(\omega,e)_\mu{}^a -
  \tfrac14\,\bigl(D+\tfrac13 R(\omega,e)\bigr) e_\mu{}^a -
  \tfrac12\,\widetilde{R}(A)_\mu{}^a - \tfrac1{32}\,T_{\mu
    b}^-\,T^{+\,ba} + \ldots\,,  \nonumber
\end{align}
where the ellipses in the last equation denote fermionic contributions. Here and elsewhere we use the notation~$\widetilde{R}_{ab} = \tfrac12\varepsilon_{abcd}\,R^{cd}$. Furthermore,~$R(\omega,e)_\mu{}^a= R(\omega)_{\mu\nu}{}^{ab} e_b{}^\nu$ is the non-symmetric Ricci tensor, and~$R(\omega,e)$ the corresponding Ricci scalar. The uncontracted curvature~$R(\omega)_{\mu\nu}{\!}^{ab}$ is defined by
\begin{equation}
  \label{eq:R-omega}
   R(\omega)_{\mu\nu}{\!}^{ab} = \partial_\mu\omega_\nu{\!}^{ab} 
   -\partial_\nu\omega_\mu{\!}^{ab} -\omega_\mu{\!}^{ac} \,
   \omega_{\nu{c}}{\!}^{b } 
   +\omega_\nu{\!}^{ac} \,   \omega_{\mu{c}}{\!}^{b }\,. 
\end{equation}
It is also useful to present explicitly the trace of the conformal boosts gauge field, including fermions:
\begin{equation}
f_\mu{}^\mu = \tfrac16\,R(\omega,e) - D + \tfrac1{12}\,e^{-1}\varepsilon^{\mu\nu\rho\sigma}\bar{\psi}_{\mu i}\gamma_\nu\mathcal{D}_\rho\psi^i_\sigma + \tfrac14\,\bar{\psi}_{\mu i}\gamma^5\gamma^\mu\chi^i \, .
\end{equation}
Throughout, the full superconformal covariant derivative is denoted by~$D_\mu$, while~$\mathcal{D}_\mu$ denotes a derivative covariant with respect to Lorentz, dilatation, chiral~$\mathrm{SO}(1,1)$, and~$\mathrm{SU}(2)$ transformations. Due to the fact that the dilatation gauge field~$b_\mu$ transforms under conformal boosts, the superconformal d'Alembertian acting on a field with non-zero Weyl weight contains the trace of~$f_\mu{}^a$. In particular, on a scalar field~$\varphi$ of weight~$w$,
\begin{equation}
\square_c \varphi \equiv D^\mu D_\mu \varphi = \mathcal{D}^\mu D_\mu \varphi + w\,f_\mu{}^\mu\varphi + \text{fermions} \, . 
\end{equation} 

For completeness we present the definitions of all the supercovariant
curvature tensors,
\begin{align}
  \label{eq:curvatures}
  R(P)_{\mu \nu}{}^a  =&\; 2 \, \mathcal{D}_{[\mu} \, e_{\nu]}{}^a
  - \tfrac1{2}\,\bar{\psi}_{i[\mu}\gamma^5\gamma^a\psi_{\nu]}{}^i \, ,
  \nonumber\\[.2ex] 
  R(Q)_{\mu \nu}{}^i = & \; 2 \, \mathcal{D}_{[\mu} \psi_{\nu]}{}^i -
  \mathrm{i}\,\gamma_{[\mu} \phi_{\nu]}{}^i +
  \tfrac{1}{16}\mathrm{i}\,T_{ab}\,
  \gamma^{ab}\gamma_{[\mu} \psi_{\nu]}{}^i \, , \nonumber\\[.2ex] 
  R(D)_{\mu \nu} = & \;2\,\partial_{[\mu} b_{\nu]} - 2\,f_{[\mu}{}^a
  e_{\nu]a} -
  \tfrac{1}{2}\mathrm{i}\,\bar{\psi}_{i[\mu}\gamma^5\phi_{\nu]}{}^i +
  \tfrac{3}{4}\,\bar{\psi}_{i[\mu}\gamma^5\gamma_{\nu]} \chi^i \, ,
  \nonumber\\[.2ex] 
  R(A)_{\mu \nu} = &\; 2 \, \partial_{[\mu} A_{\nu]} +
  \tfrac12\mathrm{i}\,\bar{\psi}_{i[\mu}{}\phi_{\nu]}{}^i +
  \tfrac34\,\bar{\psi}_{i[\mu}\gamma_{\nu]}\chi^i \, ,
  \nonumber\\[.2ex] 
  R(\mathcal{V})_{\mu \nu}{}^i{}_j =& \;
  2\, \partial_{[\mu}\mathcal{V}_{\nu]}{}^i{}_j +
  \mathcal{V}_{[\mu}{}^i{}_k \, \mathcal{V}_{\nu]}{}^k{}_j  \nonumber \\ 
  &-  2\mathrm{i}\,\bar{\psi}_{j[\mu}\gamma^5\phi_{\nu]}{}^i +
  3\,\bar{\psi}_{j[\mu}\gamma^5\gamma_{\nu]} \chi^i +
  \tfrac12\delta^i{}_j\bigl(2\mathrm{i}\,\bar{\psi}_{k[\mu}\gamma^5\phi_{\nu]}{}^k
  -3\,\bar{\psi}_{k[\mu}\gamma^5\gamma_{\nu]} \chi^k \bigr) \, ,
  \nonumber \\[.2ex] 
  R(M)_{\mu \nu}{\!}^{ab} =& \; 2 \,\partial_{[\mu} \omega_{\nu]}{}^{ab}
  - 2\, \omega_{[\mu}{}^{ac}\omega_{\nu]c}{}^b - 4 f_{[\mu}{}^{[a}
  e_{\nu]}{}^{b]} +
  \tfrac12\mathrm{i}\,\bar{\psi}_{i[\mu}\gamma^5\gamma^{ab}\phi_{\nu]}{}^i \\
  &
  -\tfrac18\mathrm{i}\,\bar{\psi}_{\mu\,i}\gamma^5\psi_\nu{}^i\,T^{ab} -
  \tfrac34\bar{\psi}_{i[\mu}\gamma^5\gamma_{\nu]}\gamma^{ab}\chi^i -
  \bar{\psi}_{i[\mu}\gamma^5\gamma_{\nu]}R(Q)^{ab\,i} \, ,
  \nonumber\\[.2ex] 
  R(S)_{\mu\nu}{\!}^i  =& \; 2 \,\mathcal{D}_{[\mu}\phi_{\nu]}{}^i +
  2\mathrm{i}\,f_{[\mu}{}^a\gamma_a \psi_{\nu]}{}^i +
  \tfrac1{16}\mathrm{i}\,\Slash{D}T_{ab}\,\gamma^{ab}
  \gamma_{[\mu}\psi_{\nu]}{}^i \nonumber\\
  & +\tfrac32
  \mathrm{i}\,\gamma_a\psi_{[\mu}{}^i\bar{\psi}_{\nu]j} 
  \gamma^5\gamma^a\chi^j -
  \tfrac14\mathrm{i}\,R(\mathcal{V})_{ab}{}^i{}_j\gamma^{ab}
  \gamma_{[\mu}\psi_{\nu]}{}^j   -
  \tfrac12\mathrm{i}\,R(A)_{ab}\gamma^5\gamma^{ab}
  \gamma_{[\mu}\psi_{\nu]}{}^i
  \, , \nonumber\\[.2ex] 
  R(K)_{\mu\nu}{\!}^a =&\; 2\,\mathcal{D}_{[\mu}f_{\nu]}{}^a 
  - \tfrac14\,\bar{\phi}_{i[\mu}\gamma^5\gamma^a\phi_{\nu]}{}^i
  \nonumber\\
  & \!\!\!-\tfrac18\Bigl[\mathrm{i}\,
  \bar{\psi}_{i[\mu}\gamma^5D_b T^{ba}\,\psi_{\nu]}^i  
  + 6 e_{[\mu}{}^a\bar{\psi}_{\nu]i}\gamma^5\Slash{D}\chi^i
%  \nonumber\\
%  &\;\qquad 
  - 3 D \bar{\psi}_{i[\mu}\gamma^5\gamma^a\psi_{\nu]}{}^i 
  +  8\bar{\psi}_{i[\mu}\gamma^5\gamma_{\nu]}D_b R(Q)^{ba\,i} \Bigr] \nonumber \,. 
\end{align}
It is convenient to introduce two modified curvatures by including suitable covariant terms,
\begin{equation}
\begin{split}
  \label{eq:modified-R}
  \mathcal{R}(M)_{ab}{}^{cd} \equiv&\; R(M)_{ab}{}^{cd} 
  + \tfrac1{32}\bigl(T_{ab}^-\,T^{+cd} + T_{ab}^+\,T^{-cd}\bigr) \, , \\ 
  \mathcal{R}(S)_{ab}{}^i{}_\pm \equiv&\; R(S)_{ab}{}^i{}_\pm 
  - \tfrac38\,T_{ab}^\pm\,\chi^i_\pm \, .
\end{split}
\end{equation}
By making use of the conventional constraints and the Bianchi identities of the superconformal algebra, one can show that the modified curvature 
$\mathcal{R}(M)_{ab}{}^{cd}$ satisfies the following relations:
\begin{equation}
\begin{split}
  \label{eq:curvature-rel}
  \mathcal{R}(M)_{\mu\nu}{}^{ab}\,e^\nu{}_b =&\; \widetilde{R}(A)_\mu{}^a + \tfrac32\,D\,e_\mu{}^a \, , \\ 
  \tfrac14\,\varepsilon_{ab}{}^{ef}\,\varepsilon^{cd}{}_{gh}\,\mathcal{R}(M)_{ef}{}^{gh} =&\; \mathcal{R}(M)_{ab}{}^{cd} \, , \\
  \varepsilon_{cdea}\,\mathcal{R}(M)^{cd\,e}{}_b =&\; \varepsilon_{becd}\,\mathcal{R}(M)_a{}^{e\,cd} = 2\,\widetilde{R}(D)_{ab} = 2\,R(A)_{ab} \, .
\end{split}
\end{equation}
Note that this modified curvature does not satisfy the pair exchange property,
\begin{equation}
  \label{eq:lack-of-pair=exchange}
\mathcal{R}(M)_{ab}{}^{cd} - \mathcal{R}(M)^{cd}{}_{ab} = 
4\,\delta_{[a}{}^{[c}\,\widetilde{R}(A)_{b]}{}^{d]} \, .
\end{equation}

In addition to the Weyl multiplet, we must also consider matter multiplets. Even though we restrict ourselves to minimal supergravity in most of this paper, the superconformal formulation of the theory is equivalent to the usual Poincar\'{e} formulation only upon including suitable compensating multiplets~\cite{deWit:1980lyi,Lauria:2020rhc}. These multiplets act as gauge compensators and allow one to gauge-fix the extra superconformal symmetries. After eliminating the auxiliary fields, one is left with a supergravity theory invariant under the super-Poincar\'{e} group which describes a physical metric, graviphoton field, and gravitini. For this procedure, two compensating multiplets are required. One is necessarily a vector multiplet, which contains the vector field that will become the graviphoton in the Poincar\'{e} theory. The Euclidean version of this multiplet involves two real scalar fields~$X_+$ and~$X_-$, one symplectic-Majorana spinor~$\Omega^i$, the gauge field~$W_\mu$ with field strength~$F_{\mu\nu}$, and an auxiliary field~$Y^{ij}$ obeying the pseudo-reality condition~$(Y^{ij})^* \equiv Y_{ij} = \varepsilon_{ik}\varepsilon_{jl}\,Y^{kl}$. We refer the reader to~\cite{deWit:2017cle} for more details. 

The other compensating multiplet can be either a hypermultiplet, a tensor multiplet, or the so-called non-linear multiplet, and different choices lead to different formulations of Poincar\'{e} supergravity. These formulations have been discussed at the two-derivative level\footnote{We are not aware of a rigorous and thorough discussion of the various formulations of Poincar\'{e} supergravity based on different compensators when higher-derivative terms are present.} in~\cite{deWit:1982na}. The hypermultiplet involves the scalars~$A_i{}^\alpha$ which are local sections of an~$\mathrm{Sp}(1) = \mathrm{SU}(2)$ bundle, together with the symplectic-Majorana fermions~$\zeta^\alpha$. There exists a covariantly constant anti-symmetric tensor~$\Omega_{\alpha\beta}$ (and its complex conjugate~$\Omega^{\alpha\beta}$ satisfying the reality condition~$\Omega_{\alpha\gamma}\,\Omega^{\beta\gamma} = \delta_\alpha{}^\beta$), which in principle depends on the scalars. The~$\mathrm{Sp}(1)$ connection is provided by the~$\mathrm{SU}(2)$ R-symmetry gauge field~$\mathcal{V}_\mu{}^i{}_j$. The sections~$A_i{}^\alpha$ are pseudo-real, i.e. they are subject to the constraint~$\varepsilon^{ij} \Omega_{\alpha\beta}A_i{}^\alpha  = A^j{}_\beta \equiv (A_j{}^\beta)^\ast$. The target-space is a hyper-K\"ahler cone whose metric is encoded in the so-called hyper-K\"ahler potential. In this paper, we only consider the case where the hyper-K\"ahler cone is flat. To ensure that the Poincar\'{e} theory based on a hypermultiplet compensator is gauged, we must include a coupling to the compensating vector multiplet~\cite{Lauria:2020rhc}. Of course, this requires that the tensor~$\Omega_{\alpha\beta}$, as well as related quantities, are invariant under the gauge group. In addition the covariant derivatives must be covariantized also with respect to the gauge group.

For completeness, we also consider the possibility of using a tensor multiplet as gauge compensator in view of the discussion in Appendix~\ref{app:R2}. The Euclidean version comprises a triplet of scalar fields~$\texttt{L}_{ij} = \varepsilon_{ik} \varepsilon_{jl}\, \texttt{L}^{kl}$, a symplectic-Majorana spinor~$\varphi^i$, an anti-symmetric tensor gauge field~$E_{\mu\nu}$ and two real scalars~$G_\pm$. The supercovariant field strength of~$E_{\mu\nu}$ is~$E^\mu = \tfrac12\,e^{-1}\varepsilon^{\mu\nu\rho\sigma}\partial_\nu E_{\rho\sigma}$ modulo fermions. Just as for the hypermultiplet, we obtain a gauged supergravity theory in the Poincar\'{e} frame provided we include a coupling to the compensating vector multiplet. The field content of the three multiplets just mentionned is summarized in Table~\ref{table:w-weights-matter-4D} together with their Weyl and chiral weights.

%%%%%%%%%%%%%%%
\begin{table}[t]
\begin{center}
\begin{tabular*}{12.5cm}{@{\extracolsep{\fill}}|c|cccc|cccc|cc| }
\hline
 & \multicolumn{4}{c|}{vector multiplet} & \multicolumn{4}{c|}{tensor multiplet} & 
 \multicolumn{2}{c|}{hypermultiplet} \\
 \hline \hline
 field & $X_\pm$ & $W_\mu$  & $\Omega^i{\!}_\pm$ & $Y^{ij}$& $\texttt{L}^{ij}$&$E_a$
 &$\varphi^i_\pm$ & $G_\pm$ & 
 $A_i{}^\alpha$ & $\zeta^\alpha{\!}_\pm$ \\[.5mm] \hline
$w$  & $1$ & $0$ & $\tfrac32$ & $2$ &$2$&$3$&$\tfrac52$ &$3$ &
 $1$ &$\tfrac32$
\\[.5mm] \hline
$c$  & $\mp1$ & $0$ & $\mp\tfrac12$ & $0$ &$0$&$0$& $\pm\tfrac12$
&$\pm1$ & $0$ &$\pm\tfrac12$ 
\\[.5mm] \hline
$\gamma_5$   & && $\pm$  &&&& $\pm$&   &  & $\pm$ \\ \hline
\end{tabular*}
\vskip 2mm
\renewcommand{\baselinestretch}{1}
\parbox[c]{12.5cm}{\caption{\footnotesize 
      Weyl weights $w$, chiral $\mathrm{SO}(1,1)$ weights $c$, 
      and chirality $\gamma_5$ for the fields in the Euclidean
      vector multiplet, tensor multiplet, and
      hypermultiplet. \label{table:w-weights-matter-4D}  }} 
\end{center}
\end{table}
%%%%%%%%%%%%%%%

%%%%%%%%%%%%%%%%%%%%%%%%%%%%%%%%%%%%%%%%
\section{An alternative off-shell formulation and the \texorpdfstring{$R^2$}{R2} invariant}
\label{app:R2}
%%%%%%%%%%%%%%%%%%%%%%%%%%%%%%%%%%%%%%%%

Besides the Weyl-squared and~$\mathbb{T}$ invariants introduced in Section~\ref{sec:superconf}, it turns out that there are additional four-derivative invariants that contain terms quadratic in the Ricci scalar~$R$. Those are built from tensor multiplets~\cite{deWit:2006gn,Kuzenko:2015jxa,Hegde:2019ioy}. It is natural to wonder what is their effect on the form of the PHD action~\eqref{eq:PHD} and how they affect the two-derivative solutions. We already mentioned that there exists an alternative formulation of Poincar\'{e} gauged supergravity that makes use of a tensor multiplet compensator instead of the hypermultiplet used previously~\cite{deWit:1982na}. Before analyzing the~$R^2$ invariant proper, let us recall how this alternative formulation works at the two-derivative level. To make use of existing results in the papers just mentioned, we work in Lorentzian signature for the remainder of this section. \\

The bosonic terms of the superconformally invariant Lagrangian density for the charged compensating tensor multiplet take the form~\cite{deWit:2006gn}:
\begin{align}
e^{-1}\mathcal{L}_\mathrm{T} =&\; -\frac12\,\texttt{L}^{-1}|\mathcal{D}_\mu \texttt{L}_{ij}|^2 + \texttt{L}\,\Bigl(\frac13 R + D\Bigr) + \texttt{L}^{-1}\bigl(E_\mu E^\mu + |G|^2\bigr) - \texttt{L}^{-1} E^\mu\mathcal{V}_\mu{}^i{}_j\,\texttt{L}_{ik}\,\varepsilon^{jk} \nonumber \\ 
&\; - \frac12\mathrm{i}\,e^{-1}\varepsilon^{\mu\nu\rho\sigma}\texttt{L}^{-3} \texttt{L}^{ij} E_{\mu\nu}\,\partial_\rho \texttt{L}_{ik}\,\partial_\sigma \texttt{L}_{jl}\,\varepsilon^{kl} \\
&\; + g\Bigl[X G + \bar{X} \bar{G} - \frac12\,Y^{ij} \texttt{L}_{ij} - \frac12\,E^\mu W_\mu\Bigr] \, , \nonumber
\end{align}
where~$g$ is the gauge coupling and~$\texttt{L}^2 \equiv \texttt{L}^{ij}\texttt{L}_{ij}$.
We now consider the two-derivative Lagrangian 
\begin{equation}
\label{eq:2der-VT}
\mathcal{L}_{2\partial} = \mathcal{L}_\mathrm{V} - \mathcal{L}_\mathrm{T} \, , 
\end{equation}
where we recall that the Lagrangian density for the compensating vector multiplet in Lorentzian signature is given by~\cite{Lauria:2020rhc}
\begin{align}
e^{-1}\mathcal{L}_\mathrm{V} =&\; -4\,|X|^2\Bigl(\frac16 R - D\Bigr) + 4\,\mathcal{D}_\mu X\,\mathcal{D}^\mu \bar{X} + \frac12\bigl(\widehat{F}_{ab}^-\bigr)^2 + \frac12\bigl(\widehat{F}_{ab}^+\bigr)^2 \\
&\;- \frac14\,X\widehat{F}_{ab}^+\,T^{ab\,+} - \frac14\,\bar{X}\widehat{F}_{ab}^-\,T^{ab\,-} - \frac12\,Y^{ij}Y_{ij} - \frac{1}{32}\,X^2 (T_{ab}^+)^2 - \frac{1}{32}\,\bar{X}^2 (T_{ab}^-)^2 \, . \nonumber
\end{align}
To go to the Poincar\'{e} frame, we impose the V-gauge~$\texttt{L}_{ij} = \delta_{ij}\texttt{L}$, the D-gauge~$\texttt{L} = $ constant, the K-gauge~$b_\mu = 0$ and the A-gauge~$X = \bar{X}$. The last two gauge choices are the same as in the formulation with a hypermultiplet compensator. The two-derivative EoMs for the auxiliary scalar fields~$D$,~$G$ and~$Y_{ij}$ then impose, respectively,
\begin{equation}
X = \frac12\,\texttt{L}^{1/2} \, , \quad G = \bar{G} = -\frac12\,g\,\texttt{L}^{3/2} \, , \quad Y_{ij} = \frac12\,g\,\texttt{L}_{ij} \, ,
\end{equation}
while the EoMs for the~$\mathrm{SU}(2)$ R-symmetry connection and the vector field~$E_\mu$ fix
\begin{equation}
E_\mu = 0 \, , \quad \mathcal{V}_\mu{}^i{}_j\,\delta_{ik}\,\varepsilon^{jk} = -\frac12\,g\,W_\mu \, .
\end{equation}
Using this in~\eqref{eq:2der-VT} gives the two-derivative Lagrangian density in the Poincar\'{e} frame,
\begin{equation}
e^{-1}\mathcal{L}_{2\partial} = -\frac12\,\texttt{L}\,R + \frac{3}{8}\,g^2\,\texttt{L}^2 - \frac12\,F_{\mu\nu}F^{\mu\nu} \, .
\end{equation}
We can now normalize the Einstein-Hilbert term by fixing the constant in the D-gauge as~$\texttt{L} = \kappa^{-2}$. Defining~$L \equiv 2\sqrt{2}\,g^{-1}\kappa$ and implementing the redefinitions in Footnote~\ref{foot:redef}, we recover precisely the two-derivative part of~\eqref{eq:PHD}. Thus, as is well-known, the formulations of Poincar\'{e} supergravity using a tensor multiplet or a hypermultiplet as a compensator are equivalent at the two-derivative level. \\

We now turn on the~$R^2$ invariant of~\cite{deWit:2006gn}. In particular, we use their equation (3.21) with~$\mathcal{H}(\texttt{L}) = c_3\,\texttt{L}^{-2}$, and add the resulting Lagrangian density to~\eqref{eq:SCHD-chi},
\begin{equation}
\label{eq:HD-tensor-full}
\mathcal{L}_{\text{HD}} = \mathcal{L}_\mathrm{V} - \mathcal{L}_\mathrm{T} + (c_1 - c_2)\mathcal{L}_{\mathrm{W}^2} + c_2\,\mathcal{L}_\text{GB} + c_3\,\mathcal{L}_{R^2} \, ,
\end{equation}
with
\begin{align}
\label{eq:R2}
\mathcal{L}_{R^2} = \texttt{L}^{-2}\Bigl[&-\frac12\,\texttt{L}^2\Bigl(\frac13 R + D\Bigr)^2 + E^2\Bigl(\frac13 R + D\Bigr) + |G|^2\Bigl(\frac16 R + 2D\Bigr) \nonumber \\
&- \mathcal{D}_a E_b\Bigl(\texttt{L}\,R(\mathcal{V})'^{ab} - \frac12\bigl(T^{-ab} G + T^{+ab}\bar{G}\bigr)\Bigr) + \frac18\Bigl(\texttt{L}\,R(\mathcal{V})'_{ab} - \frac12\bigl(T^-_{ab} G + T^+_{ab}\bar{G}\bigr)\Bigr)^2 \nonumber \\
&- \frac1{64}\Bigl(T^-_{ab} G + T^+_{ab}\bar{G}\Bigr)^2 + |\mathcal{D}_\mu G|^2 + 2\,(\mathcal{D}_{[a}E_{b]})^2\Bigr] \\
- 4&\,\texttt{L}^{-4}\,\bigl(|G|^2 + E^2\bigr)^2 \, . \nonumber
\end{align}
Here, we have imposed the V- and D-gauge given above, and defined the diagonal piece of the~$\mathrm{SU}(2)_R$ connection~$\mathcal{V}_\mu' \equiv \mathcal{V}_\mu{}^i{}_j\,\delta_{ik}\,\varepsilon^{jk}$, and the corresponding curvature~$R(\mathcal{V})_{ab}'$. The EoMs for the superconformal fields are now modified due to the presence of this new invariant. For the auxiliary scalar field~$D$, we find
\begin{equation}
\label{eq:tensor-HD-D-EoM}
0 = \texttt{L} - 4 X^2 + \bigl(12\,(c_2 - c_1) + c_3\bigr)D - c_3\,\texttt{L}^{-2}\,\Bigl(2\,|G|^2 + E^\mu E_\mu - \frac{1}{3}\,\texttt{L}^2 R\Bigr) \, .
\end{equation}
The auxiliary complex scalar~$G$ of the tensor multiplet has become dynamical in the higher-derivative theory, and its EoM reads
\begin{equation}
\begin{split}
0 =&\; \texttt{L} G + g\,\texttt{L}^2 X \\
& +c_3\Bigl(\mathcal{D}^\mu\mathcal{D}_\mu G - G\Bigl(\frac16 R + 2D\Bigr) - \frac12\,T_{ab}^+\mathcal{D}^a E^b + \frac18\,\texttt{L}\,R_{ab}' T^{+ab} - \frac1{32}(T_{ab}^+)^2\,\bar{G}  \\
& \qquad\;\; + 8\,\texttt{L}^{-2} G\,(|G|^2 + E^\mu E_\mu)\Bigr) \, .
\end{split}
\end{equation}
The EoM for the field~$\mathcal{V}_\mu'$ picks up a~$c_3$-dependent correction, 
\begin{equation}
0 = E_\nu + 2\,\Bigl(c_2 - c_1 - \frac14\,c_3\Bigr)\,\mathcal{D}^\mu R'_{\mu\nu} + 2\,c_3\,\texttt{L}^{-1}\mathcal{D}^\mu\Bigl[\mathcal{D}_\mu E_\nu + \frac18\,T_{\mu\nu}^- G + \frac18\,T_{\mu\nu}^+ \bar{G}\,\Bigr] \, ,
\end{equation}
while the~$E_\mu$-EoM in the higher-derivative theory reads
\begin{equation}
\begin{split}
0 =&\; \mathcal{V}_\nu' + \frac12\,g W_\nu - 2\,\texttt{L}^{-1} E_\nu \\
&- 2\,c_3\,\texttt{L}^{-2}\Bigl(4\,\mathcal{D}^\mu\mathcal{D}_\mu E_\nu - \mathcal{D}^\mu\Bigl[R'_{\mu\nu}\,\texttt{L} - \frac12\,T_{\mu\nu}^- G - \frac12\,T_{\mu\nu}^+ \bar{G}\,\Bigr] \\
&\qquad\qquad\quad - E_\nu\Bigl[\frac13 R + D\Bigr] + 4\,\texttt{L}^{-2} E_\nu\,(|G|^2 + E^\mu E_\mu)\Bigr) \, . 
\end{split}
\end{equation}
Lastly, the EoM for the~$U(1)_R$ connection is given by
\begin{equation}
\begin{split}
0 =&\; X^2 A_\nu + (c_2 - c_1)\,\Bigl(\nabla^\mu R(A)_{\mu\nu} + \frac1{16}\,T^+_\nu{}^\rho\,\mathcal{D}^\mu T_{\mu\rho}^- - \frac1{16}\,T^-_\nu{}^\rho\,\mathcal{D}^\mu T_{\mu\rho}^+\Bigr) \\
&- \frac18\mathrm{i}\,c_3\,\texttt{L}^{-2}\,\bigl(G\,\mathcal{D}_\nu\bar{G} - \bar{G}\,\mathcal{D}_\nu G\bigr) \, , 
\end{split}
\end{equation}
and the EoM for the (anti-)self-dual projections of the~$T$-tensor by
\begin{equation}
\begin{split}
0 =&\; X\Bigl(F_{ab}^\pm - \frac14\,X\,T_{ab}^\pm\Bigr) \\
&+ (c_2 - c_1)\Bigl(\Pi^{ef}_{\pm\,ab}\Bigl[\mathcal{D}_e\mathcal{D}^c T_{cf}^\mp + \frac12\,T^{\mp c}{}_e R_{cf}\Bigr] - \frac1{128}\,T_{ab}^\pm(T_{cd}^\mp)^2\Bigr) \\
& + c_3\,\texttt{L}^{-2}\,G^\mp\,\Pi^{cd}_{\pm\,ab}\Bigl[\mathcal{D}_cE_d - \frac14\,\texttt{L}\,R'_{cd} + \frac1{16}\bigl(T_{cd}^-\,G + T_{cd}^+\,\bar{G}\,\bigr)\Bigr] \, ,
\end{split}
\end{equation}
where~$G^- = \bar{G}$ and~$G^+ = G$. It should be clear that, unlike the case when~$c_3 = 0$ analyzed in the main text, the two-derivative solutions for the superconformal fields are \emph{not} automatically solutions to the above EoMs. The nature of the obstruction warrants some comments. Plugging in the two-derivative solutions, we find the following constraint from the~$G$-EoM,
\begin{equation}
g\,c_3\Bigl[\texttt{L}\,\Bigl(\frac16\,R + 2\,D_*\Bigr) + \frac{1}{4}\,(F_{ab})^2 - 2\,g^2\,\texttt{L}^2\Bigr] = 0 \, , 
\end{equation}
where the~$D$ scalar appearing in the equation is taken at its on-shell value
\begin{equation}
D_* = - \frac{c_3}{12\,(c_1 - c_2) - c_3}\,\Bigl(\frac{1}{2}\,g^2\,\texttt{L} - \frac{1}{3} R\Bigr) \, .
\end{equation}
Another obstruction comes from the~$T$-EoM, which gives
\begin{equation}
g\,c_3\,F_{ab} = 0 \, .
\end{equation}
Note that there is no obstruction from the other EoMs. Thus we see that the two-derivative solutions automatically solve the EoMs of the four-derivative theory only in the case where
\begin{equation}
g\,c_3 = 0 \, . 
\end{equation}
When we have both the~$R^2$ invariant and a gauging, we must instead resort to solving the EoMs perturbatively in~$c_3$. From the obstruction just mentioned, we know that the~$G$ and~$T_{ab}$ two-derivative solutions should be modified at first order in~$c_3$. So we let
\begin{equation}
G = G_0 + c_3\,G_1 \, , \qquad T_{ab}^\pm = T_{ab,0}^\pm + c_3\,T_{ab,1}^\pm \, .
\end{equation}
From the~$G$-EoM expanded to first order, we obtain
\begin{equation}
G_1 = -\frac{g}{8}\,\texttt{L}^{-1/2}\,F_{ab}^2 + \frac78\,g^3\,\texttt{L}^{3/2} \, ,
\end{equation}
where we have also used the trace of the two-derivative Einstein equation~$R = \tfrac32\,g^2\,\texttt{L}$. For the first correction to the~$T$-tensor, we find
\begin{equation}
T_{ab,1}^\pm = g^2\,\texttt{L}^{-1/2} F_{ab}^\pm \, .
\end{equation}
Using this, the action~\eqref{eq:HD-tensor-full} to first order in~$c_3$ reads
\begin{equation}
e^{-1}\mathcal{L}_{\text{HD}} = -\frac12\,\texttt{L}\,R\,\Bigl(1 - \frac{g^2 c_3}{12}\Bigr) + \frac38\,g^2\,\texttt{L}^2\,\Bigl(1 - \frac{2 g^2 c_3}{3}\Bigr) - \frac12\,F_{ab}^2\,\Bigl(1 + \frac{3g^2 c_3}{16}\Bigr) - \frac{c_3}{18}\,R^2 \, ,
\end{equation}
where we have set~$c_1 = c_2 = 0$ for clarity. Although we still have the freedom to fix the constant~$\texttt{L}$ via the D-gauge, it is clear that no choice of~$\texttt{L}$ will give rise to a canonical normalization for both the Einstein-Hilbert and the cosmological constant term. Evidently, this is remedied when considering ungauged supergravity where~$g=0$. But since we will be interested in asymptotically AdS solutions for holographic applications, the analysis of this section prompts us to set~$c_3 = 0$. 

As a final remark, we note that the peculiarities of the~$R^2$ invariant highlighted here originate from the fact that it has been constructed from a \emph{compensating} multiplet, unlike the~$\mathrm{W}^2$ and~$\mathbb{T}$ invariants based on the Weyl multiplet. While the latter contains physical fields for which it is sensible to consider higher-derivative corrections, the former should ultimately be seen as a useful tool to formulate the supergravity theory of interest in the Poincar\'{e} frame. As such, the reasons to consider higher-derivative corrections for a compensator seem unclear in the first place. Of course, in the presence of physical tensor fields, this point is moot and one should indeed include the~$R^2$ invariants of~\cite{deWit:2006gn,Kuzenko:2015jxa,Hegde:2019ioy} when considering four-derivative actions.

%%%%%%%%%%%%%%%%
\section{Supergravity spectrum}
\label{app:spec}
%%%%%%%%%%%%%%%%%%%%%%%%%%%%

In this Appendix we present the details of the calculation of the supergravity spectrum around the Euclidean AdS$_4$ vacuum solution as discussed in Section~\ref{sec:univ}. We start with the Euclidean superconformal higher-derivative action \eqref{eq:SCHD-chi} and fix the K-, V-, D- and A-gauges~\eqref{eq:K-gauge},~\eqref{eq:V-gauge},~\eqref{eq:D-gauge} and~\eqref{eq:A-gauge}. The resulting action can be written out explicitly as
\begin{equation}\begin{aligned}
	e^{-1} \mathcal{L} &= -\left(\frac{2 X^2}{3} + \frac{1}{3\kappa^2}\right) R + \left(\frac{1}{\kappa^2} - 4 X^2\right) D - \frac{g^2}{\kappa^4} - \frac{8 g^2}{\kappa^2} X^2 - 4 (\nabla_\mu X)(\nabla^\mu X) \\
	&\quad + 4 X^2 A_\mu A^\mu + \frac{X^2}{32}\left( T^+_{\mu\nu}T^{+\mu\nu} + T^-_{\mu\nu}T^{-\mu\nu}\right) - \frac{1}{2}\left(\widehat{F}^+_{\mu\nu}\widehat{F}^{+\mu\nu} + \widehat{F}^-_{\mu\nu}\widehat{F}^{-\mu\nu}\right) \\
	&\quad + \frac{X}{4}\left(\widehat{F}^+_{\mu\nu}T^{+\mu\nu} + \widehat{F}^-_{\mu\nu} T^{-\mu\nu}\right) - \frac{2 g^2}{\kappa^2}W_\mu W^\mu + \frac{1}{4\kappa^2} \mathcal{V}\ind{_\mu^j_i} \mathcal{V}\ind{^\mu^i_j}+ \frac{g}{\kappa^2} W^\mu \mathcal{V}\ind{_\mu^j_i}t\ind{^i_j} \\
	&\quad + (c_1 - c_2)\bigg{[} C_{\mu\nu\rho\sigma}C^{\mu\nu\rho\sigma} + 2 R(A)_{\mu\nu}R(A)^{\mu\nu} + 6 D^2 + \frac{1}{2} R(\mathcal{V})\ind{_{\mu\nu}^i_j} R(\mathcal{V})\ind{^{\mu\nu}_i^j} \\
	&\quad + \frac{1}{2}\mathcal{D}^\mu T^-_{\mu\nu} \mathcal{D}_\rho T^{+\rho\nu} - \frac{1}{4}T^-_{\mu\nu}R^{\mu\rho}T\ind{^+_\rho^\nu} +\frac{1}{512}T^+_{\mu\nu}T^{+\mu\nu} T^-_{\rho\sigma}T^{-\rho\sigma}\bigg{]} + c_2 e^{-1} \mathcal{L}_\text{GB}~,
\end{aligned}\end{equation}
where we have dropped total derivative terms that are irrelevant for the equations of motion as well as the irrelevant auxiliary field $Y_{ij}$, and we have defined
\begin{equation}
	\widehat{F}^\pm_{\mu\nu} \equiv F^\pm_{\mu\nu} - \frac{1}{4} X T^\pm_{\mu\nu}~.
\end{equation}

From now on, we assume that $c_1 - c_2 \neq 0$, in order to actually have non-trivial higher-derivative terms.  The equations of motion for the superconformal action with higher-derivative terms present were derived in Section~\ref{sec:Poincare}, but we now repeat them in detail in order to study the fluctuation spectrum of all fields.  First, we note that the $D$ equation of motion simply fixes $D$ to be 
\begin{equation}
	D = \frac{1}{3(c_1 - c_2)}\left(X^2 - \frac{1}{4 \kappa^2}\right)~.
\end{equation}
The equations of motion for $X$, $W_\mu$, $\mathcal{V}\ind{_\mu^i_j}$, $A_\mu$, and $T^\pm_{\mu\nu}$ are given by 
\begin{equation}\begin{aligned}
	0 &= \square X + \left( -\frac{1}{6} R - D - \frac{2 g^2}{\kappa^2} + A_\mu A^\mu - \frac{1}{64} T_{\mu\nu}T^{\mu\nu}\right)X + \frac{1}{16} F_{\mu\nu}T^{\mu\nu}~, \\
	0 &= \nabla_\mu F^{\mu\nu} - \frac{1}{2}\nabla_\mu (X T^{\mu\nu}) - \frac{2 g^2}{\kappa^2} W^\nu + \frac{g}{2\kappa^2} \mathcal{V}\ind{^\nu^j_i} t\ind{^i_j}~, \\
	0 &= \mathcal{D}_\mu R(\mathcal{V})\ind{^{\mu\nu i}_j} + \frac{1}{4\kappa^2 (c_2 - c_1)}\left( \mathcal{V}\ind{^\nu^i_j} + 2 g W^\nu t\ind{^i_j}\right)~, \\
	0 &= \nabla_\mu R(A)^{\mu\nu} + \frac{1}{16} T^{+\nu\rho}\mathcal{D}^\mu T^-_{\mu\rho} - \frac{1}{16} T^{-\nu\rho} \mathcal{D}^\mu T^+_{\mu\rho} + \frac{X^2}{c_2 - c_1} A^\nu~, \\
	0 &= \frac{1}{2}\left(\delta^\mu_{[\lambda} \delta^\nu_{\tau]} \pm \frac{1}{2}\varepsilon\ind{^\mu^\nu_\lambda_\tau}\right)\left(\mathcal{D}^\lambda \mathcal{D}^\rho T\ind{^\mp_\rho^\tau} + \frac{1}{2}R^{\lambda\rho} T\ind{^\mp_\rho^\tau}\right) \\
	&\qquad\qquad - \frac{1}{128} T^{\pm \mu\nu} (T^{\mp}_{\rho\sigma}T^{\mp \rho\sigma}) + \frac{X}{c_2 - c_1}\left(F^{\pm \mu\nu} - \frac{1}{4} X T^{\pm \mu\nu}\right)~.
\end{aligned}\end{equation}
We can use (anti-)self-duality identities to rewrite the equation of motion for $T^\pm_{\mu\nu}$ as follows:
\begin{equation}\begin{aligned}
	0 &= - \mathcal{D}_{[\mu} \mathcal{D}^\rho T^\mp_{\nu]\rho} - \frac{1}{2}R\ind{_{[\mu}^\rho} T^\mp_{\nu]\rho} - \frac{1}{4}\mathcal{D}_\rho \mathcal{D}^\rho T^\mp_{\mu\nu} - \frac{1}{8}R T^\mp_{\mu\nu} + \frac{1}{2}\left(\mathcal{D}_{[\mu} \mathcal{D}^\rho - \mathcal{D}^\rho \mathcal{D}_{[\mu}\right) T^\mp_{\nu]\rho} \\
	&\qquad\qquad - \frac{1}{128} T^{\pm \mu\nu} (T^{\mp}_{\rho\sigma}T^{\mp \rho\sigma}) + \frac{X}{c_2 - c_1}\left(F^{\pm \mu\nu} - \frac{1}{4} X T^{\pm \mu\nu}\right)~.
\end{aligned}\end{equation}
The Einstein equation can be written as
\begin{equation}\begin{aligned}
	0 &= \left(\frac{2 X^2}{3} + \frac{1}{3 \kappa^2}\right)\left(R_{\mu\nu} - \frac{1}{2} g_{\mu\nu} R\right) + g_{\mu\nu}\left(\frac{1}{2\kappa^2} - 2 X^2\right)D - g_{\mu\nu} \left(\frac{g^2}{2 \kappa^4} + \frac{4 g^2 X^2}{\kappa^2} \right) \\
	&\quad + \frac{8}{3} (\nabla_\mu X)(\nabla_\nu X) - \frac{2}{3} g_{\mu\nu} (\nabla_\rho X)(\nabla^\rho X) - \frac{4}{3} X \nabla_\mu \nabla_\nu X + \frac{4}{3} g_{\mu\nu} X \square X \\
	&\quad - (c_1 - c_2)\bigg{[} 4 R_{\mu\rho}R\ind{_\nu^\rho} - g_{\mu\nu} R_{\rho\sigma}R^{\rho\sigma} - \frac{4}{3} R_{\mu\nu} R + \frac{1}{3} g_{\mu\nu} R^2 + 2 \square R_{\mu\nu} - 4 \nabla^\rho \nabla_\mu R_{\nu\rho} \\
	&\qquad\qquad - \frac{1}{3} g_{\mu\nu} \square R + \frac{4}{3} \nabla_\mu \nabla_\nu R + \ldots \bigg{]} ~,
\end{aligned}\end{equation}
where the dots indicate the four-derivative terms that are quadratic in fields that vanish in the AdS$_4$ vacuum.

We now want to study the spectrum of field fluctuations around the AdS$_4$ vacuum.  We introduce the following notation for the field fluctuations:
\begin{equation}\begin{aligned}
	\delta X &= \phi~, \qquad	\delta W_\mu = w_\mu~, \qquad \delta \mathcal{V}\ind{_\mu^i_j} = v\ind{_\mu^i_j}~, \\
	\delta A_\mu &= a_\mu~, \qquad \delta T^\pm_{\mu\nu} = t^\pm_{\mu\nu}~, \qquad	\delta g_{\mu\nu} = h_{\mu\nu}~.
\end{aligned}\end{equation}
We also impose the standard Lorenz gauge for the vector field fluctuations:
\begin{equation}
	\nabla^\mu w_\mu = \nabla^\mu v\ind{_\mu^i_j} = \nabla^\mu a_\mu = 0~,
\end{equation}
as well as the harmonic gauge for the metric fluctuations:
\begin{equation}
	\nabla^\mu h_{\mu\nu} = \nabla_\nu h~.
\end{equation}
It is important to notice that even if $D$ is an auxiliary field that is fixed to zero for AdS$_4$, its linearized variation is non-zero:
\begin{equation}
	\delta D = \frac{1}{3 \kappa (c_1 - c_2)} \phi~.
\end{equation}
The linearized variations of the equations of motion are then given by:
\begin{equation}\begin{aligned}
\label{eq:fluct}
	X:\quad 0 &= \left(\square - \frac{1}{6 \kappa^2 (c_1 - c_2)}\right) \phi - \frac{1}{4\kappa L^2} h~, \\
	W_\mu: \quad 0 &= \left(\square + \frac{1}{L^2}\right)w_\mu + \frac{1}{4\kappa}\nabla^\nu\left(t^+_{\mu\nu} + t^-_{\mu\nu}\right) + \frac{1}{2\kappa L} v\ind{_\mu^j_i} t\ind{^i_j}~, \\
	\mathcal{V}\ind{_\mu^i_j}: \quad 0 &= \left(\square + \frac{3}{L^2} - \frac{1}{4\kappa^2 (c_1 - c_2)}\right) v\ind{_\mu^i_j} - \frac{1}{2 \kappa L (c_1 - c_2)} w_\mu t\ind{^i_j}~, \\
	A_\mu: \quad 0 &= \left(\square + \frac{3}{L^2} - \frac{1}{4 \kappa^2 (c_1 - c_2)}\right) a_\mu ~, \\
	T^\pm_{\mu\nu}: \quad 0 &= -\nabla_{[\mu}\nabla^\rho t^\mp_{\nu]\rho} - \frac{1}{4}\left(\square + \frac{4}{L^2}\right)t^\mp_{\mu\nu}  + \frac{1}{16 \kappa^2 (c_1 - c_2)} t^\pm_{\mu\nu} - \frac{1}{2\kappa (c_1 - c_2)} f^\pm_{\mu\nu}~, \\
	g_{\mu\nu}: \quad 0 &= \frac{1}{2\kappa^2}\left( - \frac{1}{2}\square h_{\mu\nu} + \frac{1}{2}\nabla_\mu \nabla_\nu h - \frac{1}{L^2} h_{\mu\nu} - \frac{1}{2L^2} g_{\mu\nu} h \right) - \frac{2}{\kappa L^2} g_{\mu\nu} \phi \\
	&\quad + \frac{2}{3\kappa}\left(g_{\mu\nu} \square - \nabla_\mu \nabla_\nu \right)\phi - (c_1 - c_2)\bigg{[} - \square^2 h_{\mu\nu} - \frac{6}{L^2} \square h_{\mu\nu} - \frac{8}{L^4} h_{\mu\nu}  \\
	&\qquad\qquad + \square \nabla_\mu \nabla_\nu  h + \frac{1}{L^2} g_{\mu\nu} \square h + \frac{2}{L^2} \nabla_\mu \nabla_\nu h + \frac{2}{L^4} g_{\mu\nu} h\bigg{]}~.
\end{aligned}\end{equation}
Above, we have defined $f^\pm_{\mu\nu}$ as the self-dual and anti-self-dual parts of the field strength $f_{\mu\nu} \equiv \nabla_\mu w_\nu - \nabla_\nu w_\mu$.

Note that when we take the trace of the linearized Einstein equation, the quartic derivative terms drop out and we are simply left with another differential equation relating $\phi$ and $h$:
\begin{equation}
	0 = \left(\square - \frac{4}{L^2}\right) \phi - \frac{3}{4\kappa L^2} h~.
\label{eq:fluct2}
\end{equation}
We can combine \eqref{eq:fluct2} with the first equation in \eqref{eq:fluct} to eliminate $h$ in favor of $\phi$ and find
\begin{equation}\label{eq:massphi}
	0 = \left(\square - m_{\phi}^2\right) \phi~, \qquad m_{\phi}^2 L^2 = \frac{ L^2}{4 \kappa^2 (c_1 - c_2)} - 2~.
\end{equation}
This in turn can be used to fix the trace mode $h$ entirely in terms of the scalar fluctuation $\phi$:
\begin{equation}
	h = \frac{4 \kappa}{3}\left(m_\phi^2 L^2 - 4\right)\phi = \frac{\kappa}{3}\left(\frac{L^2}{\kappa^2 (c_1 - c_2)} - 8\right) \phi~.
\label{eq:hphiconstraint}
\end{equation}
That is, the fields $h$ and $\phi$ are not independent from one another, and there is a single propagating scalar degree of freedom with an associated mass $m_\phi$.  This massive scalar mode is holographically dual to a scalar operator $\mathcal{O}_\phi$ with conformal dimension
\begin{equation}
	\Delta_{\mathcal{O}_\phi} = \frac{3}{2} + \sqrt{\frac{9}{4} + m_\phi^2 L^2} = \frac{3}{2} + \sqrt{\frac{1}{4} + \frac{L^2}{4 \kappa^2 (c_1 - c_2)}}~.
\label{eq:confdimphi}
\end{equation}

We now return to the linearized Einstein equation.  We first decompose the metric fluctuation into its transverse, traceless component $h_{\mu\nu}^\text{TT}$ and its trace $h$ as
\begin{equation}
	h_{\mu\nu} = h^\text{TT}_{\mu\nu} + \frac{1}{4} g_{\mu\nu} h~.
\end{equation}
We can then use the constraint \eqref{eq:hphiconstraint} to eliminate the trace mode $h$ in favor of $\phi$, which leaves us with:
\begin{equation}\begin{aligned}
	0 &= \frac{1}{2\kappa^2}\left(- \frac{1}{2}\square - \frac{1}{L^2}\right)h^\text{TT}_{\mu\nu} + (c_1 - c_2)\left(\square^2 + \frac{6}{L^2}\square + \frac{8}{L^4} \right)h^\text{TT}_{\mu\nu}	\\
	&\quad +\frac{2}{\kappa}\left(1 - \frac{32 \kappa^2 (c_1 - c_2)}{L^2}\right) \nabla_{\langle \mu} \nabla_{\nu\rangle} \phi~,
\end{aligned}\end{equation}
where we have defined the transverse, traceless differential operator
\begin{equation}
	\nabla_{\langle \mu} \nabla_{\nu\rangle} \equiv \nabla_\mu \nabla_\nu - \frac{1}{4} g_{\mu\nu} \square~.
\end{equation}
With this simplified form of the linearized Einstein equation, it is clear that there should be a change of variables that eliminates the $\phi$-dependence in the equation.  In particular, we can make the redefinition
\begin{equation}
	\psi_{\mu\nu} \equiv h_{\mu\nu}^\text{TT} + \lambda \kappa L^2 \nabla_{\langle \mu} \nabla_{\nu\rangle} \phi~,
\end{equation}
where $\lambda$ is some arbitrary constant, such that the linearized Einstein equation becomes
\begin{equation}\begin{aligned}
	0 &= -\frac{1}{4\kappa^2}\left(\square + \frac{2}{L^2}\right) \psi_{\mu\nu} + (c_1 - c_2)\left(\square^2 + \frac{6}{L^2}\square + \frac{8}{L^4}\right) \psi_{\mu\nu} \\
	&\quad + \frac{\lambda}{\kappa}\left(-\frac{3}{2} + \frac{48 \kappa^2 (c_1 -c _2)}{L^2}\right) \nabla_{\langle \mu}\nabla_{\nu\rangle}\phi +\frac{2}{\kappa}\left(1 - \frac{32 \kappa^2 (c_1 - c_2)}{L^2}\right) \nabla_{\langle \mu} \nabla_{\nu\rangle} \phi~.
\end{aligned}\end{equation}
We can then simply choose the value of $\lambda$ such that all instances of $\nabla_{\langle \mu} \nabla_{\nu \rangle} \phi$ drop out of the Einstein equation.  In particular, if we choose $\lambda = \frac{4}{3}$, then the final expression for the linearized Einstein equation is simply
\begin{equation}
	0 = -\frac{1}{4\kappa^2}\left(\square + \frac{2}{L^2}\right)\psi_{\mu\nu} + (c_1 - c_2)\left(\square^2 + \frac{6}{L^2}\square + \frac{8}{L^4} \right)\psi_{\mu\nu}~.
\end{equation}	
This can be written in a somewhat more canonical form as
\begin{equation}\label{eq:masspsi}
	\left( \square + \frac{2}{L^2} - m_{\psi}^2 \right) \left( \square + \frac{2}{L^2}\right) \psi_{\mu\nu} = 0~, \qquad m_{\psi}^2 L^2 = \frac{L^2}{4 \kappa^2 (c_1 - c_2)} - 2~.
\end{equation}
That is, the transverse traceless part of the metric fluctuations has a massless mode (with two massless propagating degrees of freedom) as well as a massive mode (with five propagating degrees of freedom) with mass $m_{\psi}$.\footnote{Note that this is consistent with the results of~\cite{Lu:2011zk,Smolic:2013gz}. The quantities $\alpha$ and $\beta$ used in these papers can be expressed in terms of the parameters in our supergravity Lagrangian as
\begin{equation}
	\alpha = - 4\kappa^2 (c_1 - c_2)~, \quad \beta = -\frac{\alpha}{3} = \frac{4 \kappa^2 (c_1 - c_2)}{3}~.\notag
\end{equation}}  The massless spin-2 mode corresponds to the stress-energy tensor in the dual SCFT which has conformal dimension $\Delta = 3$. The massive spin-2 mode is dual to a spin-2 operator $\mathcal{O}_\psi$ with conformal dimension
\begin{equation}
	\Delta_{\mathcal{O}_\psi} = \frac{3}{2} + \sqrt{\frac{9}{4} + m_\psi^2 L^2} = \frac{3}{2} + \sqrt{\frac{1}{4} + \frac{L^2}{4 \kappa^2 (c_1 - c_2)}}~.
\label{eq:confdimpsi}
\end{equation}
Importantly, the mass $m_\phi$ and $m_\psi$ of the spin-0 and spin-2 modes in \eqref{eq:massphi} and \eqref{eq:masspsi} are the same, and so are the conformal dimensions of their dual field theory operators.  This is a very non-generic situation, since general higher-derivative corrections to Einstein-Maxwell theory will result in different masses for the spin-2 and spin-0 modes~\cite{Smolic:2013gz}.  This special property of the spectrum is a direct consequence of supersymmetry.

We now move on to the fluctuations of the vector and tensor fields.  To deal with the linearized $T^\pm_{\mu\nu}$ equation of motion, we first apply a derivative to the equation and end up with:
\begin{equation}\begin{aligned}
	0 &= \left(\square + \frac{3}{L^2}\right)(\nabla^\mu t^\mp_{\mu\nu}) - 2 \nabla_\nu \nabla^\rho (\nabla^\mu t^\mp_{\mu\rho}) \\
	& \quad + \frac{1}{4 \kappa^2 (c_1 - c_2)} (\nabla^\mu t^\pm_{\mu\nu}) - \frac{1}{\kappa (c_1 - c_2)}\left(\square + \frac{3}{L^2}\right) w_\nu~.
\end{aligned}\end{equation}
The tensor fluctuations can be dualized to massive vectors by utilizing the following map:
\begin{equation}\label{eq:massivevectcdef}
	\nabla^\mu t^\pm_{\mu\nu} \to \frac{\kappa}{L^2}c^\pm_\nu~,
\end{equation}
where $c^\pm_\mu$ are massive vectors that satisfy $\nabla^\mu c^\pm_\mu = 0$, and we have chosen a convenient normalization in \eqref{eq:massivevectcdef}. This map between a tensor and a massive vector can be justified by the Hodge decomposition theorem. Recall that for a closed Riemannian manifold, a two-form~$t = \tfrac12 t_{\mu\nu} dx^\mu dx^\nu$ is uniquely decomposed into
\begin{equation}
t = d\omega_1 + \delta \omega_3 + \omega_2 \, , 
\end{equation}
where~$\delta$ is the co-differential, $\omega_1$ is a one-form,~$\omega_3$ a three-form, and~$\omega_2$ is harmonic. The divergence of~$t_{\mu\nu}$ is given by the component of the one-form
\begin{equation}
\delta t = - \star d \star t = \nabla^\mu t_{\mu\nu}\,dx^\nu \, .
\end{equation}
Using the above decomposition and the fact that $\delta$ is nilpotent and~$\omega_2$ is harmonic,  we straightforwardly obtain~$\delta t = \delta d \omega_1$. Let~$c \equiv \delta d \omega_1$, then~$c$ is a co-closed one-form, which gives  the map in \eqref{eq:massivevectcdef}. We note that in our setup we need to use the Hodge decomposition theorem on a manifold with boundary, see \cite{hodgebdry} for a review.

We can decompose the fluctuations of the spin-1 $SU(2)_R$ fields as follows
\begin{equation}
	v\ind{_\mu^i_j} = \frac{i \kappa}{L}\sum_{a=1}^3 v^{(a)}_\mu (\sigma_a)\ind{^i_j}~,
\end{equation}
where $\sigma_a$ are the Pauli matrices and $v^{(a)}_\mu$ are three transverse vectors. 

Putting everything together and combining equations appropriately, the linearized vector equations of motion are:
\begin{equation}\begin{aligned}
	0 &= \left(\square + \frac{1}{L^2}\right) w_\mu - \frac{1}{4L^2}\left(c_\mu^+ + c_\mu^-\right) - \frac{1}{L^2} v_\mu^{(3)}~, \\
	0 &= \left(\square + \frac{3}{L^2} - \frac{1}{4\kappa^2 (c_1 - c_2)}\right) v_\mu^{(1)}~, \\
	0 &= \left(\square + \frac{3}{L^2} - \frac{1}{4\kappa^2 (c_1 - c_2)}\right) v_\mu^{(2)}~, \\
	0 &= \left(\square + \frac{3}{L^2} - \frac{1}{4\kappa^2 (c_1 - c_2)}\right) v_\mu^{(3)} - \frac{1}{2\kappa^2 (c_1 - c_2)} w_\mu~, \\
	0 &= \left(\square + \frac{3}{L^2} - \frac{1}{4 \kappa^2 (c_1 -c _2)}\right) a_\mu~, \\
	0 &= \left(\square + \frac{3}{L^2} - \frac{1}{4\kappa^2 (c_1 - c_2)}\right) c^\pm_\mu - \frac{1}{\kappa^2 (c_1 - c_2)}\left(2 w_\mu + v_\mu^{(3)}\right)~.
\end{aligned}\end{equation}
After diagonalizing the corresponding mass matrix, we find the following eigenvectors, i.e. the vectors that are annihilated by $(\square + 3/L^2 - m^2)$ for some value of the mass, and their corresponding eigenvalues:
\begin{equation}\begin{aligned}
	w_\mu - 2 v_\mu^{(3)}: \quad m^2 L^2 &= 0~, \\
	a_\mu : \quad m^2 L^2 &= \gamma~, \\
	v_\mu^{(1)}: \quad m^2 L^2 &= \gamma~, \\
	v_\mu^{(2)}: \quad m^2 L^2 &= \gamma~, \\
	c_\mu^+ - c_\mu^- : \quad m^2 L^2 &= \gamma~, \\
	\frac{1}{2\gamma-1 - \sqrt{4\gamma+1}} w_\mu + \frac{1}{3-\sqrt{4\gamma+1}} v_\mu^{(3)} + c_\mu^+ + c_\mu^-: \quad m^2 L^2 &= \gamma + 1 - \sqrt{4\gamma +1}~, \\
	\frac{1}{2\gamma-1 + \sqrt{4\gamma+1}} w_\mu + \frac{1}{3+\sqrt{4\gamma+1}} v_\mu^{(3)} + c_\mu^+ + c_\mu^-: \quad m^2 L^2 &= \gamma + 1 + \sqrt{4\gamma +1}~,
\end{aligned}\end{equation}
where we have defined 
\begin{equation}\label{eq:gammadef}
\gamma \equiv \frac{L^2}{4\kappa^2 (c_1 - c_2)}
\end{equation}
This means that our fluctuation spectrum includes a single massless vector (with two propagating degrees of freedom) and six massive vectors (each with three propagating degrees of freedom).  Four of these vectors have the same mass, while the remaining two have their mass shifted due to the non-minimal couplings between vector and tensor fields in our four-derivative action.  
 
The massless vector field is dual to the R-current spin-1 operator in the field theory, with conformal dimension $\Delta = 2$.  The four vector fields with $m^2 L^2 = \gamma = \frac{L^2}{4\kappa^2 (c_1 - c_2)}$ are dual to spin-1 operators in the field theory that all have the same conformal dimension as the spin-2 operator $\mathcal{O}_\psi$, namely
\begin{equation}
	\Delta = \frac{3}{2} + \sqrt{\frac{1}{4} + m^2 L^2} = \frac{3}{2} + \sqrt{\frac{1}{4} + \frac{L^2}{4 \kappa^2 (c_1 - c_2)}} = \Delta_{\mathcal{O}_\psi}~,
\end{equation}
while the remaining two vector fields are dual to spin-1 operators whose conformal dimensions are shifted relative to $\mathcal{O}_\psi$:
\begin{equation}
	\Delta = \frac{3}{2} + \sqrt{\frac{1}{4} + m^2 L^2} = \left(\frac{3}{2} \pm 1\right) + \sqrt{\frac{1}{4} + \frac{L^2}{4 \kappa^2 (c_1 - c_2)}} = \Delta_{\mathcal{O}_\psi} \pm 1~.
\end{equation}

The linearized fluctuations above can be organized into supersymmetric multiplets. To do this it is important also to trace the charges of the linearized modes with respect to the graviphoton. We find that the linear combinations $v_\mu^{(1)} \pm i v_\mu^{(2)}$ have charges $\pm 2$ and all other bosonic fields, i.e. the other spin-1 fields as well as the spin-0 and spin-2 fields, are neutral. We find that the bosonic fluctuations above fit into one massless graviton multiplet and one massive long graviton multiplet, see Appendix A of~\cite{Klebanov:2008vq} for more details on this terminology of 4d $\mathcal{N}=2$ supergravity multiplets. These supergravity multiplets are mapped to 3d $\mathcal{N}=2$ superconformal multiplets which we discuss in more detail in Section~\ref{subsec:linspec}.

\section{Boundary counterterms and the boundary stress-tensor}
\label{app:bdry}

In this appendix, we will elaborate on the computation of the boundary stress-tensor $\tau_{\mu\nu}$, defined by
\begin{equation}
	\tau_{\mu\nu} \equiv \frac{2}{\sqrt{-h}} \frac{\delta \mathcal{L}}{\delta h^{\mu\nu}}~.
\end{equation}
In computing this, it is important that we vary both the bulk and boundary counterterms.  We will go through this computation in detail for the two-derivative action $S_{2\partial}$ and $S_\text{GB}$, the results for which appeared in \eqref{eq:tau2d} and \eqref{eq:taugb} in the main text.  Along the way, we will also show that the counterterm actions presented in the main text lead to a well-posed variational principle.  

Throughout this appendix, we will denote by $\mathcal{M}$ the bulk spacetime manifold with boundary $\partial \mathcal{M}$. The induced metric on $\partial \mathcal{M}$ is $h_{\mu\nu}$, which can be expressed as
\begin{equation}
	h_{\mu\nu} = g_{\mu\nu} - n_\mu n_\nu~,
\end{equation}
where $n$ denotes the unit normal with respect to the boundary.  For any tensor $T\ind{^{\mu_1 \ldots \mu_n}_{\nu_1 \ldots \nu_m}}$, we will use the $\perp$ symbol to represent its projection along the boundary, i.e.
\begin{equation}
	\perp\left(T\ind{^{\mu_1 \ldots \mu_n}_{\nu_1 \ldots \nu_m}}\right) = h\ind{^{\mu_1}_{\rho_1}}\ldots h\ind{^{\mu_n}_{\rho_n}} h\ind{_{\nu_1}^{\sigma_1}} \ldots h\ind{_{\nu_m}^{\sigma_m}} T\ind{^{\rho_1 \ldots \rho_n}_{\sigma_1 \ldots \sigma_m}}~.
\end{equation}
We will also denote the extrinsic curvature by
\begin{equation}
	K_{\mu\nu} \equiv \perp\left( \nabla_\mu n_\nu\right)~,
\end{equation}
and we denote by $\mathcal{R}_{\mu\nu\rho\sigma}$ the Riemann tensor corresponding to the induced boundary metric $h_{\mu\nu}$.  We will also use $D_\mu$ to denote the covariant derivative along the boundary compatible with $h_{\mu\nu}$.

We are interested in computing the boundary stress-tensor $\tau_{\mu\nu}$ for the Poincar\'e frame actions in Lorentzian signature.  The relevant ones are the two-derivative action and the Gauss-Bonnet action:
\begin{equation}\begin{aligned}
	S_{2\partial} &= \frac{1}{16\pi G_N}\int_{\mathcal{M}} d^4x\,\sqrt{-g}\,\left(R - \frac{1}{4} F_{\mu\nu} F^{\mu\nu} + \frac{6}{L^2}\right)~, \\
	S_\text{GB} &= \int_{\mathcal{M}} d^4x\,\sqrt{-g}\,E_4~,
\end{aligned}\end{equation}
where $E_4 = R_{\mu\nu\rho\sigma}R^{\mu\nu\rho\sigma} - 4 R_{\mu\nu}R^{\mu\nu} + R^2$ is the usual Gauss-Bonnet density.  These two bulk actions also have their corresponding counterterm actions:
\begin{equation}\begin{aligned}
	S_{2\partial}^\text{CT} &= \frac{1}{8\pi G_N} \int_{\partial \mathcal{M}} d^3x\,\sqrt{-h}\,\left( K - \frac{L}{2} \mathcal{R} - \frac{2}{L}\right)~, \\
	S_{\text{GB}}^\text{CT} &= 4 \int_{\partial \mathcal{M}} d^3x\,\sqrt{-h}\,\left( \mathcal{J} - 2 \mathcal{G}_{\mu\nu} K^{\mu\nu}\right)~,
\end{aligned}\end{equation}
where $\mathcal{J}$ is the trace of the tensor $\mathcal{J}_{\mu\nu}$, defined by
\begin{equation}
	\mathcal{J}_{\mu\nu} = \frac{1}{3} \left(2 K K_{\mu\rho}K\ind{^\rho_\nu} + K_{\mu\nu} K_{\rho\sigma}K^{\rho\sigma} - 2 K_{\mu\rho}K^{\rho\sigma}K_{\sigma\nu} - K^2 K_{\mu\nu}\right)~,
\end{equation}
and $\mathcal{G}_{\mu\nu} = \mathcal{R}_{\mu\nu} - \frac{1}{2} g_{\mu\nu} \mathcal{R}$ is the boundary Einstein tensor.  Armed with these bulk and boundary actions, we now need to compute their variations with respect to the boundary metric in order to determine the corresponding boundary stress-tensors.

First, we tackle the variation of the bulk actions.  Under an infinitesimal variation of the metric $g_{\mu\nu} \to g_{\mu\nu} + \delta g_{\mu\nu}$, the Riemann tensor varies as
\begin{equation}
	\delta R\ind{^\mu_{\nu\rho\sigma}} = g^{\mu\lambda}\left(\nabla_{[\rho|} \nabla_{\nu} \delta g_{\lambda|\sigma] } + \nabla_{[\rho} \nabla_{\sigma]} \delta g_{\nu\lambda} - \nabla_{[\rho|} \nabla_{\lambda} \delta g_{\nu|\sigma]} \right)~.
\end{equation}
This in turn can be used to show that the bulk Ricci scalar varies as
\begin{equation}
	\delta R = - R^{\mu\nu} \delta g_{\mu\nu} + \nabla^\mu\left( \nabla^\nu \delta g_{\mu\nu} - \nabla_\mu g^{\nu\rho} \delta g_{\nu\rho}\right)~.
\label{eq:deltar}
\end{equation}
Additionally, the Gauss-Bonnet density varies as
\begin{equation}\begin{aligned}
	\delta E_4 &= \left(-2 R\ind{^\mu_{\rho\sigma\lambda}}R^{\nu\rho\sigma\lambda} + 4 R^{\mu\rho\nu\sigma} R_{\rho\sigma} + 4 R^{\mu\rho}R\ind{^\nu_\rho} - 2 R R^{\mu\nu}\right) \delta g_{\mu\nu} \\
	&\quad + 4 \nabla_\mu \left( P^{\mu\nu\rho\sigma} \nabla_\sigma \delta g_{\nu\rho}\right)~,
\label{eq:deltae4}
\end{aligned}\end{equation}
where we have defined $P_{\mu\nu\rho\sigma}$ as the divergence-free part of the Riemann tensor:
\begin{equation}
	P_{\mu\nu\rho\sigma} \equiv R_{\mu\nu\rho\sigma} - 2 R_{\mu[\rho} g_{\sigma]\nu} + 2 R_{\nu[\rho} g_{\sigma]\mu} + R g_{\mu[\rho} g_{\sigma]\nu}~, 
\label{eq:pdef}
\end{equation}
such that $\nabla^\mu P_{\mu\nu\rho\sigma} = 0$.  The second term of \eqref{eq:deltar} and the second line of \eqref{eq:deltae4} are both total derivatives, which means that we can employ Stokes' theorem to write the variation of the bulk actions as follows:
\begin{equation}\begin{aligned}
	\delta S_{2\partial} &= \frac{1}{16 \pi G_N} \int_{\mathcal{M}} d^4x\,\sqrt{-g}\,\left( -G^{\mu\nu} + \frac{3}{L^2}g^{\mu\nu} + \frac{1}{2} F^{\mu\rho}F\ind{^\nu_\rho} - \frac{1}{8}g^{\mu\nu} F_{\rho\sigma}F^{\rho\sigma}\right) \delta g_{\mu\nu} \\
	& \quad + \frac{1}{16 \pi G_N} \int_{\partial \mathcal{M}} d^3x\,\sqrt{-h}\,n^\mu \left(\nabla^\nu \delta g_{\mu\nu} - g^{\nu\rho} \nabla_\mu \delta g_{\nu\rho}\right)~, \\
	\delta S_\text{GB} &= -2 \int_{\mathcal{M}} d^4x\,\sqrt{g}\, H^{\mu\nu} \delta g_{\mu\nu}  + 4 \int_{\partial \mathcal{M}} d^3x\,\sqrt{h}\, n_\mu P^{\mu\nu\rho\sigma} \nabla_\sigma \delta g_{\nu\rho}~,
\end{aligned}\end{equation}
where we use $H_{\mu\nu}$ to denote the Lovelock tensor:
\begin{equation}
	H_{\mu\nu} \equiv R\ind{_\mu_{\rho\sigma\lambda}}R\ind{_\nu^{\rho\sigma\lambda}} - 2 R_{\mu\rho\nu\sigma} R^{\rho\sigma} - 2 R_{\mu\rho} R\ind{_\nu^\rho} + R R_{\mu\nu} - \frac{1}{4}g_{\mu\nu}E_4~.
\end{equation}
The bulk component of the two-derivative variation vanishes for any two-derivative solution, due to the Einstein equation, and so only the boundary component of the variation remains.  For the Gauss-Bonnet term, the Lovelock tensor can be rewritten as $H\ind{^\mu_\nu} = \delta^\mu_{[\nu} R\ind{^{\rho\sigma}_{\rho\sigma}} R\ind{^{\lambda\tau}_{\lambda \tau]}}$, and since any fully antisymmetrized product over five distinct indices must vanish in four dimensions, the Lovelock tensor itself must vanish as well.  Therefore we are left only with a boundary term, as required by the well-known topological nature of the Gauss-Bonnet invariant in four dimensions.  As for this remaining boundary term, we can decompose $n_\mu P^{\mu\nu\rho\sigma}$ into its components normal and tangent to the boundary.  The result is
\begin{equation}
	n_\mu P^{\mu\nu\rho\sigma} = \,\perp\left(n_\mu P^{\mu\nu\rho\sigma}\right) + 2 \perp \left( n_\mu n_\lambda P^{\mu \nu \lambda [\sigma}\right) n^{\rho]}~.
\end{equation}
The Gauss-Codazzi relations then inform us that this can be further written as
\begin{equation}\begin{aligned}
	n_\mu P^{\mu\nu\rho\sigma} &= 2 \mathcal{G}^{\nu[\rho} n^{\sigma]} + 2 K^{\mu\nu} K\ind{_\mu^{[\rho}}n^{\sigma]} - 2 K K^{\nu[\rho} n^{\sigma]} + \left( K^2 - K_{\mu\lambda}K^{\mu\lambda}\right) h^{\nu[\rho} n^{\sigma]} \\
	&\quad + 2 D^{[\sigma} K^{\rho]\nu} + 2 D_{\mu} K^{\mu [ \sigma} h^{\rho]\nu} + 2 D^{[\rho} K h^{\sigma]\nu}~,
\end{aligned}\end{equation}
where $D_\mu$ denotes the covariant derivative with respect to the boundary metric $h_{\mu\nu}$.  Thus the total variations of the bulk two-derivative and Gauss-Bonnet actions in four spacetime dimensions are given purely by the following boundary integrals:
\begin{equation}\begin{aligned}
	\delta S_{2\partial} &= \frac{1}{16 \pi G_N} \int_{\partial \mathcal{M}} d^3x\,\sqrt{-h}\,n^\mu \left(\nabla^\nu \delta g_{\mu\nu} - g^{\nu\rho} \nabla_\mu \delta g_{\nu\rho}\right)~, \\
	\delta S_\text{GB} &= 4 \int_{\partial \mathcal{M}} d^3x\,\sqrt{h}\,\bigg{[}2 \mathcal{G}^{\nu[\rho} n^{\sigma]} + 2 K^{\mu\nu} K\ind{_\mu^{[\rho}}n^{\sigma]} + \left( K^2 - K_{\mu\lambda}K^{\mu\lambda}\right) h^{\nu[\rho} n^{\sigma]} \\
	&\quad - 2 K K^{\nu[\rho} n^{\sigma]} + 2 D^{[\sigma} K^{\rho]\nu} + 2 D_{\mu} K^{\mu [ \sigma} h^{\rho]\nu} + 2 D^{[\rho} K h^{\sigma]\nu}\bigg{]} \nabla_\sigma \delta g_{\nu\rho}~.
\label{eq:deltagbc}
\end{aligned}\end{equation}

For the boundary counterterms, we first note that the variation of the normal vector $n_\mu$ is $\delta n_\mu = \frac{1}{2} n_\mu n^\nu n^\rho \delta g_{\nu\rho}$.  From this, it follows that the extrinsic curvature varies as
\begin{equation}
	\delta K_{\mu\nu} = \frac{1}{2} K_{\mu\nu} n^\rho n^\sigma \delta g_{\rho\sigma} + 2 n\ind{_{(\mu}} K\ind{_{\nu)}^\rho} n^\sigma \delta g_{\rho\sigma} - \frac{1}{2} h\ind{_\mu^\rho}h\ind{_\nu^\sigma} n^\lambda \left( 2\nabla_{(\rho} \delta g_{\sigma) \lambda}  - \nabla_\lambda \delta g_{\rho\sigma}\right)~.
\end{equation}
The relevant part of this expression is the projection along the boundary, since all instances of $\delta K_{\mu\nu}$ end up being contracted with tensors that are purely tangent to the boundary.  This tangential component is straightforward to calculate and reads
\begin{equation}
	\perp \delta K_{\mu\nu} = h\ind{_{(\mu}^\rho} h\ind{_{\nu)}^\sigma} n^\lambda \nabla_{[\lambda} \delta g_{\rho]\sigma} - \frac{1}{2} D\ind{_{(\mu}}\left( h\ind{_{\nu)}^\rho} n^\sigma \delta g_{\rho\sigma}\right) + \frac{1}{2} K\ind{_{(\mu}^\rho} h\ind{_{\nu)}^\sigma} \delta g_{\rho\sigma}~.
\end{equation}
We also need the variation of the boundary Ricci tensor $\mathcal{R}_{\mu\nu}$.  This variation is automatically tangential to the boundary, and it is easily computed using the same procedure one would use to find the variation $\delta R_{\mu\nu}$ in the bulk.  The result is simply
\begin{equation}
	\perp \delta \mathcal{R}_{\mu\nu} =D\ind{_{(\mu}} D^\lambda \left(h\ind{_{\nu)}^\rho} h\ind{_\lambda^\sigma} \delta g_{\rho\sigma}\right) - \frac{1}{2} D_{\mu} D_{\nu}\left( h^{\rho\sigma} \delta g_{\rho\sigma}\right) - \frac{1}{2} D^\lambda D_\lambda \left(h\ind{_\mu^\rho}h\ind{_\nu^\sigma} \delta g_{\rho\sigma}\right)~.
\end{equation}
From these expressions, we can also find $\delta K$ and $\delta \mathcal{R}$ by contracting indices appropriately:
\begin{equation}\begin{aligned}
	\delta K &= h^{\mu\nu} n^\rho \nabla\ind{_{[\rho}} \delta g\ind{_{\mu]\nu}} - \frac{1}{2} D_\mu \left( h^{\mu\nu} n^\rho \delta g_{\nu\rho}\right) - \frac{1}{2} K^{\mu\nu} \delta g_{\mu\nu}~, \\
	\delta \mathcal{R} &= D_\mu D_\nu \left( h^{\mu\rho} h^{\nu\sigma} \delta g_{\rho\sigma}\right) - D^\rho D_\rho \left( h^{\mu\nu} \delta g_{\mu\nu}\right) - \mathcal{R}^{\mu\nu} \delta g_{\mu\nu}~.
\end{aligned}\end{equation}

Armed with these intermediate variations, let us first consider the two-derivative counterterm action.  Its variation is written as follows:
\begin{equation}\begin{aligned}
	\delta S_{2\partial}^\text{CT} = \frac{1}{16\pi G_N}\int_{\partial \mathcal{M}} d^3x\,\sqrt{-h}&\,\bigg{[} \left( K - \frac{L}{2}\mathcal{R} - \frac{2}{L}\right) h^{\mu\nu}\delta g_{\mu\nu} - K^{\mu\nu}\delta g_{\mu\nu}  + L \mathcal{R}^{\mu\nu}\delta g_{\mu\nu}  \\
	&\quad - n^\mu \left(\nabla^\nu \delta g_{\mu\nu} - g^{\nu\rho} \nabla_\mu \delta g_{\nu\rho}\right) - D_\mu\left( h^{\mu\nu} n^\rho \delta g_{\nu\rho}\right) \\
	&\quad - L D_\mu D_\nu \left(h^{\mu\rho} h^{\nu\sigma} \delta g_{\rho\sigma}\right) + L D^\rho D_\rho \left(h^{\mu\nu} \delta g_{\mu\nu}\right) \bigg{]}~.
\end{aligned}\end{equation}
The terms normal to the boundary are precisely opposite to the ones present in $\delta S_{2\partial}$ and thus they end up cancelling.  Additionally, we assume that the boundary $\partial \Sigma$ is compact such that we can freely integrate by parts and drop any total (boundary) covariant derivative terms.  Putting all this together, the total variation reads
\begin{equation}
	\delta\left( S_{2\partial} + S_{2\partial}^\text{CT}\right) = \frac{1}{16\pi G_N}\int_{\partial \mathcal{M}} d^3x\,\sqrt{-h}\,\bigg{[} \left( K - \frac{L}{2}\mathcal{R} - \frac{2}{L}\right) h^{\mu\nu} - K^{\mu\nu}  + L \mathcal{R}^{\mu\nu}\bigg{]}\delta g_{\mu\nu}~.
\end{equation}
Everything contracted with $\delta g_{\mu\nu}$ has no components normal to the boundary, and so we can replace $\delta g_{\mu\nu}$ in this variation with $\delta h_{\mu\nu}$.  Therefore the boundary stress-tensor is
\begin{equation}
	\tau_{\mu\nu}^{(2\partial)} = \frac{1}{8\pi G_N}\left( K_{\mu\nu} - K h_{\mu\nu} - L \mathcal{G}_{\mu\nu} + \frac{2}{L} h_{\mu\nu}\right)~.
\end{equation}

For the Gauss-Bonnet counterterm, the computation is a little more involved.  First, by carefully keeping track of all terms and simplifying as much as possible, we compute that
\begin{equation}\begin{aligned}
	\delta \mathcal{J} &= - \frac{3}{2} \mathcal{J}^{\mu\nu} \delta g_{\mu\nu} + \left(2 K K^{\nu\rho} - 2 K^{\nu\lambda}K\ind{^\rho_\lambda} + K_{\mu\lambda}K^{\mu\lambda} h^{\nu\rho} - K^2 h^{\nu\rho}\right) n^\sigma \nabla_{[\sigma} \delta g_{\nu]\rho} \\
	&\quad + \left(K^{\mu\nu}D^\rho K_{\mu\nu} + D_\mu (K K^{\mu\rho}) - K D^\rho K - D_\mu ( K^{\mu\nu}K\ind{^\rho_\nu})\right) n^\sigma \delta g_{\rho\sigma}~,
\label{eq:deltaj}
\end{aligned}\end{equation}
as well as
\begin{equation}\begin{aligned}
	\delta \left(\mathcal{G}_{\mu\nu}K^{\mu\nu}\right) &= \frac{1}{2} K \mathcal{R}^{\mu\nu} \delta g_{\mu\nu} - \frac{1}{2}\mathcal{R} K^{\mu\nu} \delta g_{\mu\nu} - \frac{3}{2} \mathcal{G}^{\mu\rho}K\ind{^\nu_\rho} \delta g_{\mu\nu} + \mathcal{G}^{\nu\rho} n^\sigma \nabla_{[\sigma} \delta g_{\nu]\rho} \\
	&\quad + \left( h\ind{_\nu^\rho} D_\sigma D_\mu K^{\mu\nu} - \frac{1}{2} h^{\rho\sigma} D_\mu D_\nu K^{\mu\nu} - \frac{1}{2}  D_{\lambda}D^\lambda K^{\rho\sigma}\right) \delta g_{\rho\sigma} \\
	&\quad + \left( - \frac{1}{2} D^\mu D^\nu K + \frac{1}{2} h^{\mu\nu}D_\rho D^\rho K\right) \delta g_{\mu\nu}~.
\label{eq:deltagk}
\end{aligned}\end{equation}
Note that in both of these expressions we freely moved derivatives around via integration by parts, keeping in mind that we want these variations to be inside a boundary integral eventually.  Combining \eqref{eq:deltaj}, \eqref{eq:deltagk}, and \eqref{eq:deltagbc}, we find that a number of terms cancel, and we are left with
\begin{equation}\begin{aligned}
	&\delta\left(S_\text{GB} +  S_\text{GB}^\text{CT}\right) = 4 \int_{\partial \mathcal{M}} d^3x\,\sqrt{-h}\,\bigg{[} - \frac{3}{2} \mathcal{J}^{\mu\nu} \delta g_{\mu\nu} + \frac{1}{2} \mathcal{J} h^{\mu\nu} \delta g_{\mu\nu} \\
	&\quad + \left(3 \mathcal{G}^{\mu\rho} K\ind{^\nu_\rho} - \mathcal{G}_{\rho\sigma}K^{\rho\sigma}h^{\mu\nu} + \mathcal{R} K^{\mu\nu} - \mathcal{R}^{\mu\nu} K\right)\delta g_{\mu\nu} \\
	&\quad + \left( K^{\mu\nu} D^\rho K_{\mu\nu} + D_\mu (K K^{\mu\rho}) - K D^\rho K - D_\mu (K^{\mu\nu}K\ind{^\rho_\nu})\right)n^{\sigma} \delta g_{\rho\sigma} \\
	&\quad + \left( D_\lambda D^\lambda K^{\rho\sigma} + h^{\rho\sigma}D_{\mu\nu} K^{\mu\nu} - 2 D^\sigma D_\mu K^{\mu\rho} + D^\rho D^\sigma K - h^{\rho\sigma} D_\lambda D^\lambda K \right) \delta g_{\rho\sigma} \\
	&\quad + \left(2 D\ind{^{[\sigma}} K\ind{^{\rho]\nu}} + 2 D_\mu K\ind{^{\mu[\sigma}} h\ind{^{\rho]\nu}} + 2 D\ind{^{[\rho}} K h\ind{^{\sigma]\nu}}\right) \nabla_\sigma \delta g_{\nu\rho}\bigg{]}~.
\label{eq:vargb}
\end{aligned}\end{equation}
From here, we can use the following Gauss-Codazzi relation:
\begin{equation}
	X^{\mu\nu\rho} \nabla_\rho \delta g_{\mu\nu} = D_\rho \left(X^{\mu\nu\rho}\delta g_{\mu\nu}\right) + 2 X^{(\mu\rho)\sigma}K_{\rho\sigma} n^\nu \delta g_{\mu\nu} - \left(D_\rho X^{\mu\nu\rho}\right)\delta g_{\mu\nu}~,
\end{equation}
which applies for any tensor $X^{\mu\nu\rho}$ that has no components normal to the boundary, i.e. $\perp\left(X^{\mu\nu\rho}\right) = X^{\mu\nu\rho}$.  We will continue to drop all total derivative terms in the boundary action, so this identity lets us trade all instances of $\nabla_\rho \delta g_{\mu\nu}$ in the full Gauss-Bonnet variation \eqref{eq:vargb} for terms that involve no derivatives acting on the metric fluctuation.  After a great deal of simplification, this identity yields
\begin{equation}
	\delta\left(S_\text{GB} +  S_\text{GB}^\text{CT}\right) = \int_{\partial \mathcal{M}} d^3x\,\sqrt{-h}\,\bigg{[} -6 \mathcal{J}^{\mu\nu} + 2 \mathcal{J} h^{\mu\nu} + 4 \mathcal{P}^{\mu\rho\nu\sigma}K_{\rho\sigma}\bigg{]} \delta g_{\mu\nu}~,
\label{eq:vargb2}
\end{equation}
where $\mathcal{P}_{\mu\nu\rho\sigma}$ is the divergence-free part of the boundary Riemann tensor $\mathcal{R}_{\mu\nu\rho\sigma}$, which we can obtain from $P_{\mu\nu\rho\sigma}$ in \eqref{eq:pdef} by replacing all instances of the Riemann tensor with the boundary Riemann tensor and all instances of the metric with the boundary metric.  The total variation \eqref{eq:vargb2} is therefore such that everything contracted with $\delta g_{\mu\nu}$ is tangent to the boundary.  We can therefore freely replace $\delta g_{\mu\nu}$ with $\delta h_{\mu\nu}$ in this expression, and so the resulting boundary stress-tensor is
\begin{equation}
	\tau_{\mu\nu}^{(\text{GB})} = 12 \mathcal{J}_{\mu\nu} - 4 \mathcal{J} h_{\mu\nu} - 8 \mathcal{P}_{\mu\rho\nu\sigma} K^{\rho\sigma}~.
\end{equation}

%%%%%%%%%%%%%%%%%%%%%%%%%%%%%%%%%%%%%
\section{Euclidean off-shell BPS variations}
\label{app:susy-vars}
%%%%%%%%%%%%%%%%%%%%%%%%%%%%%%%%%%%%%

Here we present the Euclidean off-shell BPS variations for the fermions of the superconformal multiplets used in the main text and reviewed in Appendix~\ref{app:sugra}. We then analyze the consequences of demanding that a given field configuration is fully supersymmetric.
 
The (chiral) fermions we must consider are the gravitini~$\psi_\mu{}^i{\!}_\pm$ and the auxiliary~$\chi^i_\pm$ in the Weyl multiplet, the gaugino~$\Omega^i_\pm$ in the vector multiplet, and the hyperino~$\zeta^\alpha_\pm$ in the hypermultiplet. In a bosonic background, their transformations under Q- and S-supersymmetry are given by~\cite{deWit:2017cle}
\begin{align}
\delta\psi_\mu{}^i{\!}_\pm =&\; 2\,\mathcal{D}_\mu \epsilon_\pm^i + \tfrac1{16}\mathrm{i}\,T_{ab}^\mp\,\gamma^{ab} \gamma_\mu \epsilon_\mp^i - \mathrm{i}\,\gamma_\mu \eta_\mp^i \, , \nonumber \\[1mm]
%%%    
\delta\chi_\pm^i=&\; \tfrac1{24}\mathrm{i}\,\gamma^{ab}\,\Slash{\mathcal{D}}T_{ab}^\mp\,\epsilon_\mp^i + \tfrac1{6}R(\mathcal{V})^\mp_{ab}{}^i{}_j \,\gamma^{ab} \epsilon_\pm^j \mp \tfrac13 R(A)^\mp_{ab} \,\gamma^{ab}\epsilon_\pm^i +  D\,\epsilon_\pm^i +\tfrac1{24}\,T_{ab}^\mp\,\gamma^{ab} \eta_\pm^i \, , \nonumber \\[1mm]
%%% 
\delta\Omega^{i}_\pm =&\, -2\mathrm{i}\,\Slash{\mathcal{D}}X_\pm\,\epsilon^i_\mp - \tfrac12\bigl[F_{ab} - \tfrac14\,X_\mp T_{ab}\,\bigr]\gamma^{ab}\epsilon_\pm^i + \varepsilon_{kj}\,Y^{ik}\epsilon^j_\pm + 2\,X_\pm\,\eta^i_\pm \, , \\[1mm]
\delta\zeta^\alpha_\pm =&\, -\mathrm{i}\,\Slash{\mathcal{D}}A_i{}^\alpha\,\epsilon^i_\mp - 2g\,X_\mp\,t^\alpha{}_\beta\,A_i{}^\beta\,\epsilon^i_\pm + A_i{}^\alpha\,\eta^i_\pm \, , \nonumber
\end{align}
where $t^\alpha{}_\beta$ generates the gauging. To analyze the consequences of preserving supersymmetry, one must be mindful of the presence of the S-supersymmetry parameter~$\eta^i_\pm$ in the above variations. An efficient way to deal with this is to introduce auxiliary fermions that transform inhomogeneously under S-supersymmetry~\cite{LopesCardoso:2000qm}. Explicitly, we introduce the following S-compensating fermions:
\begin{equation}
\zeta^{i\,\mathrm{V}}_\pm \equiv \tfrac12\,(X_\pm)^{-1}\,\Omega^{i}_\pm \, ,\quad \text{and} \quad \zeta^{i\,\mathrm{H}}_\pm \equiv -\chi_\mathrm{H}^{-1}\,\varepsilon^{ij}\,\Omega_{\alpha\beta}\,A_j{}^\alpha\,\zeta_\pm^\beta \, ,
\end{equation}
where
\begin{equation}
\chi_\mathrm{H} = \tfrac12\,\varepsilon^{ij}\Omega_{\alpha\beta}\,A_i{}^\alpha A_j{}^\beta \, .
\end{equation}
Their variations are:
\begin{equation}
\begin{split}
\delta\zeta^{i\,\mathrm{V}}_\pm =&\, -\mathrm{i}\,X_\pm^{-1}\Slash{\mathcal{D}} X_\pm\,\epsilon^i_\mp - \tfrac14\,X_\pm^{-1}\bigl[F_{ab}^{\mp} - \tfrac14\,X_\mp T_{ab}^\mp\,\bigr]\gamma^{ab}\epsilon^i_\pm + \tfrac12\,X_\pm^{-1}\,\varepsilon_{kj}\,Y^{ik}\epsilon^j_\pm + \eta^i_\pm \, , \\[1mm]
\delta\zeta^{i\,\mathrm{H}}_\pm =&\, -\tfrac12\,\mathrm{i}\,\Slash{k}\,\epsilon^i_\mp - \mathrm{i}\,\Slash{k}^{\,i}{}_j\,\epsilon^j_\mp + 2g\,X_\mp\,\mu_{kj}\,\varepsilon^{ik}\,\epsilon^j_\pm + \eta^i_\pm \, ,
\end{split}
\end{equation}
where we introduced the moment map
\begin{equation}
\mu_{ij} \equiv \chi_{\mathrm{H}}^{-1}\,A_i{}^\alpha\,\Omega_{\alpha\beta}\,t^\beta{}_\gamma\,A_j{}^\gamma \, ,
\end{equation}
and an~$SU(2)_R$ singlet and triplet through the equation
\begin{equation}
\tfrac12\,k_\mu\,\delta^i{}_j + k_\mu{}^i{}_j = -\chi_\mathrm{H}^{-1}\,\varepsilon^{ik}\,\Omega_{\alpha\beta}\,A_k{}^\alpha \mathcal{D}_\mu A_j{}^\beta \, .
\end{equation}
Note that~$k_\mu = \chi_\mathrm{H}^{-1}\,\mathcal{D}_\mu \chi_\mathrm{H}$, while the explicit expression of~$k_\mu{}^i{}_j$ will not be needed.\\

With the help of~$\zeta^{i\,\mathrm{V}}_\pm$ and~$\zeta^{i\,\mathrm{H}}_\pm$ we can build S-invariant combinations of fermions. We can then derive the consequences of requiring that the BPS variations of such combinations vanish in order to preserve supersymmetry. The transformations we analyze are
\begin{align}
\label{eq:compensators}
\delta\bigl(\zeta^{i\,\mathrm{V}}_\pm - \zeta^{i\,\mathrm{H}}_\pm\bigr) =&\; \tfrac12\mathrm{i}\,\bigl(X_\pm^{-2}\,\chi_{\mathrm{H}}\bigr)^{-1}\,\Slash{\mathcal{D}}\bigl(X_\pm^{-2}\,\chi_{\mathrm{H}}\bigr)\,\epsilon^i{\!}_\mp + \mathrm{i}\,\slash{k}{\,}^i{}_j\,\epsilon^j{\!}_\mp  + \tfrac12\,X_\pm^{-1}\,\varepsilon_{kj}\,Y^{ik}\epsilon^j_\pm \nonumber \\
& - \tfrac14\,X_\pm^{-1}\bigl[F_{ab}^{\mp} - \tfrac14\,X_\mp T_{ab}^\mp\,\bigr]\gamma^{ab}\,\epsilon^i_\pm  - 2g\,X_\mp\,\mu_{kj}\,\varepsilon^{ik}\,\epsilon^j_\pm \, , \\
\label{eq:hyper}
\delta\bigl(\zeta^\alpha_\pm - A_i{}^\alpha \zeta^{i\,\mathrm{H}}_\pm\bigr) =&\, -\mathrm{i}\,\chi_{\mathrm{H}}^{1/2}\,\Slash{\mathcal{D}}\bigl(\chi_{\mathrm{H}}^{-1/2} A_i{}^\alpha\bigr)\,\epsilon^i_\mp + \mathrm{i}\,\slash{k}{\,}^i{}_j A_i{}^\alpha\epsilon^j_\mp \nonumber \\
&\, - 2g\,X_\mp\bigl(t^\alpha{}_\beta\,A_j{}^\beta + \mu_{kj}\,\varepsilon^{ik}\,A_i{}^\alpha\bigr)\,\epsilon^j_\pm\, , \\
\label{eq:dilatino}
\delta\bigl(\chi^i{\!}_\pm - \tfrac1{24}\,T_{ab}^\mp\gamma^{ab}\zeta^{i\,\mathrm{H}}_\pm\bigr) =&\; D\,\epsilon^i_\pm+ \tfrac1{24}\mathrm{i}\,\chi_\mathrm{H}^{-1/2}\,\mathcal{D}_\mu\bigl(\chi_\mathrm{H}^{1/2}T_{ab}^\mp\bigr)\,\gamma^{ab}\gamma^\mu\,\epsilon^i_\mp + \tfrac1{24}\mathrm{i}\,T_{ab}^\mp\,k_\mu{}^i{}_j\,\gamma^{ab}\gamma^\mu\,\epsilon^j_\mp \nonumber \\
& + \tfrac16\,R(\mathcal{V})^\mp_{ab}{}^i{}_j\,\gamma^{ab}\,\epsilon^j_\pm \mp \tfrac13\,R(A)^\mp_{ab}\,\gamma^{ab}\,\epsilon^i_\pm - \tfrac1{12}\,g\,X_\mp\,T_{ab}^\mp\,\mu_{kj}\,\varepsilon^{ik}\,\gamma^{ab}\,\epsilon^j_\pm \, .
\end{align}
In addition, rather than working with the gravitini themselves, it will be more convenient to work with the Q-supersymmetry curvature given in~\eqref{eq:curvatures}. Note that this amounts to studying the integrability condition of the Q-supersymmetry parameter. Compensating for the S-transformation, the relevant BPS variation is
\begin{equation}
\begin{split}
\label{eq:RQ}
\delta\bigl(R(Q)_{ab}{}^i{\!}_\pm - \tfrac1{16}T_{cd}^\mp\,\gamma^{cd}\,\gamma_{ab}\,\zeta^{i\,\mathrm{H}}_\pm\bigr) =&\; \tfrac14\mathrm{i}\,\chi_{\mathrm{H}}^{1/2}\,\Slash{\mathcal{D}}\bigl(\chi_{\mathrm{H}}^{-1/2}T_{ab}^\mp\bigr)\,\epsilon^i_\mp + \tfrac1{8}\mathrm{i}\,T^\mp_{[a}{}^c\,\gamma_{b]c}\,\slash{k}\,\epsilon^i_\mp \\
&\;+ R(\mathcal{V})^\mp_{ab}{}^i{}_j\,\epsilon^j_\pm - \tfrac12\,\mathcal{R}(M)_{ab}{}^{cd}\gamma_{cd}\,\epsilon^i_\pm \\
&\;- \tfrac18\,g\,T_{cd}^\mp\,\gamma^{cd}\,\gamma_{ab}\,X_\mp\,\mu_{kj}\,\varepsilon^{ik}\,\epsilon^j_\pm + \ldots \, ,
\end{split}
\end{equation}
where the~\ldots~are terms proportional to~$k_\mu{}^i{}_j$. We do not write them explicitly as we will see that they vanish on the supersymmetric configurations considered below. 
The last BPS variation to consider comes from the supercovariant derivative of the auxiliary fermion~$\zeta^{i\,\mathrm{H}}_\pm$. This condition is needed to ensure that the S-supersymmetry has been compensated properly~\cite{LopesCardoso:2000qm}. It is given by
\begin{align}
\label{eq:compensator-deriv}
&\,\delta\bigl(D_\mu\zeta^{i\,\mathrm{H}}_\pm - \tfrac14\,\Slash{k}\gamma_\mu\zeta^{i\,\mathrm{H}}_\pm - \mathrm{i}\,\varepsilon^{ij}X_\mp\,\mu_{jk}\,\gamma_\mu\,\zeta^{k\,\mathrm{H}}_\mp \bigr) = \nonumber \\
&\;\; - \mathrm{i}\,f_\mu{}^a\,\gamma_a\,\epsilon^i_\mp + \tfrac18\mathrm{i}\,R(\mathcal{V})^\mp_{ab}{}^i{}_j\,\gamma^{ab}\gamma_\mu\,\epsilon^j_\mp \mp \tfrac14\mathrm{i}\,R(A)^\mp_{ab}\,\gamma^{ab}\,\gamma_\mu\,\epsilon^i_\mp - \tfrac1{16}\mathrm{i}\,g\,X_\mp\,\mu_{kj}\,\varepsilon^{ik}\,T^\mp_{ab}\,\gamma^{ab}\gamma_\mu\,\epsilon^j_\mp \nonumber \\
&\;\; - \tfrac1{32}\,\chi_{\mathrm{H}}^{-1/2}\mathcal{D}_\nu\bigl(\chi_{\mathrm{H}}^{1/2}T_{ab}\bigr)\,\gamma^\nu\gamma^{ab}\gamma_\mu\,\epsilon^i_\pm + 2\mathrm{i}\,g^2 X_+\,X_-\,\mu_{kj}\,\mu^{ik}\,\gamma_\mu\,\epsilon^j_\mp \\ 
&\;\; -\tfrac12\mathrm{i}\,\mathcal{D}_\mu k_\nu\,\gamma^\nu\,\epsilon^i_\mp +\tfrac18\mathrm{i}\,k_\nu\,k_\rho\,\gamma^\nu\gamma_\mu\gamma^\rho\,\epsilon^i_\mp - g\,k_\mu X_\mp\,\mu_{kj}\,\varepsilon^{ik}\,\epsilon^j_\pm + \ldots \, , \nonumber 
\end{align}
Again, the~\ldots~stands for terms proportional to $k_\mu{}^i{}_j$ which we will not need. Recall that~$f_\mu{}^a$ is the gauge field of special conformal transformations, which contains the Ricci scalar and tensor once the conventional constraints are imposed, see Appendix~\ref{app:sugra}.

%%%%%%%%%%%%%%%%%%%%%%%%%%%%
\subsection*{Full-BPS conditions}
\label{app:full-BPS}
%%%%%%%%%%%%%%%%%%%%%%%%%%%%

For a bosonic field configuration to be full-BPS, we require the variations above to vanish for \emph{any} spinor parameter~$\epsilon_\pm^i$. It will turn out this can be satisfied only on a unique background, pure (Euclidean) AdS$_4$. Since $\epsilon_\pm^i$ is unconstrained, we can decompose the BPS variations according to the number of~$\gamma$-matrices they contain. From~\eqref{eq:compensators} we find the conditions
\begin{equation}
\mathcal{D}_\mu\bigl(X_\pm^{-2}\,\chi_\mathrm{H}\bigr) = 0 \, , \quad k_\mu{}^i{}_j = 0 \, , \quad F^\mp_{ab} = \tfrac14\,X_\mp\,T_{ab}^\mp \, , \quad Y_{ij} = 4\,g\,X_+\,X_-\,\mu_{ij} \, .
\end{equation}
From~\eqref{eq:hyper} we obtain
\begin{equation}
\mathcal{D}_\mu\bigl(\chi_\mathrm{H}^{-1/2}A_i{}^\alpha\bigr) = 0 \, , \qquad t^\alpha{}_\beta\,A_j{}^\beta + \varepsilon^{ik}\mu_{kj}\,A_i{}^\alpha = 0 \, ,
\end{equation}
and from~\eqref{eq:dilatino} we get
\begin{equation}
\mathcal{D}^a\bigl(\chi_\mathrm{H}^{1/2}\,T_{ab}^\mp\bigr) = 0 \, , \quad R(A)_{ab}^\mp = 0 \, , \quad D = 0 \, , \quad R(\mathcal{V})^\mp_{ab}{}^i{}_j = \tfrac12\,g\,X_\mp\,T_{ab}^\mp\,\varepsilon^{ik}\mu_{kj} \, .
\end{equation}
Observe that the requirement of maximal supersymmetry imposes that~$D$ vanishes \emph{off-shell}, or in other words that~$D=0$ at any order in a derivative expansion of the action. From~\eqref{eq:RQ} we find the additional conditions
\begin{equation}
\mathcal{D}_c T_{ab}^\mp = \tfrac12\,k_d\,\bigl(\delta^d{}_c\,T_{ab}^\mp - 2\,\delta^d{}_{[a} T_{b]c}^\mp + 2\,\eta_{c[a}T_{b]}^{\mp\,d}\,\bigr) \, , \quad \mathcal{R}(M)_{ab}{}^{cd} = 0 \, , \quad g\,X_\mp\,T_{ab}^\mp\,\varepsilon^{ik}\mu_{kj} = 0 \, .
\end{equation}
At this stage, it is convenient to fix the K- and D-gauge as in the main text~\eqref{eq:K-gauge} and~\eqref{eq:D-gauge}. Since~$\chi_\mathrm{H}$ is only charged under dilatations, it is clear that this gauge choice fixes
\begin{equation}
\label{eq:KD-gauge}
k_\mu = 0 \, .
\end{equation}
In this gauge, the additional conditions we obtain from~\eqref{eq:compensator-deriv} simplify rather drastically to a single condition on the scalar curvature
\begin{equation}
R = 24\,g^2\,X_+\,X_-\,\mu^{ij}\mu_{ij} \, .
\end{equation}
The above gauge choice also simplifies the analysis of the BPS conditions. First, since the~$A_\mu$ connection is constrained to be pure gauge we can locally set it to zero, and this implies that the scalars~$X_\pm$ must be constant. Second, when~$g \neq 0$, the vanishing of the tensor~$X_\mp\,T_{ab}^\mp$ implies that the~$\mathcal{V}_\mu{}^i{}_j$ connection is also pure gauge, and that the field strength~$F_{ab}$ vanishes. This in turn forces the scalars~$A_i{}^\alpha$ to be constant. We can fix their value by choosing the V-gauge~\eqref{eq:V-gauge}, and this gauge choice fixes the moment map in terms of the gauging generators
\begin{equation}
\mu_{ij} = \varepsilon_{ik}\,t^k{}_j \, .
\end{equation}
The remaining off-shell BPS conditions then impose that the space-time is conformally flat with constant curvature, and that the auxiliary triplet~$Y_{ij}$ is constant:
\begin{equation}
\label{eq:full-BPS}
Y_{ij} = 4\,g\,X_+\,X_-\,\varepsilon_{ik}\,t^k{}_j \, , \quad R = 48\,g^2\,X_+\,X_- \, , \quad C_{abcd} = 0 \, .
\end{equation}

We end the full-BPS analysis by noting that the above conditions are stronger than, and therefore imply, the equations of motion for the superconformal fields~$A_\mu$,~$\mathcal{V}_\mu{}^i{}_j$ and~$T_{ab}$ derived in the main text from the four-derivative action~\eqref{eq:SCHD-chi}. On the other hand, the~$Y$ equation of motion~\eqref{eq:Y-EoM} gives us the precise value of the constant scalars~$X_\pm = \tfrac12\,\kappa^{-1}$, which in turn fixes the scalar curvature on-shell,\footnote{Note that we have \emph{not} implemented the redefinitions of Footnote~\ref{foot:redef} in this discussion in order to use the conventions of the (gauged) supergravity literature for the BPS variations. Hence the Ricci scalar appears constant and positive in these conventions.}
\begin{equation}
R = 12\,L^{-2} \, ,
\end{equation}
with~$L^{-2} = g^2\,\kappa^{-2}$. Alternatively, we could use the above as the definition for~$L^{-2}$ which would fix the scalars via~\eqref{eq:full-BPS} and imply the equation of motion for the~$Y^{ij}$ triplet.

%%%%%%%%%%%%%%%%%%%%%%%%%%%%%%%%%%%%%%%%%%

\bibliography{HD-AdS4-Holo}
\bibliographystyle{JHEP}

\end{document}